\documentclass{aa}  
\usepackage{dcolumn}
\usepackage{multirow}
\usepackage{subcaption}
\usepackage{graphicx}
\usepackage{txfonts}
\usepackage[colorlinks=true,
            urlcolor=blue,
            linkcolor=blue,
            citecolor=blue]
            {hyperref}
\newcommand{\Ha}{H$\alpha$}
\newcommand{\Hb}{H$\beta$}
\newcommand{\Hg}{H$\gamma$}

\newcommand{\mcc}[1]{\multicolumn{1}{c}{#1}}

\newcommand{\mcr}[1]{\multicolumn{1}{r}{#1}}
\newcommand{\kms}{km\,s$^{-1}$}

\begin{document}

    \title{The Great Slump: Mrk\,926 reveals discrete and varying Balmer line satellite components during a drastic phase of decline
    \thanks{Based on observations obtained with the Hobby-Eberly Telescope.}
    \thanks{The reduced spectra and the photometric light curve are only available in electronic form at the CDS via anonymous ftp to \url{cdsarc.u-strasbg.fr} (130.79.128.5) or via \url{http://cdsweb.u-strasbg.fr/cgi-bin/qcat?J/A+A/}.}}

   \author{Wolfram Kollatschny \inst{1}, 
           Martin W. Ochmann \inst{1,2}, 
           Shai Kaspi \inst{3},  
           Claas Schumacher \inst{1},
           Ehud Behar\inst{4}, 
           Doron Chelouche \inst{5}, 
           Keith Horne \inst{6},
           Bj\"orn M\"uller \inst{1,7},
           Stephen E. Rafter \inst{4},
           Rolf Chini   \inst{2,8},
           Martin Haas \inst{2}
           and
           Malte A. Probst \inst{1}
          }

   \institute{Institut f\"ur Astrophysik, Universit\"at G\"ottingen,
              Friedrich-Hund Platz 1, D-37077 G\"ottingen, Germany\\
              \email{wkollat@astro.physik.uni-goettingen.de}
        \and
        Astronomisches Institut, Ruhr-Universit\"at Bochum,
        Universit\"atsstrasse 150, 44801 Bochum, Germany
        \and
        School of Physics \& Astronomy and the Wise Observatory,
        The Raymond and Beverly Sackler Faculty of Exact Sciences
        Tel-Aviv University, Tel-Aviv 69978, Israel
        \and
        Technion, Institute of Technology, Haifa 3200003, Israel        
        \and
        Physics Department and the Haifa Research Center for Theoretical Physics and
        Astrophysics, University of Haifa, Haifa 3498838, Israel
        \and
        SUPA School of Physics and Astronomy, University of St.~Andrews, St.~Andrews, Fife, KY16 9SS, UK
        \and    
        Max-Planck-Institut f\"ur Sonnensystemforschung, Justus-von-Liebig-Weg 3, D-37077 G\"ottingen, Germany
        \and
        Instituto de Astronom\'{i}a, Universidad Cat\'{o}lica del Norte, Avenida Angamos 0610, Antofagasta, Chile
}

   \date{Received August 12, 2021 ; Accepted October 11, 2021}
   \authorrunning{Kollatschny et al.}
   \titlerunning{Balmer line variability in Mrk\,926}

 
  \abstract
   {}
   {Mrk\,926 is known to be a highly variable active galactic nucleus. Furthermore, it is known to show very broad-line profiles. We intended to study the continuum and line profile variations of this object with high temporal resolution in order to determine its broad-line region (BLR) structure and to derive its black hole mass.}
   {We carried out a  high-cadence spectroscopic variability campaign of Mrk\,926 with the 10m HET telescope, aided by photometric V-band data taken with the C18 telescope at the Wise Observatory, over a period of about five months. We extracted spectroscopic continuum and line light curves, and computed cross-correlation functions (CCFs) as well as velocity-resolved CCFs with respect to the combined spectroscopic and photometric V-band light curve.
   }  
   {The continuum luminosity of Mrk\,926 showed a drastic decrease during our campaign. The luminosity dropped to less than 50\% of its original luminosity within only 2.5 months. Furthermore, the spectra of Mrk\,926 show complex and very broad Balmer line profiles, including outer Balmer satellites ranging from $\pm5000$ to $\pm 13\,000$ \kms.
   The integrated \Ha{}, \Hb{}, and \ion{He}{i}\,$\lambda 5876$  line light curves are delayed relative to the continuum light curve.
   The \Ha{} and \Hb{} lines show two velocity-delay structures in the central part of their line profile (within $\pm 5000$\, \kms{}), at $\sim 10$ and $\sim 57$ light-days and at $\sim 5$ and $\sim 48$  light-days, respectively. These structures might be interpreted as the upper and lower halves of an ellipse in the velocity-delay plane, which might be the signature of a line-emitting ring, inclined by $\sim50^\circ$ to the line of sight and orbiting the black hole at radii, $R$, of 33.5 and 26.5 light-days.
   We determined continuum luminosities, $\log( \lambda\,L_\lambda/{\rm erg~s}^{-1}),$ of 43.68 to 44.13, which are in  good agreement with the established $R_\text{BLR}-L_\text{AGN}$ relation.
   Adopting delays of 33.5 and 26.5 days for \Ha{} and \Hb{}, respectively, we derive a black hole mass of $(1.1 \pm 0.2) \times 10^8 M_{\odot}$; this indicates a low Eddington ratio, which decreased from 8 to 3 percent during our campaign.
   The Balmer satellite components show a higher correlation coefficient  with respect to the continuum than the central line profile, and their response to the continuum variations is on the order of only $3-5$ days. We attribute this to the central line segment and the Balmer satellites having different, spatially distinct regions of origin.
   } 
    {}

   \maketitle
%

\section{Introduction}
Active galactic nuclei (AGN) are variable in all frequency bands on timescales of days to years. It is generally accepted that the variability of the central ionizing continuum source causes line intensity variations in the broad emission lines. Furthermore, changes in the kinematics and structure in the broad-line region (BLR) as well as obscuration can induce variations in the line profiles. However, many details of the innermost BLR, such as size, structure, and kinematics, are poorly understood. In addition, most of the variability campaigns carried out in the past were devoted to Seyfert galaxies that show broad emission lines with full widths at half maximum (FWHMs) of 1000 to 4000 km\,s$^{-1}$, and only very few reverberation campaigns have been carried out for AGN with very broad lines \citep[FWHM $\gg$ 4000 \kms{}; e.g., 3C390.3:][and other AGN therein]{sergeev02, gezari07}. Broad-line reverberation studies exist for more than 50 Seyfert galaxies \citep[e.g.,][and references therein]{peterson02, kaspi05, denney09, barth11, bentz13, shapovalova13, shen16, du18}. However, detailed velocity-resolved reverberation mapping studies have been carried out for only about a dozen Seyfert galaxies (e.g., Mrk\,110: \cite{kollatschny03}; Arp\,151: \cite{bentz08}; Mrk\,335, Mrk\,1501, 3C\,120, PG\,2130: \cite{grier13};  3C120: \cite{kollatschny14}; Mrk\,335, Mrk\,1044, Mrk\,142: \cite{xiao2018}; HE1136-2304: \cite{kollatschny18}; NGC\,5548:
\cite{horne21} and references therein).

\begin{figure}[h!]
\centering
\includegraphics[width=0.32\textheight,angle=0]{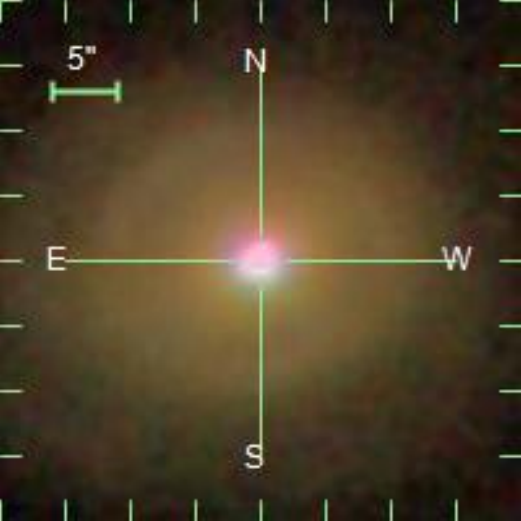}
\caption[]{Composite SDSS image of Mrk\,926 taken from the SDSS SkyServer\footnotemark. The crosshairs are centered on the nucleus.}
\label{fig:mrk926_SDSS_image}
\end{figure}

 Mrk\,926 can be classified as either a strong Seyfert 1 galaxy or a quasar based on its luminosity during the campaign presented here (see Sect.~\ref{sec:eddington_ratio}). Its Balmer and \ion{He}{i} lines have widths (FWHM) of more than 5000 km s$^{-1}$. In a study on the long-term variability of very broad-line AGN \citep{kollatschny06}, Mrk\,926 additionally exhibited strong H$\beta$ line variability amplitudes. The nuclear activity in Mrk\,926 was detected by \cite{ward78} when they took a spectrum of the optical counterpart of a strong X-ray source. Mrk\,926 is a compact source on direct images in the optical \citep{garnier96}. We show an SDSS \citep[DR16:][]{sdss20} image of Mrk\,926 in Fig.~\ref{fig:mrk926_SDSS_image}. Based on SDSS r-band images, it has an isophotal diameter (major axis) of 57.12 arcsec. This corresponds to 53\,kpc. The deVaucouleurs diameter amounts to 6.00 arcsec, corresponding to\,5.56 kpc\footnote{\url{https://ned.ipac.caltech.edu/}}.
Mrk\,926 is the brighter galaxy member (R.A.~=~23:04:43.5, Dec.~=~$-$08:41:09 (2000), $z=0.04701$) in a double system  with PGC\,1000273. This close starburst companion is located one arcmin to the south. 

\footnotetext{\url{https://skyserver.sdss.org}}
\cite{osterbrock82} published the first emission-line profiles of Mrk\,926. They reported that the lines did not vary between July 1978 and December 1979. We were able to show that Mrk\,926 varied by a factor of 2 in the continuum as well as by a factor of 2.5 in the Balmer lines over a period of eight years between 1990 and 1997 \citep{kollatschny06}. In a follow-up paper \citep{kollatschny10}, we focused on line profile variations in Mrk\,926 as the line profiles of AGN and their variations can give us information about the structure of the line-emitting region. We intensively studied the variations in individual line segments of the Balmer lines and the changes in the profiles during a variability campaign in 2005 and another shorter campaign conducted
in 2004. 

Here, we selected Mrk\,926 (MCG-2-58-22) as a target for an even more detailed, combined spectroscopic and photometric variability study of an AGN that shows very broad emission lines. The paper is structured as follows: In Sect.~\ref{sec:observations_reduction} we describe the observations and data reduction. The results of the variability and reverberation mapping study are presented in Sect.~\ref{sec:results}. We discuss the results in Sect.~\ref{sec:discussion} and give a summary in Sect.~\ref{sec:summary}. Throughout this paper, we assume a $\Lambda$ cold dark matter cosmology with a Hubble constant of $H_0$~=~73~km s$^{-1}$ Mpc$^{-1}$, $\Omega_{\Lambda}$~=~0.73, and $\Omega_{\rm M}$~=~0.27. With a redshift of  $z=0.04701,$ this results in a luminosity distance of $D_{\rm L}$~=~200.2 Mpc~=~6.18 $\times 10^{26}$~cm using the cosmology calculator developed by \cite{wright06}.

\section{Observations and data reduction}\label{sec:observations_reduction}
\subsection{Spectroscopic observations with HET}\label{sec:spectroscopic_observations}
We took optical spectra of Mrk\,926 with the 9.2\,m Hobby-Eberly Telescope (HET) located at the McDonald Observatory at 31 epochs between 2010 July 19 and November 19. Thus, we obtained 31 spectra over a period of 122.7 days. The date of the campaign was fixed by the allocated observing time slots. The average interval between the observations was 4.1 days. Excluding the first two observations, the average time interval between observations reduces to 2.7 days with a majority of consecutive observations only 1-2 days apart. The log of our spectroscopic observations is given in Table~\ref{log_of_HET_obs}. All spectroscopic observations were taken with identical instrumental setups. We used the Marcario Low Resolution Spectrograph (LRS) attached at the prime focus of HET. The detector was a 3072x1024 15 $\mu$m pixel Ford Aerospace charge-coupled device (CCD) with 2x2 binning. The spectra cover the wavelength range from 4160\,\AA\ to 6930~\AA\ (LRS grism 2 configuration) in the rest frame of the galaxy (and 4350\,\AA\ to 7250~\AA\ observed frame) with a resolving power of 650 at 5000\,\AA\ (7.7\,\AA\ FWHM). Nearly all observations were taken with an exposure time of 20 minutes, which in most cases yielded a S/N of $> 100 $ per resolution element in the continuum. The slit width was fixed to 2\arcsec\hspace*{-1ex}.\hspace*{0.3ex}0 projected on the sky at an optimized position angle to minimize differential refraction. Furthermore, all observations were taken at the same airmass owing to the particular design of the HET. Typical seeing values ranged from 1\arcsec\hspace*{-1ex}.\hspace*{0.3ex}4 to 1\arcsec\hspace*{-1ex}.\hspace*{0.3ex}9. 

In addition to the spectra of Mrk\,926, necessary calibration images including bias, flat, and arc frames (CdHgAr and Xe), as well as spectrophotometric standard stars (Hiltner\,600) for flux calibration were taken. We reduced the spectra (bias subtraction, cosmic ray correction, flat-field correction, 2D-wavelength calibration, night sky subtraction, flux calibration, etc.) in a homogeneous way with standard \textsc{IRAF}\footnote{\url{https://iraf.net/}} reduction packages \citep{tody86, tody93} like we did in previous publications \citep[e.g.,][]{kollatschny01}. In addition, we corrected for atmospheric absorption in the oxygen B band by fitting the continuum of the standard star Hiltner\,600 and using this fit to normalize its spectrum. In this way, we created a template of the oxygen absorption that we scaled to each object spectrum individually. The spatial resolution per binned pixel  was  0\arcsec\hspace*{-1ex}.\hspace*{0.3ex}472. We extracted four columns in our object spectra. This conforms to the slit width.
Great care was taken to ensure high-quality intensity and wavelength calibrations to keep the intrinsic measurement errors very low \citep[e.g.,][]{kollatschny01, kollatschny10}.
The spectra were not always taken under photometric conditions. Therefore, we calibrated all spectra to the same absolute forbidden narrow-line fluxes under the assumption of a non-varying narrow-line flux on timescales of years. We chose [\ion{O}{iii}]\,$\lambda$4959 and [\ion{O}{iii}]\,$\lambda$5007, as well as [\ion{O}{i}]\,$\lambda$6300/63, [\ion{N}{ii}]\,$\lambda$6548/84, and [\ion{S}{ii}]\,$\lambda$6717/31  as calibration lines for the blue and red section of the spectrum, respectively. The blue part of the spectrum was scaled to an [\ion{O}{iii}]\,$\lambda$5007 flux of $3.14 \times 10^{-13} \rm erg\,s^{-1}\,cm^{-2}$. The same flux value has been used before for Mrk\,926 in \cite{kollatschny06} and \cite{kollatschny10}. \cite{durret88} and \cite{morris88} derived similar [\ion{O}{iii}]\,$\lambda$5007 fluxes from their Mrk\,926 spectra: $2.05 \times 10^{-13} \rm erg\,s^{-1}\,cm^{-2}$ and $3.31 \times 10^{-13} \rm erg\,s^{-1}\,cm^{-2}$, respectively. We note that hereafter all continuum flux densities and line fluxes at wavelengths  $>~6500$~\AA\ (observed frame) refer to the intercalibration with respect to [\ion{O}{i}]\,$\lambda$6300/63,  [\ion{N}{ii}]\,$\lambda$6548/84, and [\ion{S}{ii}]\,$\lambda$6717/31.
The accuracy of the flux calibration was tested for all forbidden emission lines in the spectra. We calculated difference spectra for all epochs with respect to the mean spectrum of our variability campaign. Corrections for small spectral shifts ($<$ 0.5 \AA ) and small scaling factors were implemented by minimizing the residuals of the narrow emission lines in the  difference spectra. A relative flux accuracy of 1 -- 2\% was obtained for most of our spectra.\\
\begin{table}[!h]
\tabcolsep+6mm
\caption{Log of spectroscopic HET observations of Mrk\,926.}
\centering
\begin{tabular}{ccc}
\hline \hline 
\noalign{\smallskip}
Mod. Julian Date & UT Date & Exp. time \\
            &         &  [sec.]   \\
\noalign{\smallskip}
\hline 
\noalign{\smallskip}
55396.41  &       2010-07-19      &      1200   \\
55397.42  &       2010-07-20      &      1200   \\      
55441.31  &       2010-09-02      &      1200   \\
55443.30  &       2010-09-04      &      1200   \\
55444.30  &       2010-09-05      &      1200   \\
55445.28  &       2010-09-06      &      1200   \\
55446.29  &       2010-09-07      &      1200   \\
55451.27  &       2010-09-12      &      1200   \\
55452.26  &       2010-09-13      &      1200   \\
55453.26  &       2010-09-14      &      1200   \\
55467.23  &       2010-09-28      &      1800   \\
55470.20  &       2010-10-01      &      1200   \\
55471.21  &       2010-10-02      &      1200   \\
55472.22  &       2010-10-03      &      1200   \\
55477.20  &       2010-10-08      &      1200   \\
55480.20  &       2010-10-11      &      1800   \\
55481.20  &       2010-10-12      &       951   \\
55483.20  &       2010-10-14      &      1300   \\
55494.16  &       2010-10-25      &      1200   \\
55496.15  &       2010-10-27      &      1200   \\
55497.16  &       2010-10-28      &      1200   \\
55498.17  &       2010-10-29      &      1200   \\
55499.16  &       2010-10-30      &      1200   \\
55501.14  &       2010-11-01      &      1200   \\
55505.12  &       2010-11-05      &      1200   \\
55506.12  &       2010-11-06      &      1200   \\
55507.11  &       2010-11-07      &      1200   \\
55508.11  &       2010-11-08      &      1200   \\
55514.12  &       2010-11-14      &      1200   \\
55515.10  &       2010-11-15      &      1200   \\
55519.10  &       2010-11-19      &      1200   \\
\hline 
\vspace{-.7cm}
\end{tabular}
\label{log_of_HET_obs}
\end{table}
\subsection{Photometric observations at the Wise Observatory}\label{sec:photometric_observations}

Mrk\,926 was observed with a Bessell V-band filter \citep{bessell90} at the C18 telescope \citep{brosch08} of the Wise Observatory in Israel, from 2010 August 28 to December 10. Observations were carried out daily with gaps due to bad weather on some nights. The SBIG STL-6303 CCD was used with 3072$\times$2048  9\,$\mu$m pixel, 
and plate-scale of 1.47 arcsec\,pix$^{-1}$, which gives a field of view of  75$\times$50 arcmin. Each night several exposures of 5 min each were obtained. The V-band filter has a bandwidth ranging from 4700\,\AA\ to 6500\,\AA\ and an effective central wavelength of 5415\AA. The contribution of the total \Hb{} flux to the V band can be estimated to be $<10$\%. Since the variable part of \Hb{} only accounts for 30 -- 40 \% of the total \Hb{} flux during our campaign (see Table~\ref{variab_statistics}) the variable \Hb{} contribution to the V band is $<4$\% and is therefore negligible.

Data reduction was carried out in the standard way for bias, dark, and flat-field, using standard \textsc{IRAF} routines. In order to measure the instrumental magnitude of the objects in each image we used the DAOPHOT package \citep{stetson87} as implemented in IRAF. We derived the point-spread function for each image and measured all available objects in the image. We then used the DAOSTAT program \citep{netzer96} to choose the non-variable stars in the field and use them to inter-calibrate by differential photometry the instrumental magnitude of Mrk\,926 between the images and to get its light curve. Measurements from the same night were averaged to obtain a single measurement for each night in order to decrease the uncertainty on each point in the light curve.

\begin{figure*}[!htp]
\centering
    \begin{minipage}[l]{\textwidth}
        \centering
        \includegraphics[width=0.43\textheight,angle=-90]{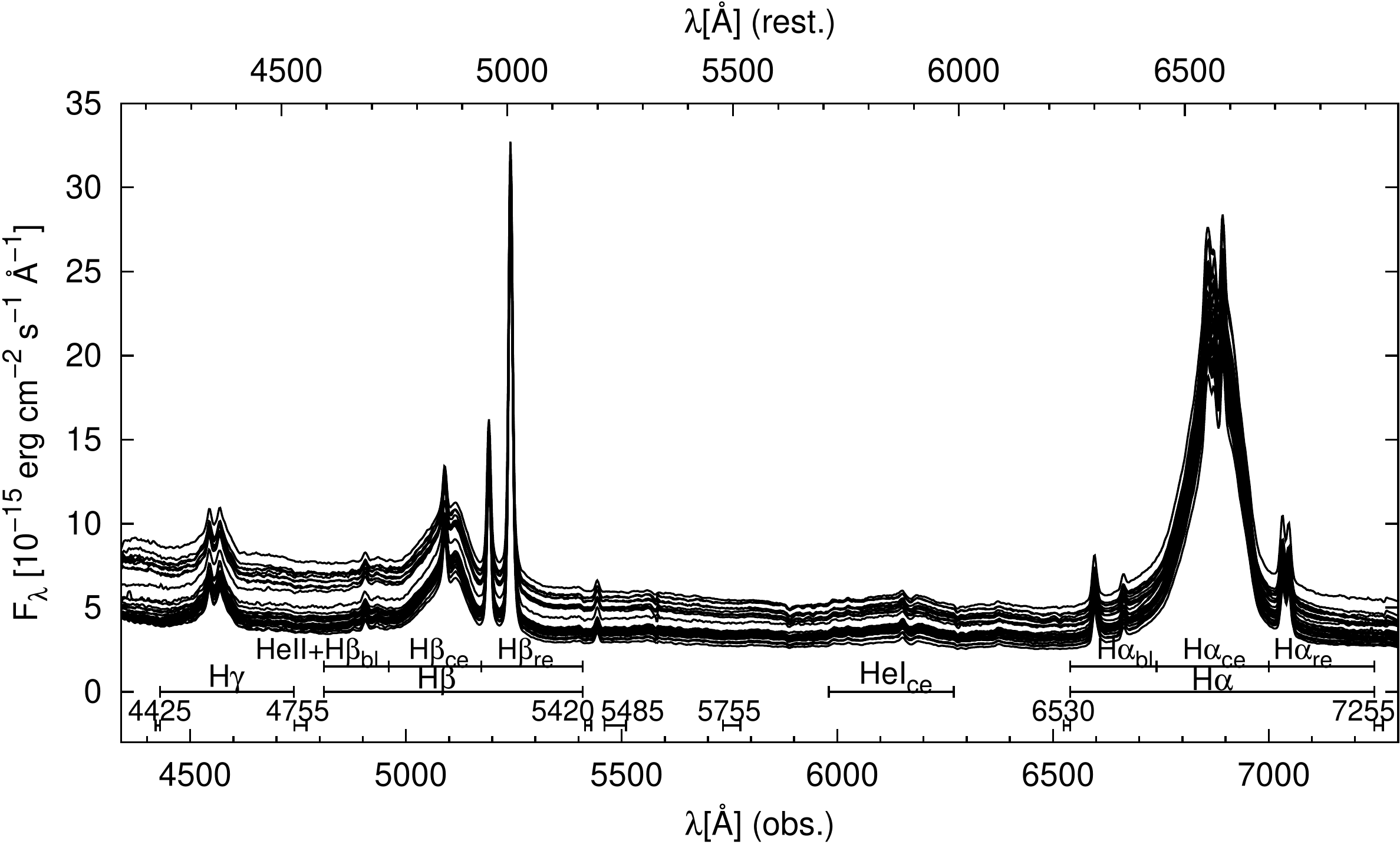}
        \caption{Reduced and [\ion{O}{iii}]\,$\lambda$5007-intercalibrated optical spectra of Mrk\,926 taken with the HET telescope during our variability campaign in 2010. The continuum bands and the broad-line integration intervals are marked according to Table~\ref{tab:cont_boundaries}. The spectra shown here are not corrected for Galactic foreground extinction.}
        \label{ochmspectraall.pdf}
    \end{minipage}
\vfill
    \begin{minipage}[l]{\textwidth}
        \centering
        \vspace*{0.4cm}
        \includegraphics[width=0.43\textheight,angle=-90]{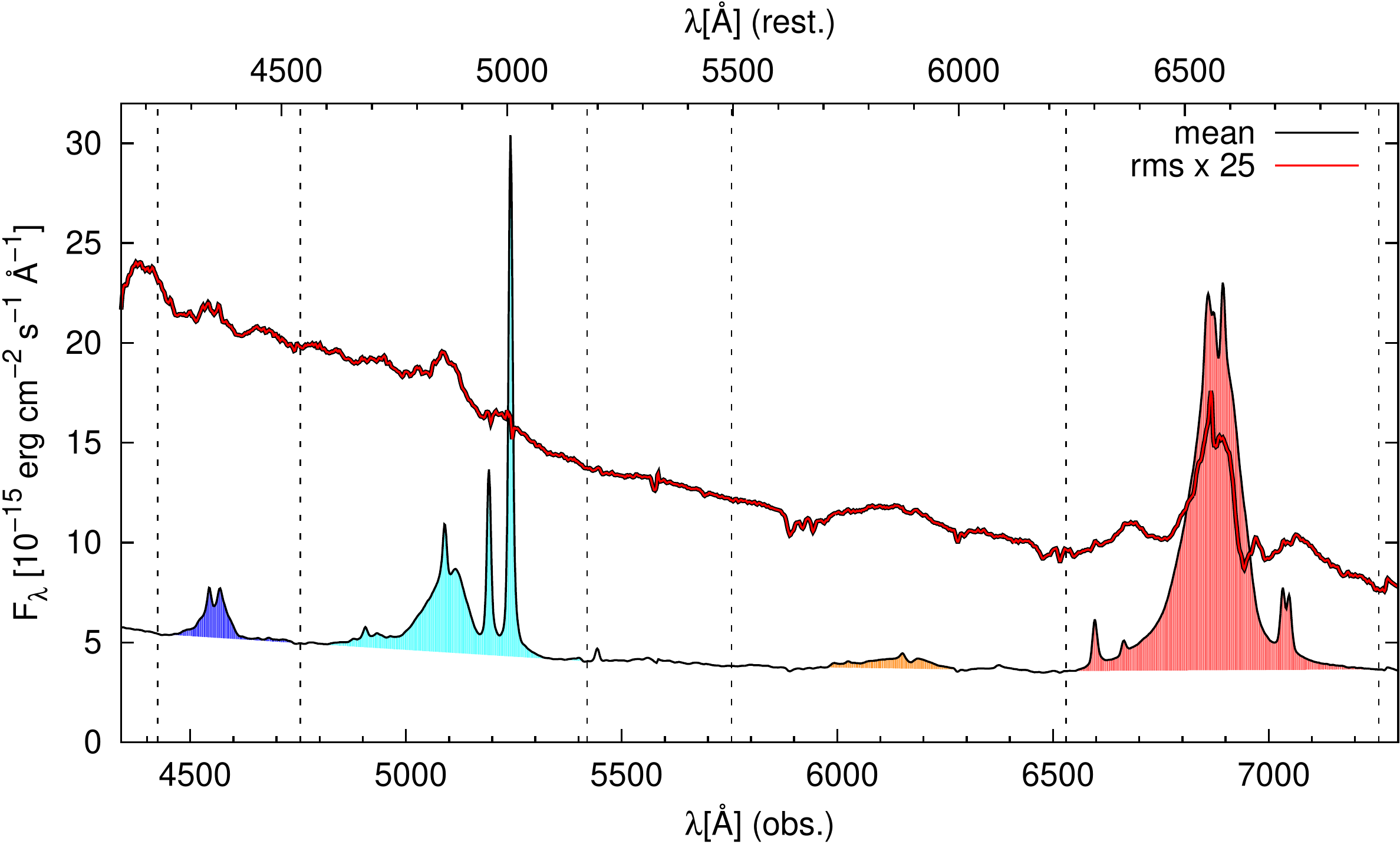}
        \caption{Combined mean spectrum (black) and rms spectrum (red; see Sect.~\ref{sec:continuum_spectral_variations} for details) of Mrk\,926 from our variability campaign. The rms spectrum was scaled to allow for a direct comparison. The central wavelengths of the pseudo-continua used for linear continuum subtraction are denoted by the vertical dashed lines. The line integration limits are marked by the shaded areas.}
        \label{ochmavg_rms_spectra.pdf}
    \end{minipage}
\end{figure*}

\section{Results}\label{sec:results}
\subsection{Continuum and spectral line variations}\label{sec:continuum_spectral_variations}
We present all the reduced [\ion{O}{iii}]\,$\lambda$5007-intercalibrated optical spectra of our variability campaign from Mrk\,926 in Fig.~\ref{ochmspectraall.pdf}. They clearly show variations in the continuum intensities. The mean and rms spectra of Mrk\,926 are shown in Fig.~\ref{ochmavg_rms_spectra.pdf}.\footnote{The presented mean and rms spectra are a combination of the blue and red intercalibrations. We merged them at 6345\ \AA\ and used a small scaling factor to adapt the redward calibration to the intercalibration with respect to [\ion{O}{iii}]\,$\lambda$5007.} The rms spectrum indicates the variable part of the emission lines. It was scaled to allow for a better comparison with the mean spectrum and to enhance weaker line structures. The continuum regions and integration limits of the broad emission lines used in our present study are given in Table~\ref{tab:cont_boundaries} and shown at the bottom of the spectra in Fig.~\ref{ochmspectraall.pdf}. We selected the continuum regions by inspecting the mean and rms spectra for regions that are free of both strong emission and absorption lines. 

A continuum region at 5100 \AA\ (rest frame) is often used in studies of the variable continuum flux in AGN. Normally, this region is free of strong emission lines and close to the [\ion{O}{iii}]\,$\lambda$5007 flux calibration line. However, in our case this region is contaminated by H$\beta$ because of the very broad line profile widths in Mrk\,926. Thus, we settled for a continuum range slightly more shifted to the red part of the spectrum (5180 \AA\ in the rest frame). Finally, we determined the continuum intensities at seven wavelength ranges (at 4425, 4755, 5420, 5485, 5755, 6530 and 7255 \AA{} in the observed frame). We integrated the broad emission-line intensities of the Balmer and \ion{He}{i} lines between the wavelength boundaries given in Table~\ref{tab:cont_boundaries}. The red wing of the \ion{He}{ii}\,$\lambda 4686$ line and the blue wing of H$\beta$ are superimposed because of their very broad line widths. There is no clear minimum between the two lines. Therefore, we assigned the emission-line flux shortward of 4960\,\AA\ to a mix of the \ion{He}{ii}\,$\lambda 4686$ line and blue H$\beta$ wing, and the flux longward of 4960\,\AA\ to H$\beta$. Before integrating the emission-line flux, we subtracted a linear pseudo-continuum defined by the continua given in Table~\ref{tab:cont_boundaries} (Col.\,3). The results of the continuum and the integrated line intensity measurements -- including the narrow-line components -- are given in Tables~\ref{HET_cont_intens} and~\ref{em_integline_intens}. The light curves of the integrated emission-line fluxes of \Ha{}, \Hb{}, \Hg{}, \Hb{}$_\textrm{blue}$+\ion{He}{ii}\,$\lambda 4686$, and \ion{He}{i}\,$\lambda 5876_\textrm{center}$ are given in Fig.~\ref{ochmLClin.pdf}. For further analysis, we also measured the narrow-line components of each broad line in the mean spectrum. The de-blending process of the narrow and broad components in the Balmer and \ion{He}{i}\,$\lambda$ 5876 lines is described in more detail in Sect.~\ref{sec:mean_rms_profiles}.  The resulting narrow-line fluxes are presented in Table~\ref{NEL-intensities}.
\begin{table}[!bp]
\centering
\tabcolsep+1.5mm
\caption{Continuum boundaries, line integration limits, and pseudo-continua.}
\begin{tabular}{lcc}
\hline \hline 
\noalign{\smallskip}
Cont./Line                   & Wavelength Range & Pseudo-Continuum \\
\noalign{\smallskip}
(1)                           & (2)                             & (3) \\
\hline 
\noalign{\smallskip}
Cont.~4425 (4225)                                   & 4420 -- 4430      & \\
Cont.~4755 (4540)                                   & 4742 -- 4770      & \\
Cont.~5420 (5180)                                   & 5415 -- 5430      & \\
Cont.~5485 (5240)                                   & 5460 -- 5510      & \\
Cont.~5755 (5500)                                   & 5735 -- 5775      & \\
Cont.~6530 (6240)                                   & 6525 -- 6540      & \\
Cont.~7255 (6930)                                   & 7245 -- 7265      & \\
\hline
H$\gamma$                                           & 4430 -- 4740      & 4425 -- 4755 \\
\Hb{}                                               & 4810\ -- 5410     & 4755 -- 5420 \\
H$\beta_\textrm{blue}$+\ion{He}{ii}\,$\lambda4686$  & 4810 -- 4960      & 4755  -- 5420  \\
H$\beta_\textrm{center}$                            & 4960  -- 5175     & 4755  -- 5420  \\
H$\beta_\textrm{red}$                               & 5175  -- 5410     & 4755  -- 5420  \\
\ion{He}{i}\,$\lambda5876_\textrm{center}$          & 5980  -- 6270     & 5755  -- 6530  \\
\Ha{}                                               & 6540  -- 7245     & 6530  -- 7255  \\
H$\alpha_\textrm{blue}$                             & 6540  -- 6740     & 6530  -- 7255  \\
H$\alpha_\textrm{center}$                           & 6740  -- 7000     & 6530  -- 7255  \\
H$\alpha_\textrm{red}$                              & 7000  -- 7245     & 6530  -- 7255  \\
\noalign{\smallskip}
\hline 
\end{tabular}
\label{tab:cont_boundaries}
\tablefoot{The continuum wavelength values in brackets correspond to the rest frame of Mrk\,926. All wavelengths are given in \AA.}
\end{table}
\begin{table}[!h]
    \centering
    \tabcolsep+2.5mm
    \caption{Observed narrow-emission-line fluxes as measured in the mean spectrum shown in Fig.~\ref{ochmavg_rms_spectra.pdf}.}
    \begin{tabular}{lr|lr}
        \hline \hline 
        \noalign{\smallskip}
        Line & \mcc{Flux} & Line & \mcc{Flux}\\
        \noalign{\smallskip}
        \hline 
        \noalign{\smallskip}
        H$\gamma_{\rm narrow}$                          & 7.4 $\pm$  1.0  &   $\left[\ion{O}{i}\right]\,\lambda$6363  & 8.0 $\pm$ 1.5   \\
        \ion{He}{ii}\,$\lambda 4686$                    & 6.3 $\pm$ 0.9   &   $\left[\ion{N}{ii}\right]\,\lambda$6548 & 18.8 $\pm$ 3.0  \\ 
        H$\beta_\textrm{narrow}$                        & 15.7 $\pm$ 1.8  &   H$\alpha_\textrm{narrow}$               & 42.4 $\pm$ 3.4  \\
        $\left[\ion{O}{iii}\right]\,\lambda$4959        & 102. $\pm$ 2.1  &   $\left[\ion{N}{ii}\right]\,\lambda$6584 & 54.0 $\pm$ 3.4  \\ 
        $\left[\ion{O}{iii}\right]\,\lambda$5007        & 314. $\pm$ 3.3  &   $\left[\ion{S}{ii}\right]\,\lambda$6717 & 41.1 $\pm$ 2.1  \\ 
        \ion{He}{i}\,$\lambda 5876$                     & 3.1 $\pm$ 0.7   &   $\left[\ion{S}{ii}\right]\,\lambda$6731 & 37.0 $\pm$ 2.0  \\ 
        $\left[\ion{O}{i}\right]\,\lambda$6300          & 33.8 $\pm$ 2.7  &                                           &              \\ 
        \hline
    \end{tabular}
     \label{NEL-intensities}
    \tablefoot{In units of 10$^{-15}$ erg cm$^{-2}$ s$^{-1}$.}
\end{table}

\subsubsection{Intercalibration of HET and Wise light curves} \label{sec:combined_lightcurve}
\begin{figure*}[!htp]
\centering
\vspace{-3mm}
 \includegraphics[width=15.5cm,angle=0]{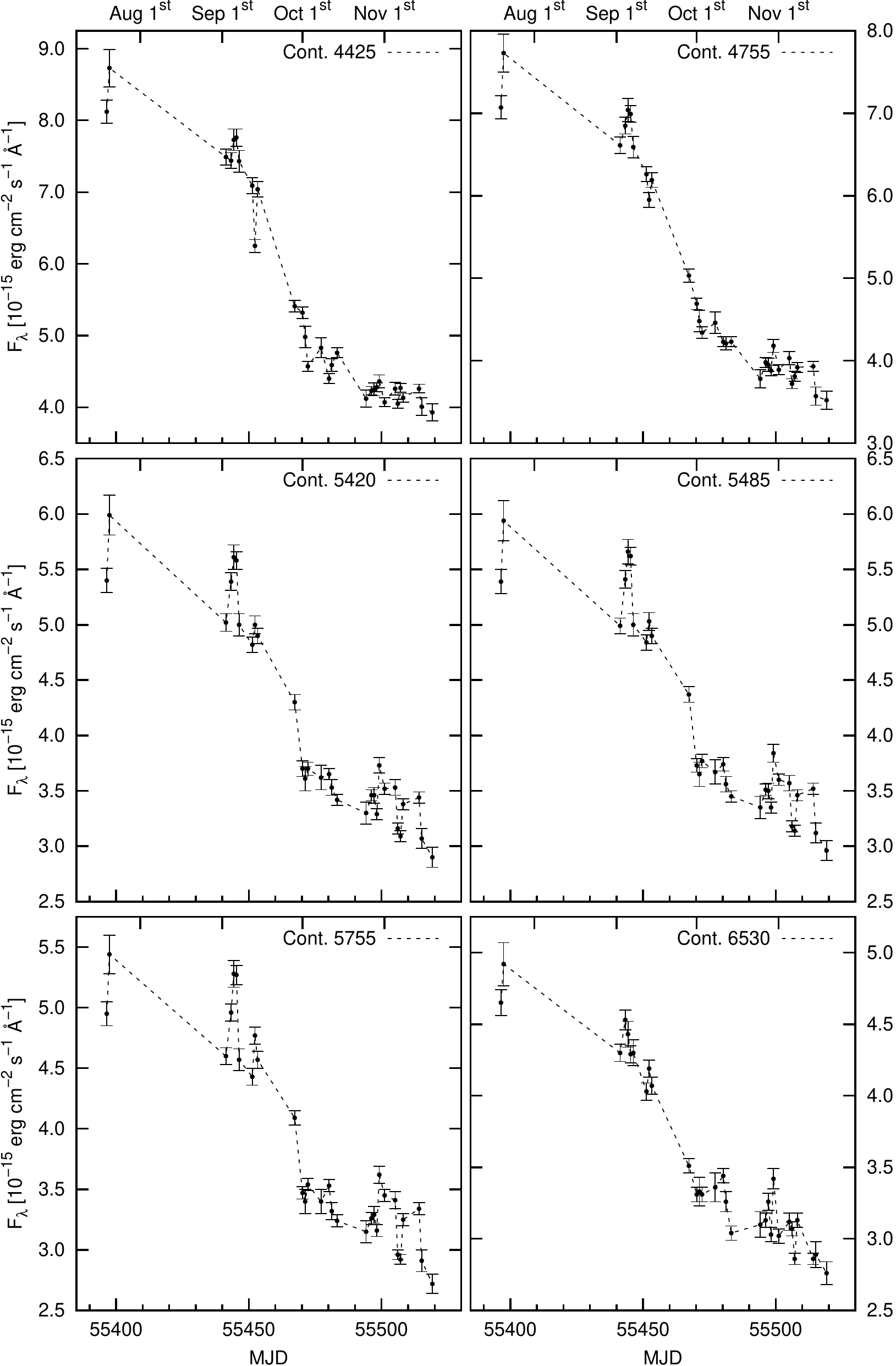}
 \caption{Light curves of the continuum flux densities at
 4425, 4755, 5420, 5485, 5755, and 6530\,\AA\ (observed frame; in units of 10$^{-15}$ erg cm$^{-2}$ s$^{-1}$\,\AA$^{-1}$) of
 our HET variability campaign from 2010 July 19 to November 19.}
  \label{ochmLCcont.pdf}
\end{figure*}

\begin{figure*}[!htp]
\centering
 \includegraphics[width=0.40\textheight,angle=-90]{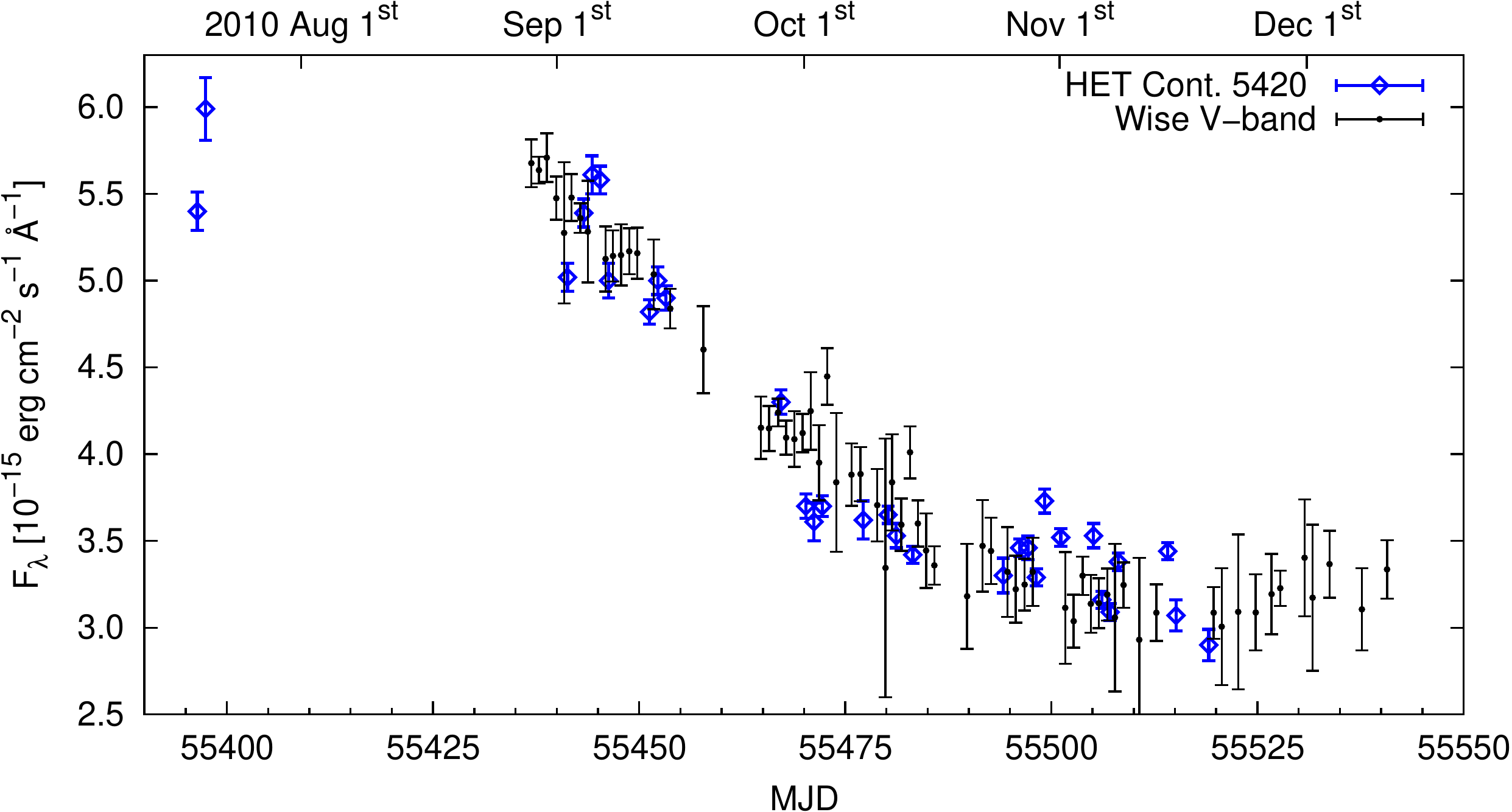}
 \caption{Combined spectroscopic (HET continuum flux densities at 5420\,\AA, or 5180\,\AA\ in rest frame; blue) and photometric (V-band continuum data obtained at the Wise Observatory; black) continuum light curve of our campaign (see Sect.~\ref{sec:continuum_spectral_variations} for details).}
  \label{ochmLCcombiWisHet.pdf}
\end{figure*}

\begin{figure*}[!htp]
\centering
\vspace{-3mm}
 \includegraphics[width=15.5cm,angle=0]{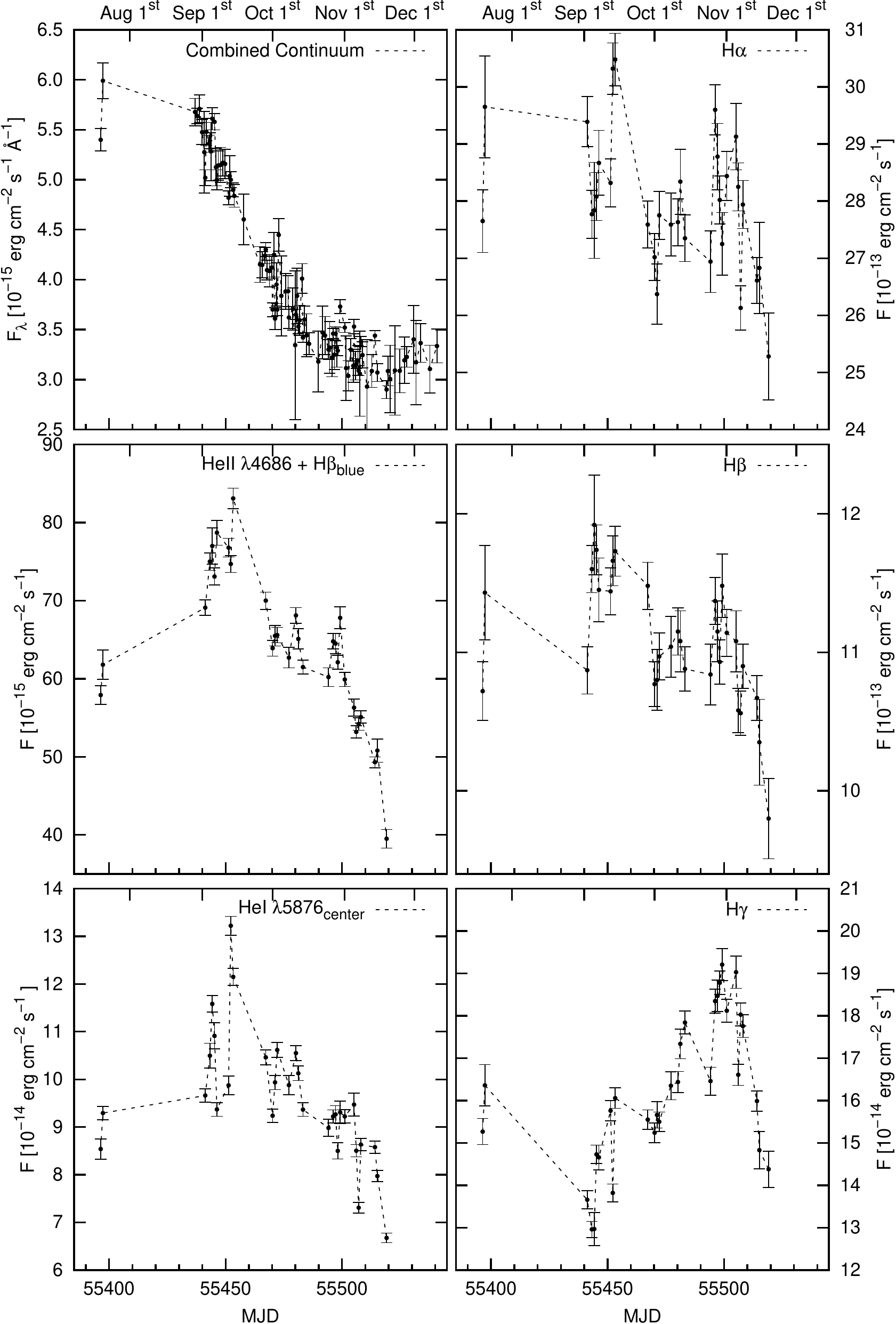}
 \caption{Light curves of the combined continuum flux densities at 5420\,\AA\ (observed frame; in units of 10$^{-15}$ erg cm$^{-2}$ s$^{-1}$\,\AA$^{-1}$) as well as of the integrated emission-line fluxes of  \Ha{}, \Hb{}, \Hg{}, \ion{He}{ii}\,$\lambda 4686$ + \Hb{}$_{\rm blue}$, and \ion{He}{i}\,$\lambda 5876$ for our HET variability campaign from 2010 July 19 to November 19.}
  \label{ochmLClin.pdf}
\end{figure*}

We present the light curves of the continuum flux densities at 4425, 4755, 5420, 5485, 5744, and 6530\,\AA\ (observed frame) in Fig.~\ref{ochmLCcont.pdf}. In order to increase the temporal resolution, we created a combined continuum light curve based on the HET continuum flux densities at 5420\, \AA{} and on the photometric V-band continuum light curve obtained at the Wise Observatory. We followed the procedure described in \citet{grier12} and \citet{zetzl18} by applying a multiplicative scale factor and an additional flux adjustment component to put the Wise Observatory instrumental magnitude light curve on the same scale as the HET continuum light curve in order to correct for different instrumentations, differences in the aperture sizes and therefore different host galaxy contributions, and constant contributions from emission lines within the bandwidth of the V-band filter. The optimal scale factor and flux adjustment component were determined by finding the set of parameters for which the cumulative difference of (nearly) simultaneous spectroscopic and photometric continuum measurements reached its minimum. The resulting combined light curve is given in Table~\ref{tab:combined_continuum} and shown in Fig.~\ref{ochmLCcombiWisHet.pdf}. Excluding the first two data points, the time interval between observations amounts to only 1.1 days. For further analysis, we corrected this light curve for Galactic foreground extinction ($A_{\rm V}=0.113$; \citealt{schlafly11}) and subtracted the V-band host galaxy contribution ($1.35 \times 10^{-15}$ erg cm$^{-2}$ s$^{-1}$\,\AA$^{-1}$; see Sect.~\ref{sec:host_galaxy}). For comparison, we also show the combined continuum light curve in Figs.~\ref{ochmLClin.pdf}, \ref{ochmHa_segments_20200929.pdf}, and \ref{ochmHb_segments_20200929.pdf}.\\


\subsection{Host galaxy contribution to the optical continuum flux}\label{sec:host_galaxy}
\begin{figure}[h!]
\centering
    \begin{subfigure}[t]{0.45\textwidth}
    \centering
    \includegraphics[width=9.5cm,angle=0]{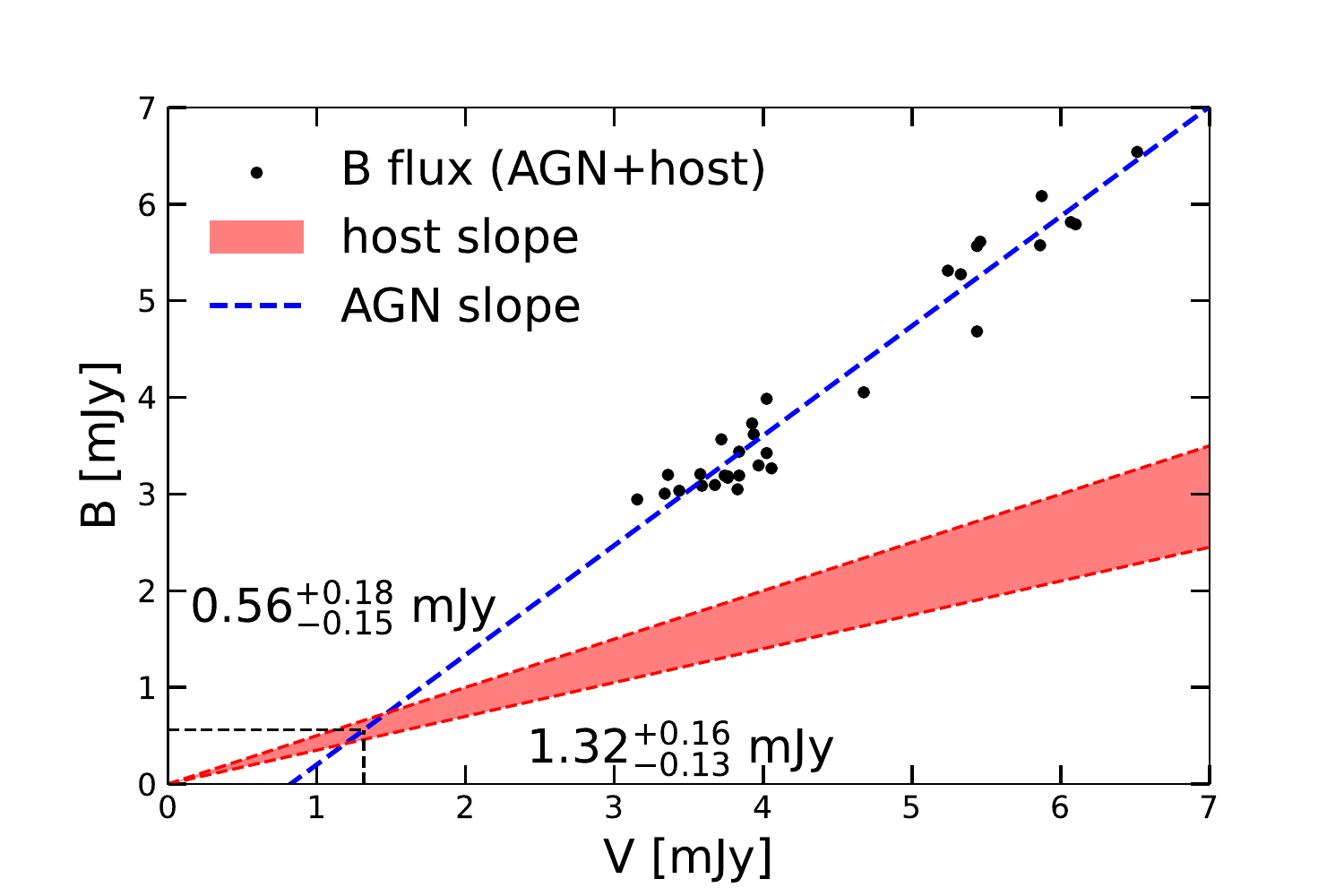}
    \label{ochmFVG_BvsV.pdf}
    \end{subfigure}
\vfill
    \begin{subfigure}[t]{0.45\textwidth}
    \centering
    \includegraphics[width=9.5cm,angle=0]{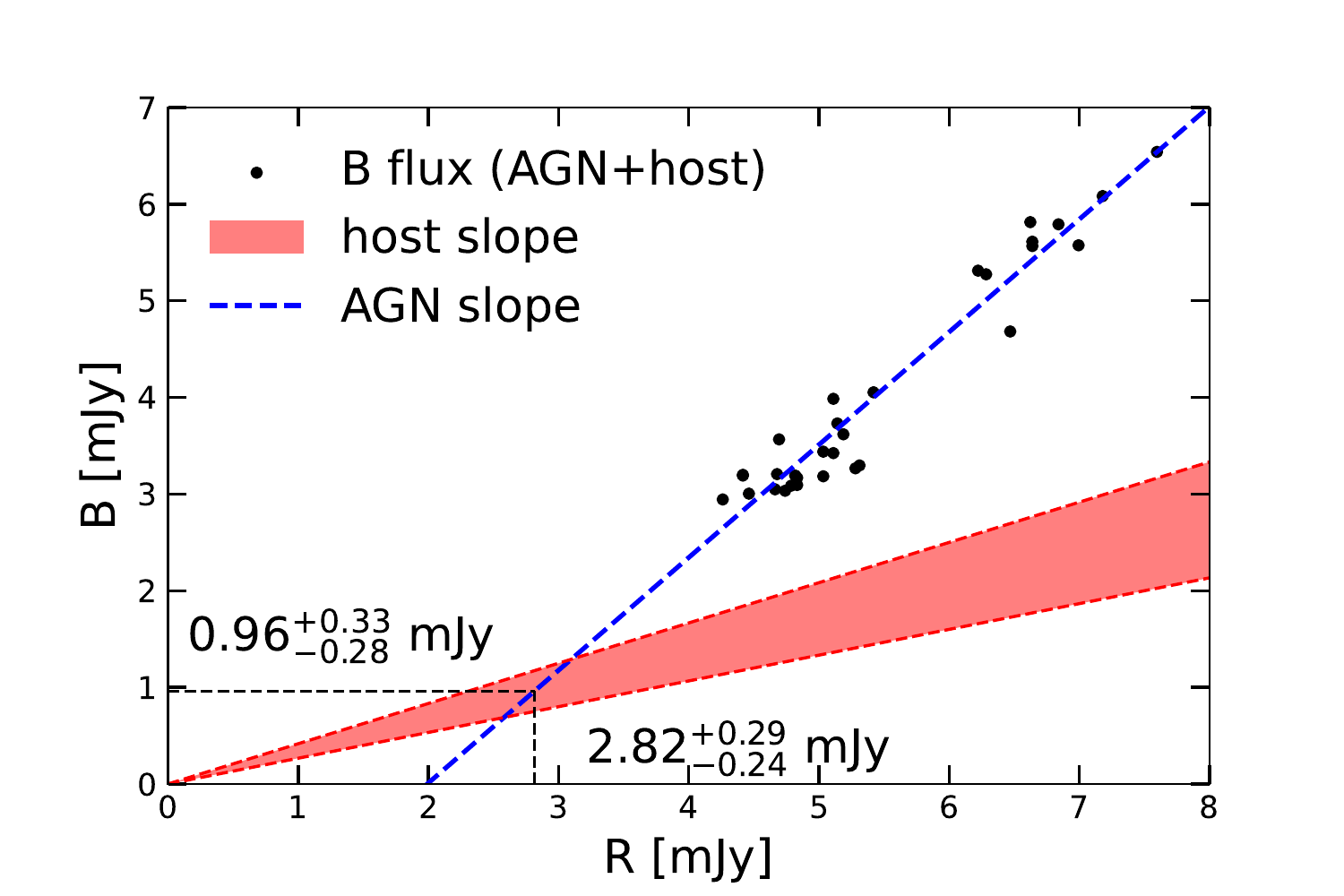}
    \label{ochmFVG_BvsR.pdf}
    \end{subfigure}
\caption{Extinction-corrected B versus V (\textit{upper panel}) and B versus R (\textit{lower panel}) flux variations of Mrk\,926. The dashed blue  lines (AGN slope) represent the best linear fit to the B versus V and B versus R fluxes, respectively. The red shaded area shows the range of host slopes. The dashed black lines indicate the B and V as well as the B and R values of the host galaxy. }
\label{ochmFVG.pdf}
\end{figure}

The observed flux of the variable AGN component is contaminated by the flux contribution of the host galaxy. This contribution is nearly constant for observations taken with identical aperture, and only minor deviations arise due to slight variations from the exact pointing as well as different seeing conditions. However, these effects are taken into account by our error estimates.
The overall host galaxy contribution varies from band to band as the stellar component of the host galaxy has a different flux distribution than the AGN component.  To estimate the relative contribution of the host galaxy flux, we used the flux variation gradient (FVG) method \citep{choloniewski81, winkler92, haas11, zetzl18}. This method allows the varying AGN flux in our aperture to be disentangled from the host galaxy contribution. We took the continuum flux densities at 4425, 5420, and 6530\,\AA\ (observed frame) as proxies for B-, V-, and R-band fluxes, respectively, since these continuum ranges are close to the maxima of the B-, V-, and R-filter curves (B-band filter: 4300 \AA{}, V-band filter: 5400 \AA{},  R-band filter: 6100 \AA{}). In this way, we also excluded the contribution of emission lines (see Fig.~\ref{ochmspectraall.pdf}). Figure~\ref{ochmFVG.pdf} shows the B versus V and B versus R fluxes (black solid circles) of Mrk\,926 based on the HET spectra (aperture: 2 x 2 arcsec). The B, V, and R values in this figure have been corrected for Galactic foreground extinction (A$_{B}$ = 0.149, A$_{V}$ = 0.113, A$_{R}$ = 0.089; \cite{schlafly11}) and are given in units of mJy.
The converted light curves before and after Galactic foreground extinction are given in Table~\ref{tab:extinct_corr_continua}. The blue dashed line in Fig.~\ref{ochmFVG.pdf} gives the best linear fit to the B versus V and B versus R fluxes.
The red shaded area marks the range of host slopes for nearby AGN as determined by \cite{sakata10}. The intersection point between the AGN and host galaxy slopes allows the contribution of the host galaxy fluxes in the B, V, and R bands to be determined; the resulting host galaxy contributions in the individual bands are indicated by the dashed black lines.

We derived a B-band host galaxy contribution of 0.76\,mJy (mean of 0.56\,mJy and 0.96\,mJy). The corresponding values are 1.32\,mJy for the V band and 2.82\,mJy for the R band. By subtracting the host galaxy contribution for each individual band, we isolate the AGN flux in all bands. All flux values (minimal and maximal host+AGN, host contribution as well as true minimal and maximal AGN flux) are listed in Table~\ref{bvrhostflux}. Based on the HET spectra, the host galaxy contribution corresponds to 12 -- 26 \%, 20 -- 42 \%, and 37 -- 66 \% for the B, V, and R bands, respectively.  Altogether, the spectra of Mrk\,926 are dominated by emission from the nucleus. The weak host contribution in the B and V bands and, especially, the very strong and broad emission lines prevent us from subtracting a reliable host galaxy template since the stellar signature of the host galaxy is suppressed.

\begin{table}[!h]
\tabcolsep+1.5mm
\caption{Extinction-corrected B, V, and R values for the combined host galaxy plus AGN fluxes as well as for the host galaxy and AGN fluxes alone as determined by the FVG method.  When a range is given, it corresponds to the minimum and maximum flux.}
\centering
\begin{tabular}{l|c|c|c}
\hline \hline 
\noalign{\smallskip}
Flux Component               & B band & V band  & R band \\
\noalign{\smallskip}
\hline
\noalign{\smallskip}
& \multicolumn{3}{c}{[mJy]}\\
(1)                   & (2)  & (3)   & (4) \\
\noalign{\smallskip}
\hline
\noalign{\smallskip}
Host+AGN (BvsR)  &2.94 -- 6.54 &       &4.26 -- 7.60\\
Host+AGN (BvsV)  &2.94 -- 6.54  &3.15 -- 6.51   &\\
\hline
Host (BvsR)      &0.96  &       &2.82\\
Host (BvsV)      &0.56  &1.32   &\\
\hline
AGN (BvsR)       &1.98 -- 5.58  &       &1.44 -- 4.78\\
AGN (BvsV)       &2.39 -- 5.98  &1.83 -- 5.19   &\\
\hline \hline 
\noalign{\smallskip}
& \multicolumn{3}{c}{[10$^{-15}$\,erg\,s$^{-1}$\,cm$^{-2}$\,\AA$^{-1}$]}\\
(1)                   & (2)  & (3)   & (4) \\
\noalign{\smallskip}
\hline
\noalign{\smallskip}
Host+AGN (BvsR)  &4.51 -- 10.01  &       &3.00 -- 5.34\\
Host+AGN (BvsV)  &4.51 -- 10.01  &3.22 -- 6.65   &\\
\hline
Host (BvsR)      &1.47  &       &1.98\\
Host (BvsV)      &0.86  &1.35   &\\
\hline
AGN (BvsR)       &3.03 -- 8.54  &       &1.01 -- 3.36\\
AGN (BvsV)       &3.66 -- 9.16  &1.87 -- 5.30   &\\
\hline
\end{tabular}
\label{bvrhostflux}
\end{table}

\subsection{Variability statistics}
The following statistics are based on the spectroscopic
and photometric observations from HET and the Wise Observatory, respectively. The continua were corrected for Galactic foreground extinction (see Sect.~\ref{sec:continuum_spectral_variations}), and the lines were corrected for contribution from the narrow lines (see Table~\ref{NEL-intensities}).
In Table~\ref{variab_statistics}, we present the minimum and maximum fluxes, $F_\text{min}$ and $F_\text{max}$, peak-to-peak amplitudes, $R_\text{max}$ = $F_\text{max}$/$F_\text{min}$, the mean flux over the period of observations, $<F>$, the standard deviation, $\sigma_F$, and the fractional variation,
\begin{equation}
F_{\rm var} = \sqrt{{\sigma_F}^2 - \Delta^2} / <F> 
,\end{equation}
as defined by \cite{rodriguez97}. The quantity $\Delta^2$ is the mean square value of the uncertainties $\Delta_\text{i}$ associated with the fluxes $F_\text{i}$. The $F_{\rm var}$ uncertainties are defined in \cite{edelson02}. The peak-to-peak amplitudes and the fractional variations in the continuum decrease as a function
of wavelength.
\begin{table}[h!]
    \centering
    \tabcolsep+1.3mm
    \caption{Variability statistics ($F_\text{min}$, $F_\text{max}$, $R_\text{max}$,
    $<F>$, $\sigma_\text{F}$, and F$_\text{var}$) for the continuum (after correction for Galactic foreground extinction) and for the broad emission lines (after correction for the contribution of narrow-line components; see Table~\ref{NEL-intensities}).}
    \begin{tabular}{lrrrrrrr}
        \hline \hline 
        \noalign{\smallskip}
        Cont./Line & \multicolumn{1}{c}{$F_\text{min}$} & \multicolumn{1}{c}{$F_\text{max}$} & \multicolumn{1}{c}{$R_\text{max}$} & \multicolumn{1}{c}{$<F>$} & \multicolumn{1}{c}{$\sigma_\text{F}$} & \multicolumn{1}{c}{$F_\text{var}$} \\
        \noalign{\smallskip}
        \noalign{\smallskip}
        \hline
        \noalign{\smallskip}
        Comb. V band            & 3.22 & 6.65 & 2.07 & 4.41 & 0.97 & 0.21\\
        \noalign{\smallskip}
        \hline 
        \noalign{\smallskip}
        Cont.~4425  (4225)            & 4.51 & 10.01 & 2.22 & 6.22 & 1.77 & 0.28\\     
        Cont.~4755  (4540)            & 4.04 & 8.87 & 2.20 & 5.67 & 1.52 & 0.27\\ 
        Cont.~5420  (5180)            & 3.22 & 6.65 & 2.07 & 4.50 & 1.01 & 0.22\\
        Cont.~5485  (5240)            & 3.28 & 6.59 & 2.01 & 4.54 & 0.99 & 0.22\\
        Cont.~5755  (5500)            & 3.02 & 6.04 & 2.00 & 4.23 & 0.90 & 0.21\\
        Cont.~6530  (6240)            & 3.00 & 5.34 & 1.78 & 3.85 & 0.68 & 0.17\\
        Cont.~7255  (6930)            & 3.20 & 5.03 & 1.57 & 3.93 & 0.54 & 0.14\\     
        \noalign{\smallskip}
        \hline
        \noalign{\smallskip}
        \Hg{}                                               & 122.2 & 184.7 & 1.51 & 154.6 & 17.6  & 0.11\\                      
        \Hb{}                                               & 542. & 754. & 1.39 & 670. & 46.         & 0.07\\                     
        H$\beta_{\rm blue}$+\ion{He}{ii}\,$\lambda$4686     & 33.2 & 76.80 & 2.31 & 57.8 & 9.6     & 0.17\\
        H$\beta_{\rm center}$                               & 419. & 556. & 1.33 & 502. & 30.       & 0.06\\                     
        H$\beta_{\rm red}$                                  & 89.2 & 126.8 & 1.42 & 109.5 & 8.5    & 0.08\\
        HeI\,$\lambda 5876_{\rm center}$                    & 63.7 & 129.1 & 2.03 & 92.7 & 13.2    & 0.14\\                 
        \Ha{}                                               & 2293. & 2813. & 1.23 & 2562. & 118.   & 0.04\\                    
        H$\alpha_{\rm blue}$                                & 131.3 & 192.4 & 1.47 & 159.2 & 14.7  & 0.09\\                   
        H$\alpha_{\rm center}$                              & 2082. & 2456. & 1.18 & 2286. & 93.3  & 0.04\\                 
        H$\alpha_{\rm red}$                                 & 79.6 & 166.9 & 2.10 & 116.1 & 23.9   & 0.21\\
        \hline 
        \noalign{\smallskip}
    \end{tabular}
\label{variab_statistics}
\tablefoot{In units of 10$^{-15}$\,erg\,s$^{-1}$\,cm$^{-2}$\,\AA$^{-1}$ for the continuum and in units of 10$^{-15}$\,erg\,s$^{-1}$\,cm$^{-2}$ for the broad emission lines. The fluxes in this table are observed-frame fluxes. The continuum wavelength values in brackets correspond to the rest frame of Mrk 926.}
\end{table}

\subsection{Mean and rms line profiles}\label{sec:mean_rms_profiles}
\begin{figure*}[!htp]
\centering
\includegraphics[width=0.9\textwidth,angle=0]{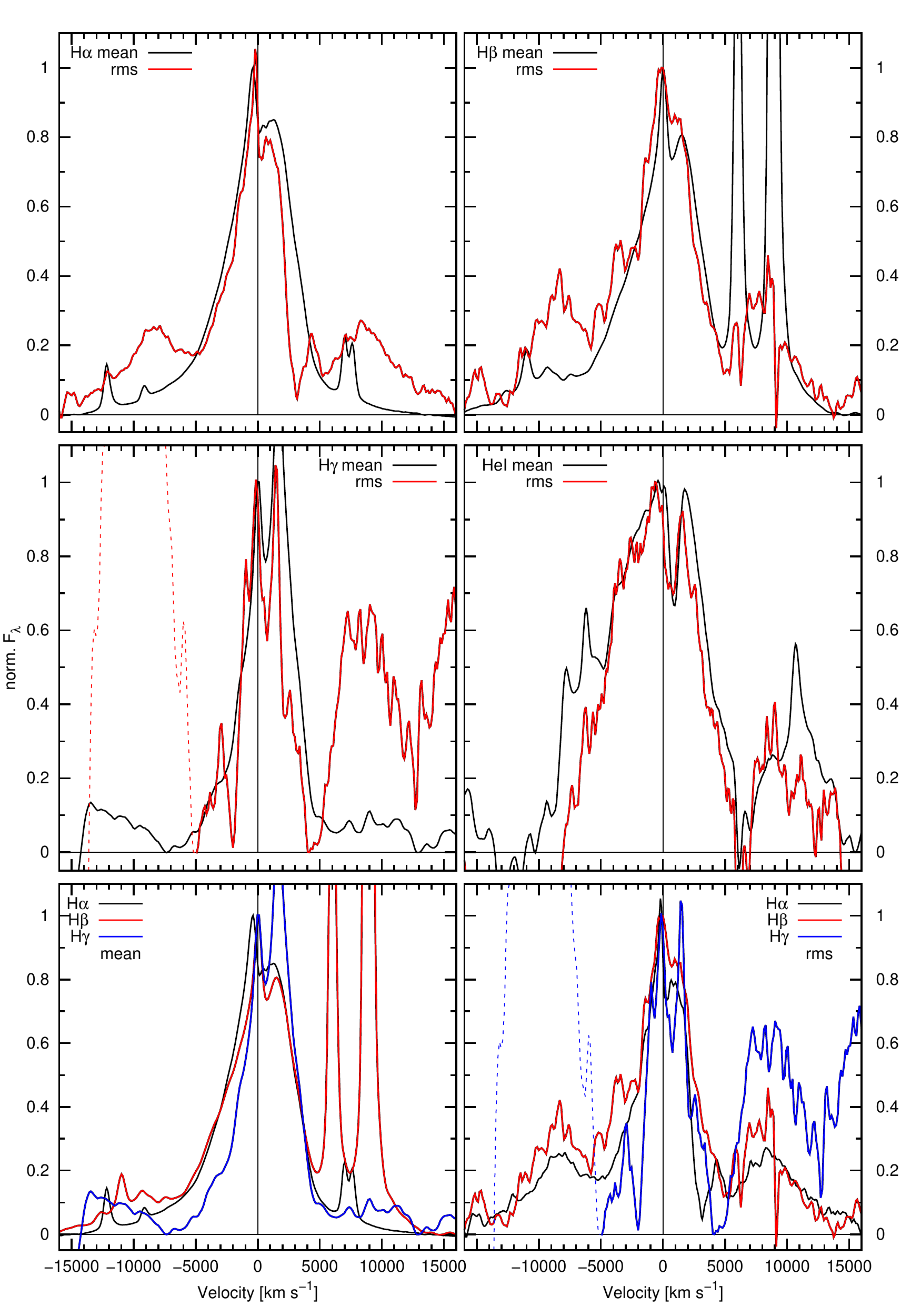}
\caption{Normalized mean and rms  profiles of the Balmer lines and  HeI\,$\lambda 5876$ in velocity space. The dashed part of the \Hg{} line profile denotes the part that is heavily affected by uncertainties in the flux calibration due to boundary effects at the blue end of the spectrum.}
\label{ochmprofmeanrms.pdf}
\end{figure*}
We determined normalized mean and rms profiles of the Balmer and \ion{He}{I} lines in Mrk\,926 after subtracting a linear pseudo-continuum from each mean and rms broad-line profile. In order to obtain the true mean broad-line profiles, we subtracted the narrow Balmer and \ion{He}{I} line components in velocity-space by means of a scaled  {[\ion{O}{iii}]\,$\lambda$5007 } template based on the observed [\ion{O}{iii}]\,$\lambda$5007 line profile in our mean spectrum. Furthermore, we subtracted the $\left[\text{NII}\right]$6548/6584  emission lines that are superimposed on the mean \Ha{} profile using the same template and sticking to a line-ratio of 1:3.
The rms profiles illustrate the line profile variations during our campaign. Therefore, the constant narrow components disappear in these rms profiles. Finally, we normalized both the mean and rms profiles such that the maximum line flux was set to unity. The final normalized mean and rms profiles of the Balmer lines \Ha{}, \Hb{}, \Hg{}, as well as of the \ion{He}{i}\,$\lambda 5876$ line in velocity space are shown in Fig.~\ref{ochmprofmeanrms.pdf}. The subtraction of all narrow Balmer and \ion{He}{I} components as well as of $\left[\text{NII}\right]$6548/6584 and the normalization of the true broad-line profiles allows the mean and rms profiles to be compared with each other in a more accurate way. 

The Balmer lines exhibit -- besides to their central component -- an additional inner red component at 1200 $\pm$  300 \kms{} in the mean as well as rms profiles (see Fig.~\ref{ochmprofmeanrms.pdf}). Furthermore, the Balmer lines show additional outer broad-line components in their rms profiles (Fig.~\ref{ochmprofmeanrms.pdf}) at $\pm 5000$ to $\pm 13\,000$ \kms. These Balmer satellites are clearly separated from the central component ($\pm$ 5000 \kms{}) and exhibit an amplitude that amounts to 25 -- 40\% of that of the maximum rms amplitude in \Ha\ and \Hb, respectively. These components are not recognizable in the mean profiles. The \Hg{} profile is contaminated by the [{O}{III}]\,$\lambda$4363 line and a strong residual is present in the rms profile. This is due to \Hg{} being at the blue end of the spectrum where the flux calibration for the first few pixels is not optimal.
Furthermore, it is not possible to fix a line-free continuum region on the blue side. These boundary effects heavily affect the blue part of the \Hg{} line profile. Therefore, the H$\gamma$ profile can only be used to derive some general trends. Despite the boundary effects mentioned before, the rms profile of \Hg{} confirms the presence of Balmer satellites at $\pm 5000$ to $\pm 13\,000$ \kms. The red wing of \ion{He}{i}\,$\lambda 5876$ might also indicate the presence of such a line satellite, however, a clear detection cannot be stated due to the interference by absorption in the line wings. In addition, the central \ion{He}{i}\,$\lambda 5876$ profile is contaminated by NaD absorption.

We determined the line widths (FWHM) of the mean and rms line profiles of all the Balmer lines and \ion{He}{i}\,$\lambda 5876$.  Additionally, we parameterized the line widths of the mean profiles by their line dispersion $\sigma_\text{line}$ \citep{fromerth00, peterson04}. However, due to the very broad, partially overlapping and complex rms line profiles with additional outer components, it was not possible do determine reliable $\sigma_\text{line}$ (rms) values. Likewise, by cause of \Hg{} being affected by boundary effects, the overlapping \ion{He}{ii}\,$\lambda$4686 and H$\beta_{\rm blue}$ profiles, as well as strong absorption in \ion{He}{i}\,$\lambda 5876$, a determination of $\sigma_\text{line}$ (mean) for these lines was deemed unreliable. Therefore, we determined the line dispersion $\sigma_\text{line}$ (mean) solely for the \Ha{}  mean profile over a velocity interval from $-15\,000$ to $+15\,000$ \kms{}. 

The resulting widths (FWHM) for all emission lines, and the line dispersion of \Ha{} ($\sim 3500$ \kms{}) are given in Table~\ref{tab:line_widths}. The \Ha{} and \Hb{} lines show similar mean profiles with line widths (FWHM) of $\sim 5200$ \kms as well as similar rms profiles with line widths (FWHM) of $\sim 4000$ \kms. The \Hg{} line exhibits a profile similar to that of \Ha{} and \Hb{}, but the line width and line dispersion are smaller than for the other Balmer lines. We note, however, that these results are less reliable due to the aforementioned boundary effects. The \ion{He}{i}\,$\lambda 5876$ line (mean and rms) is by a factor of about 1.5 broader than the Balmer lines.

\begin{table}[!hbp]
\centering
\tabcolsep+2mm
\caption{FWHM of the mean and rms line profiles as well as the line dispersion, $\sigma_\text{line}$,  of the mean profiles.}
\begin{tabular}{lccc}
\hline \hline 
\noalign{\smallskip}
Line & \mcc{FWHM (mean)} & \mcc{FWHM (rms)} & \mcc{$\sigma_\text{line}$ (mean)} \\
     & \mcc{[\kms{}]} & \mcc{[\kms{}]} & \mcc{[\kms{}]} \\
\noalign{\smallskip}
\hline 
\noalign{\smallskip}
\Ha{}                               & 5370 $\pm$  400 & 3890 $\pm$ 500  &  3460 $\pm$ 1000  \\
\Hb{}                               & 5110 $\pm$  400 & 4230 $\pm$ 400  &   --  \\
\Hg{}                               & 4300 $\pm$ 1000 & 3260 $\pm$ 800  &   --  \\
\ion{He}{i}\,$\lambda 5876$         & 8500 $\pm$ 1000 & 7650 $\pm$ 800  &   --  \\
\noalign{\smallskip}
\hline 
\end{tabular}
\label{tab:line_widths}
\end{table}

\subsection{CCF analysis of the  broad emission lines}\label{sec:1D_CCFs}
\begin{figure*}[!htp]
    \centering
    \begin{subfigure}[b]{0.48\textwidth}
         \centering
         \begin{minipage}[t]{0.48\textwidth}
            \includegraphics[width=\textwidth]{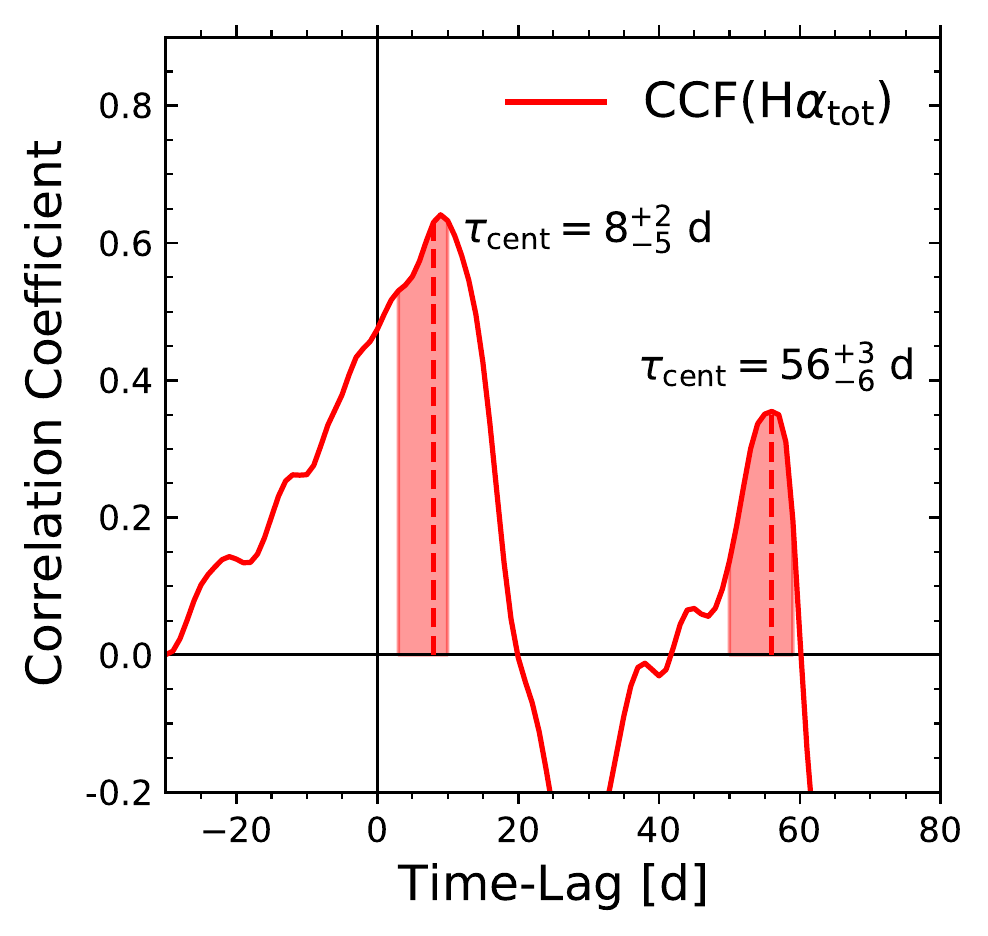}
         \end{minipage}
         \hfill
         \begin{minipage}[t]{0.48\textwidth}
            \includegraphics[width=\textwidth]{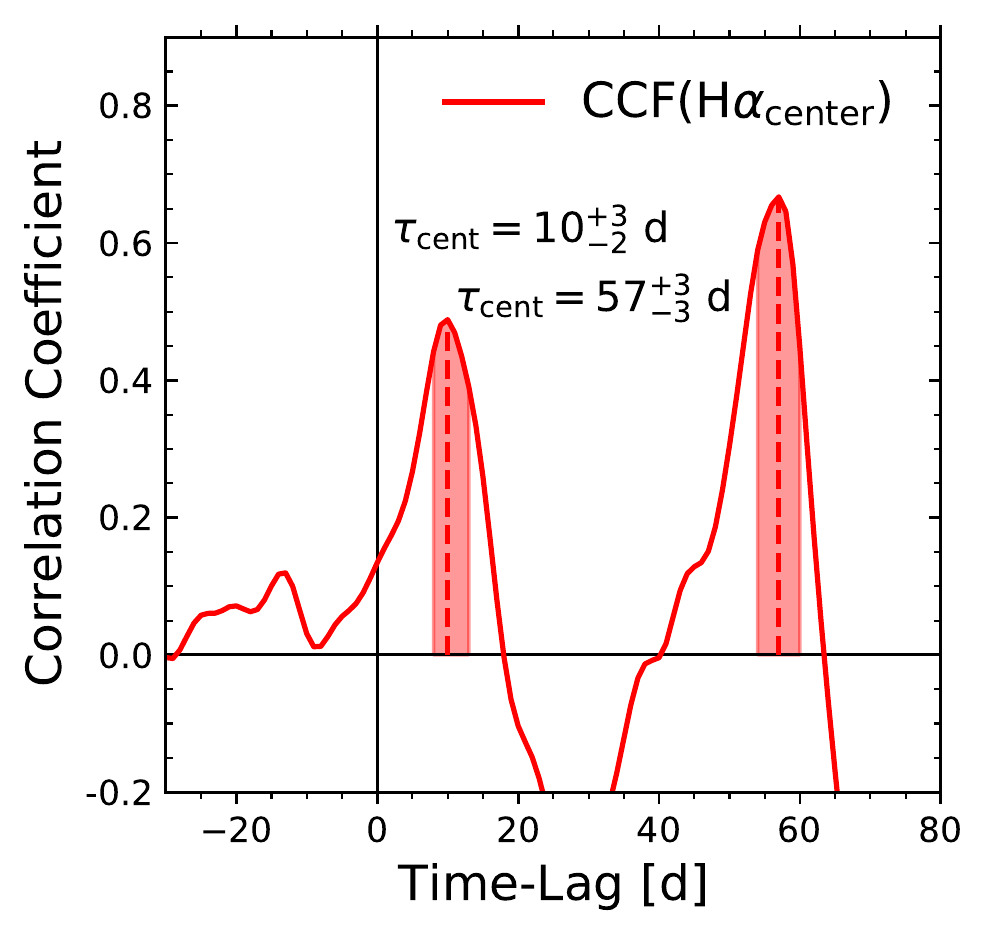}
         \end{minipage}
         \vfill
         \begin{minipage}[t]{0.48\textwidth}
            \includegraphics[width=\textwidth]{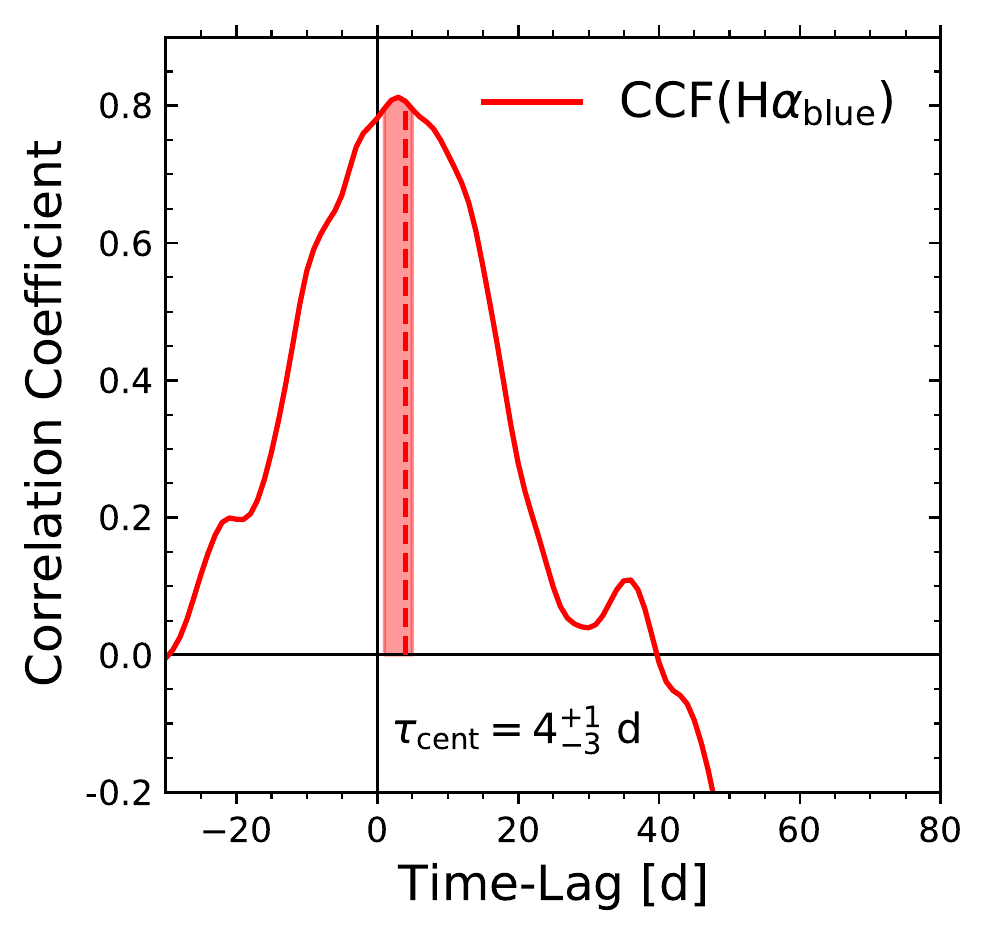}
         \end{minipage}
         \hfill
         \begin{minipage}[t]{0.48\textwidth}
            \includegraphics[width=\textwidth]{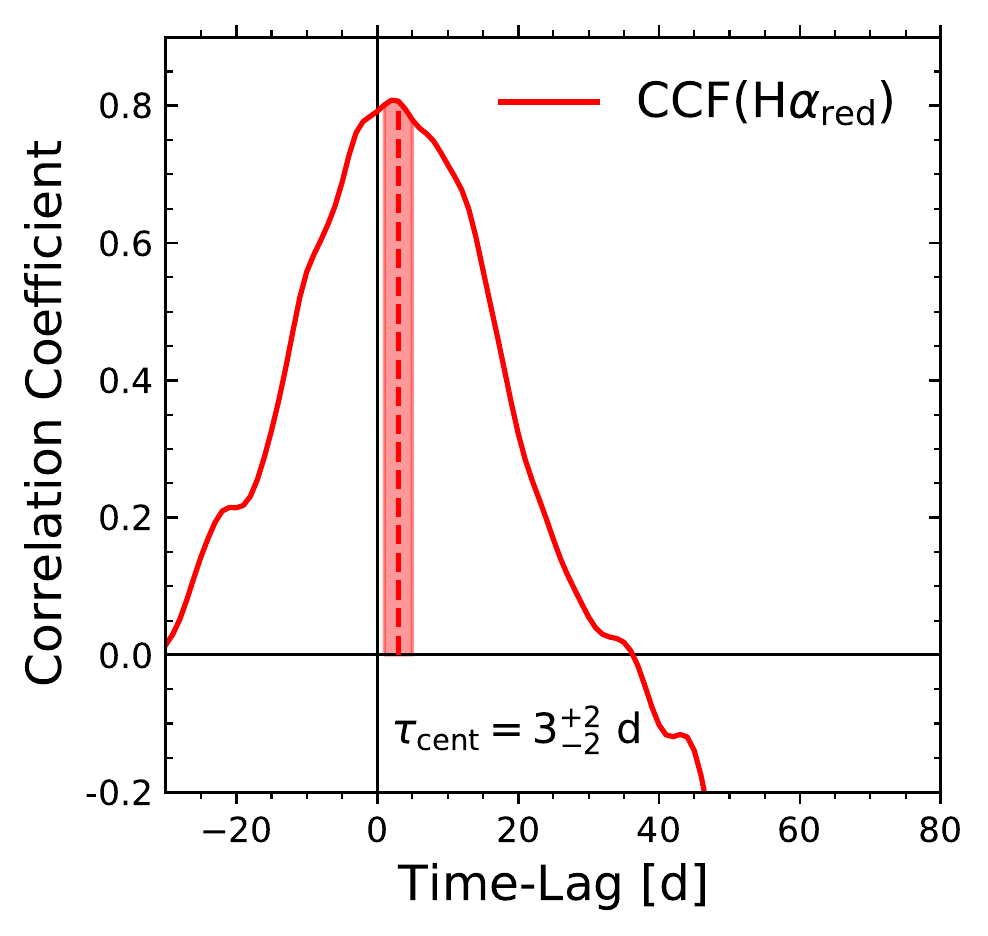}
         \end{minipage}
        \caption{}
    \end{subfigure}
    \hfill
    \begin{subfigure}[b]{0.48\textwidth}
         \centering
         \begin{minipage}[t]{0.48\textwidth}
            \includegraphics[width=\textwidth]{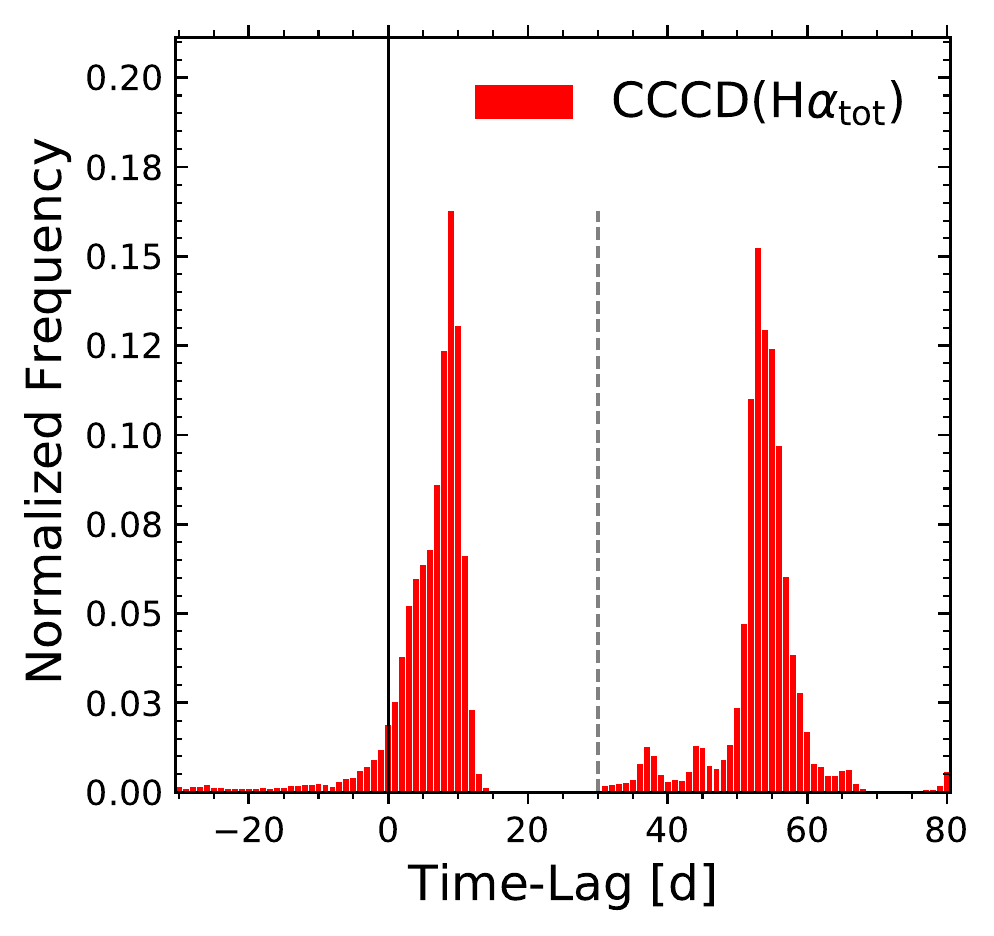}
         \end{minipage}
         \hfill
         \begin{minipage}[t]{0.48\textwidth}
            \includegraphics[width=\textwidth]{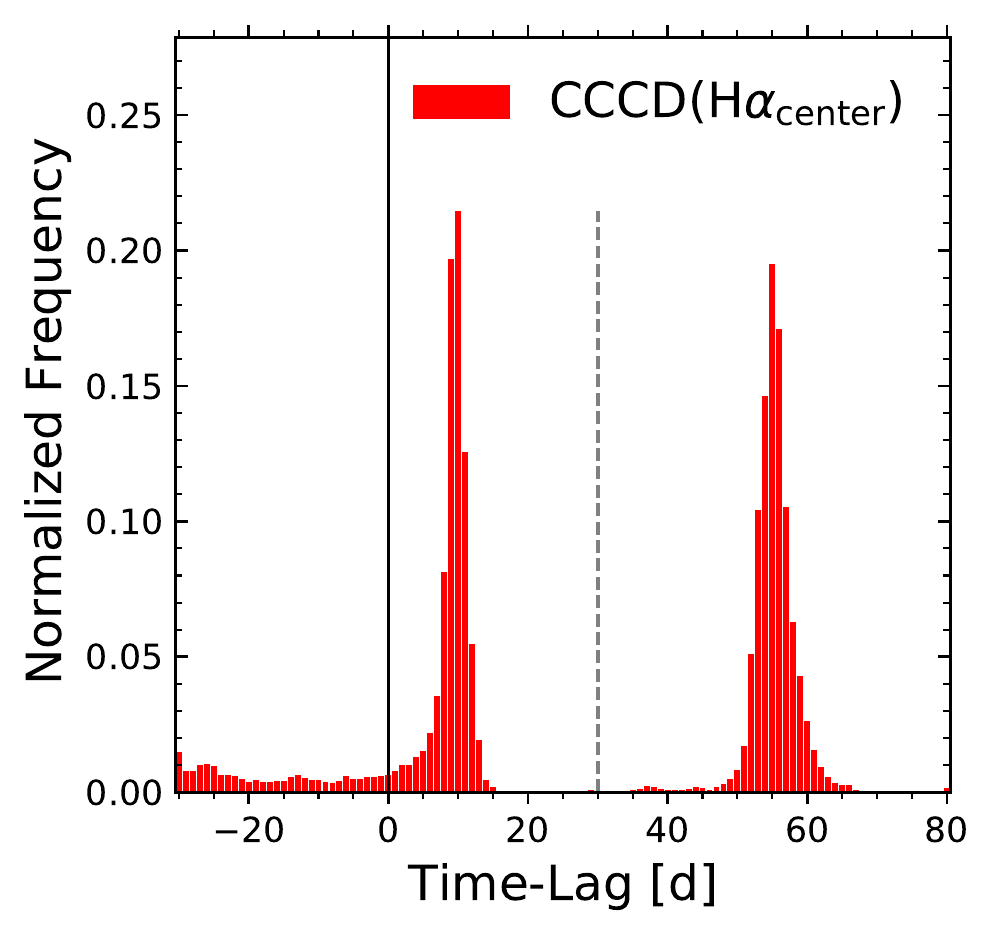}
         \end{minipage}
         \vfill
         \begin{minipage}[t]{0.48\textwidth}
            \includegraphics[width=\textwidth]{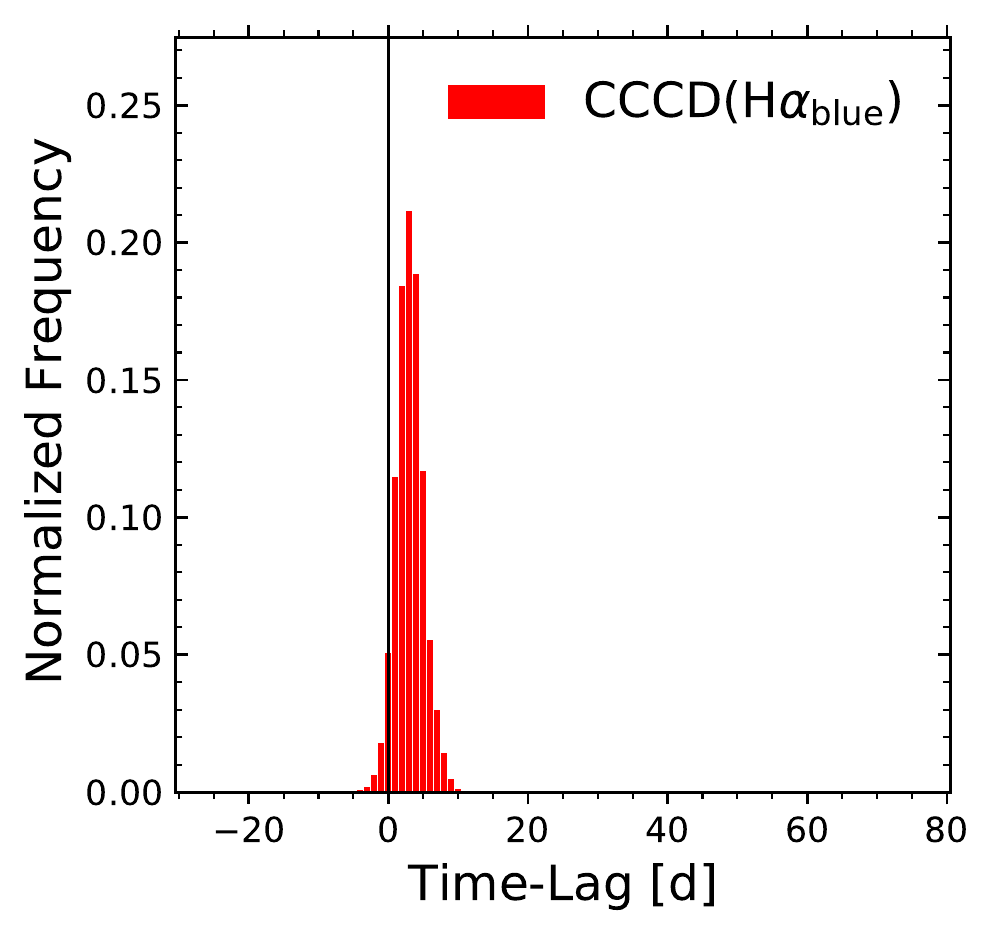}
         \end{minipage}
         \hfill
         \begin{minipage}[t]{0.48\textwidth}
            \includegraphics[width=\textwidth]{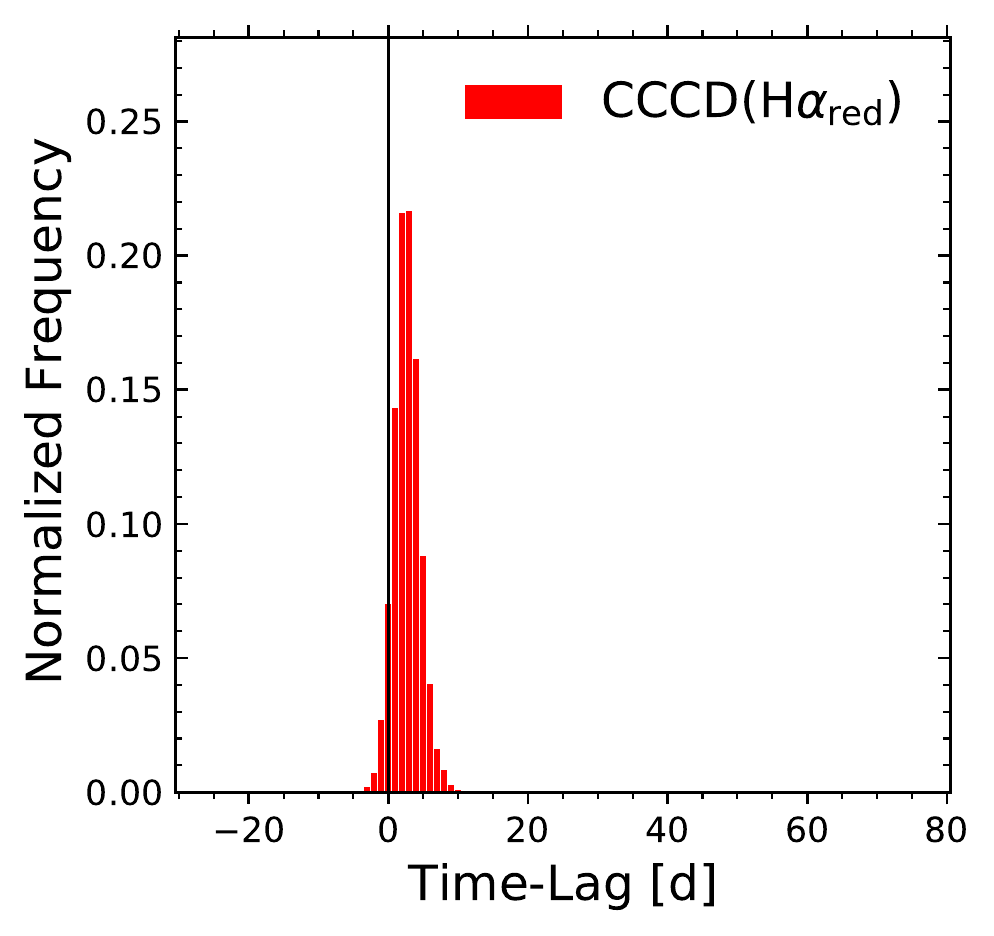}
         \end{minipage}
        \caption{}
    \end{subfigure}
    \vfill
    \begin{subfigure}[b]{0.48\textwidth}
         \centering
         \begin{minipage}[t]{0.48\textwidth}
            \includegraphics[width=\textwidth]{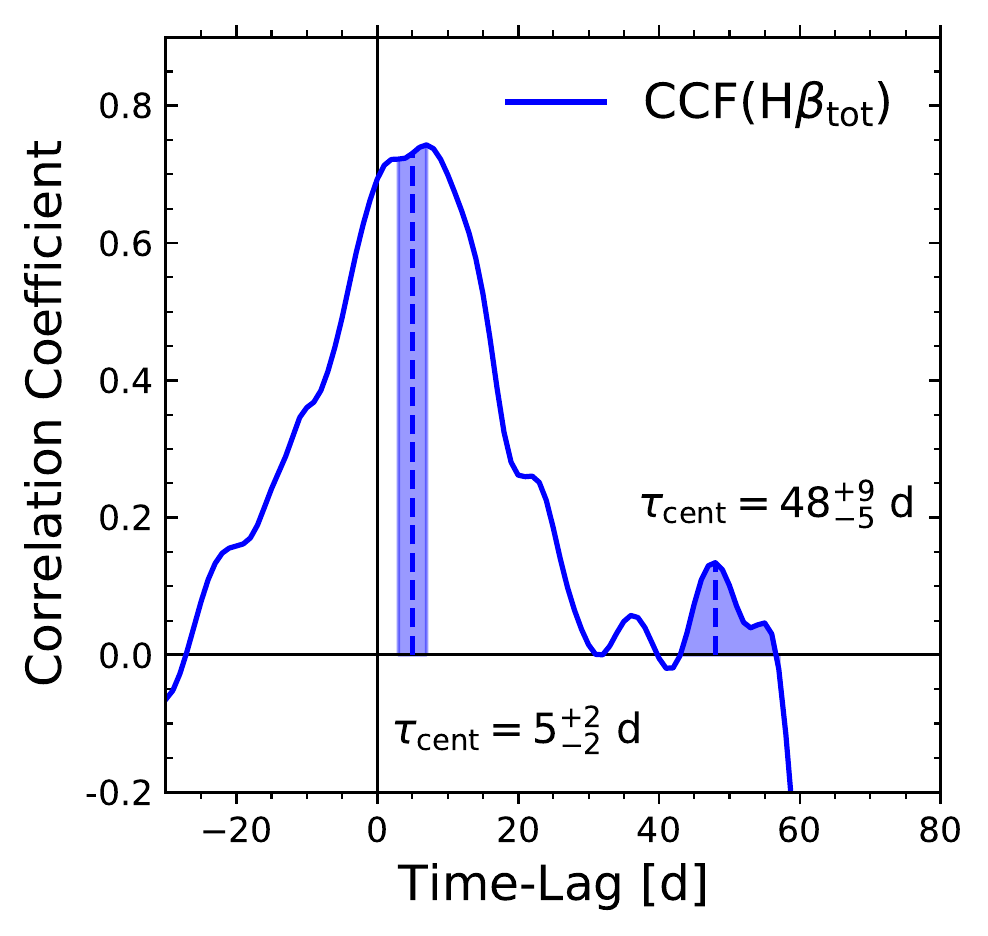}
         \end{minipage}
         \hfill
         \begin{minipage}[t]{0.48\textwidth}
            \includegraphics[width=\textwidth]{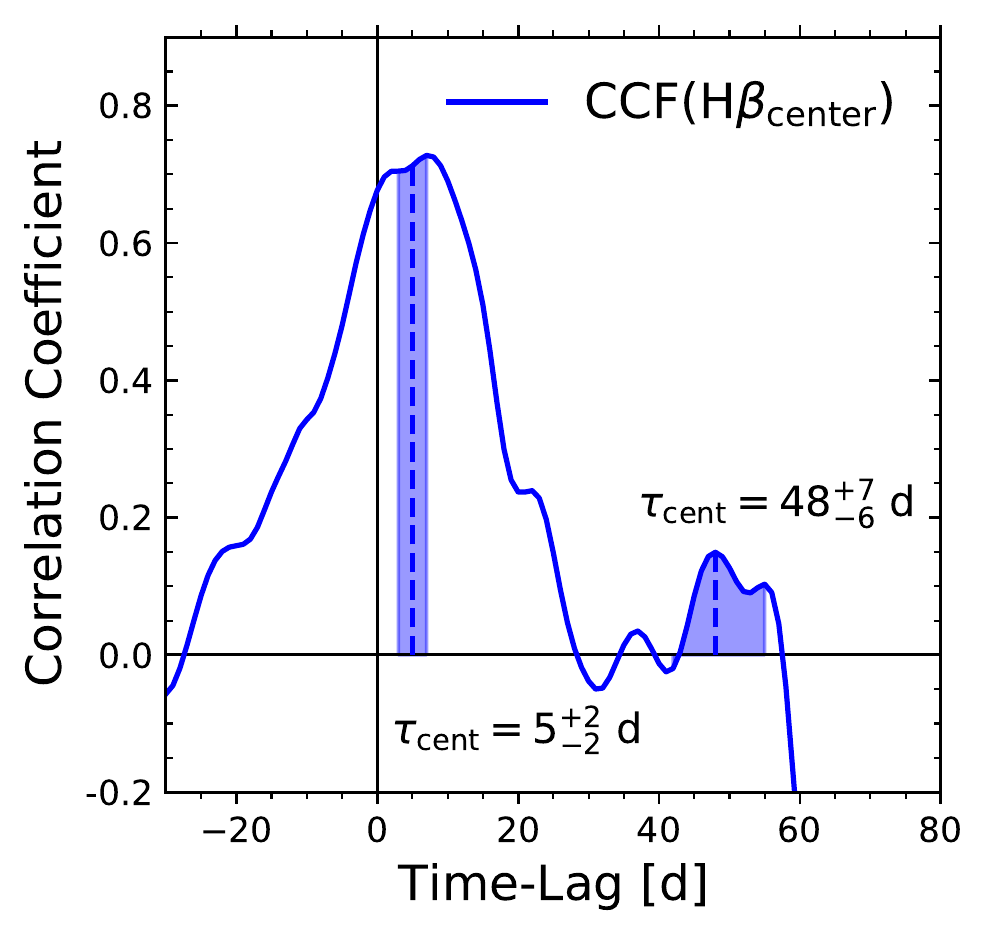}
         \end{minipage}
         \vfill
         \begin{minipage}[t]{0.48\textwidth}
            \includegraphics[width=\textwidth]{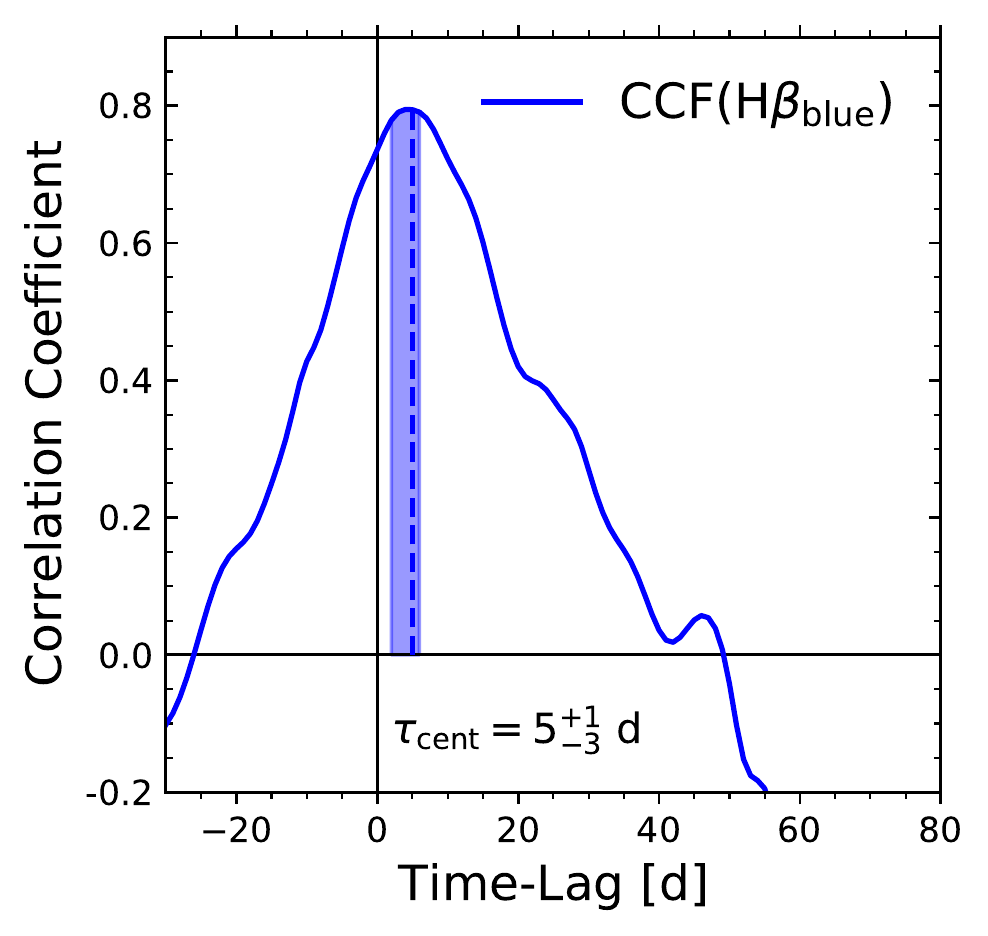}
         \end{minipage}
         \hfill
         \begin{minipage}[t]{0.48\textwidth}
            \includegraphics[width=\textwidth]{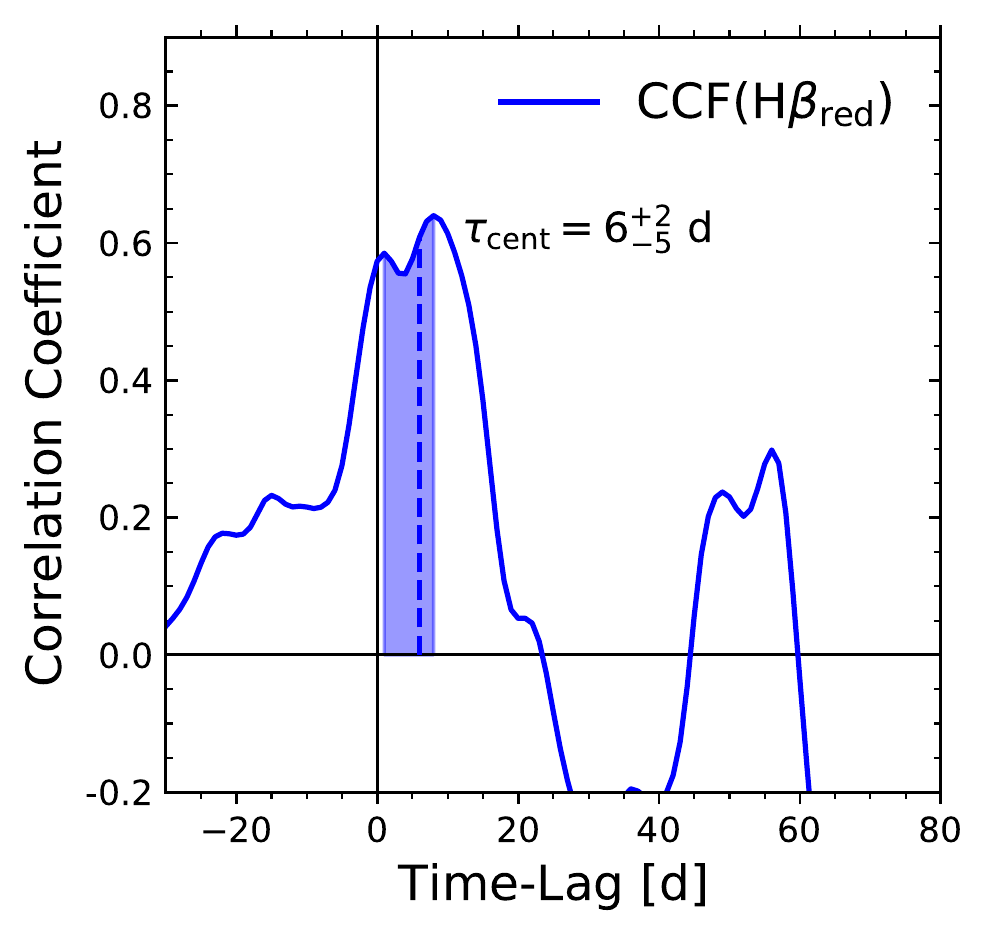}
         \end{minipage}
        \caption{}
        \label{fig:test}
    \end{subfigure}
    \hfill
    \begin{subfigure}[b]{0.48\textwidth}
         \centering
         \begin{minipage}[t]{0.48\textwidth}
            \includegraphics[width=\textwidth]{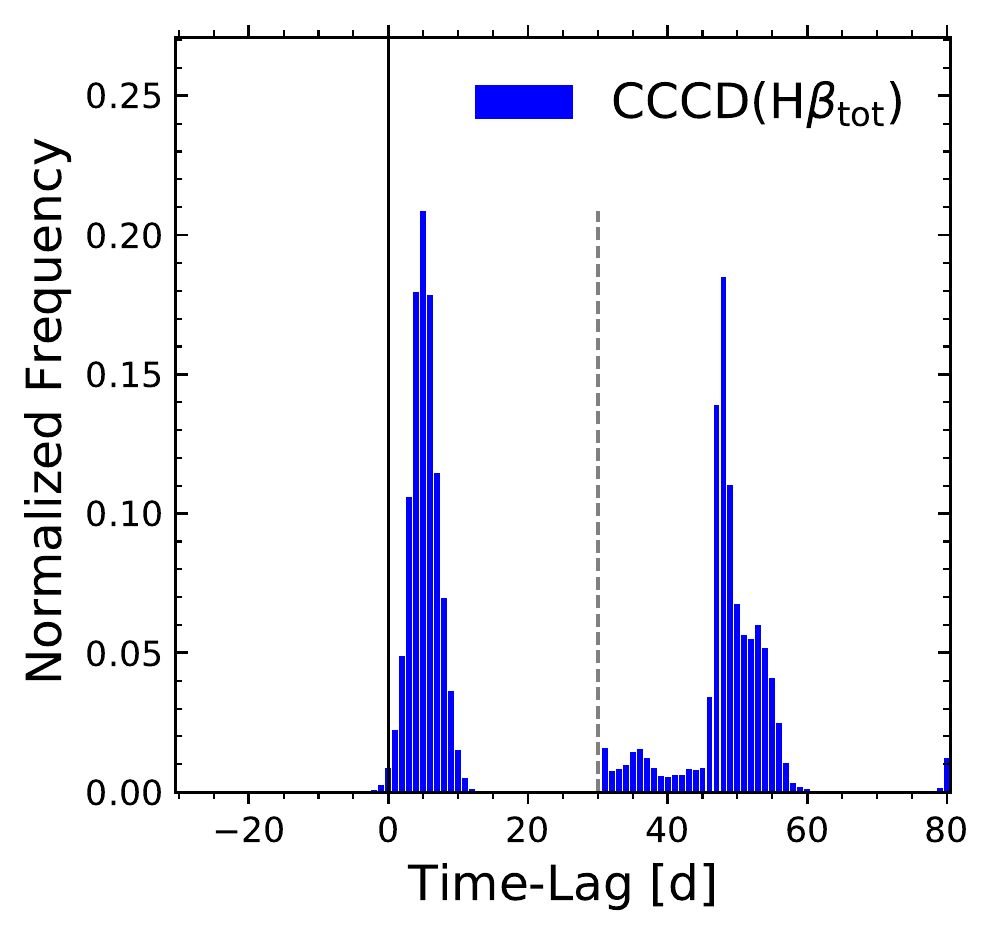}
         \end{minipage}
         \hfill
         \begin{minipage}[t]{0.48\textwidth}
            \includegraphics[width=\textwidth]{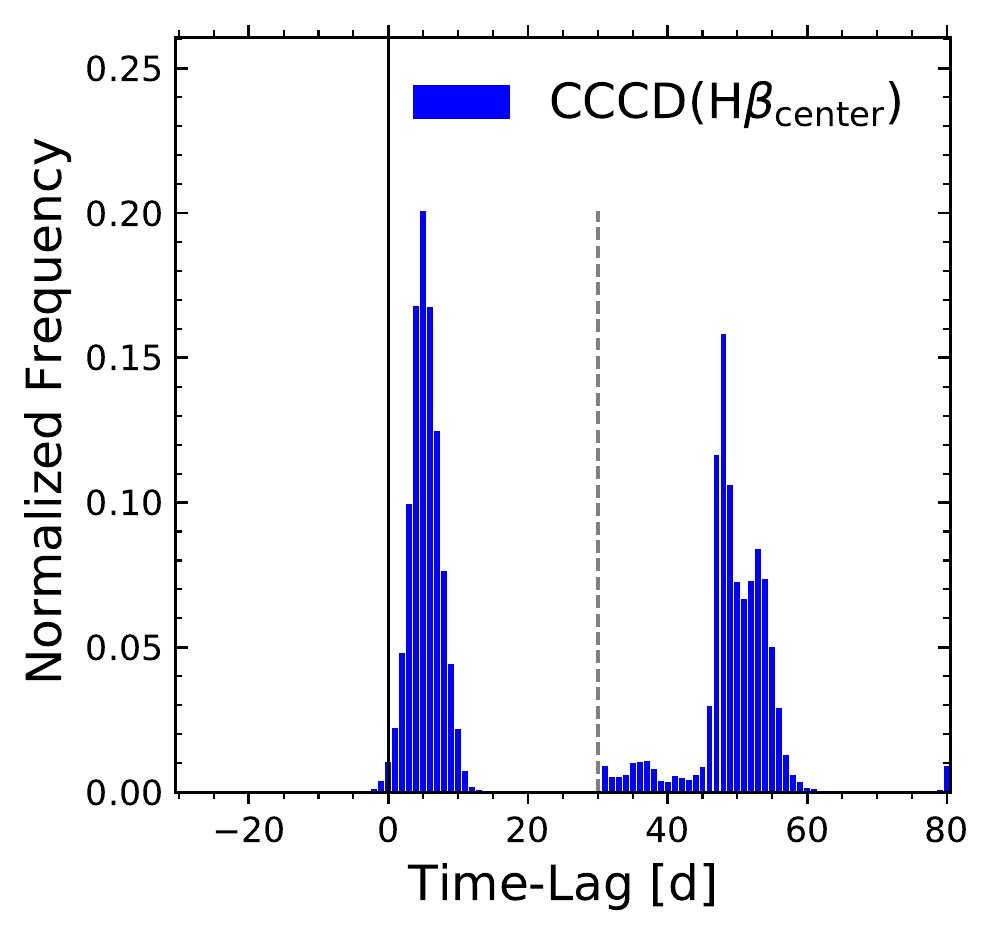}
         \end{minipage}
         \vfill
         \begin{minipage}[t]{0.48\textwidth}
            \includegraphics[width=\textwidth]{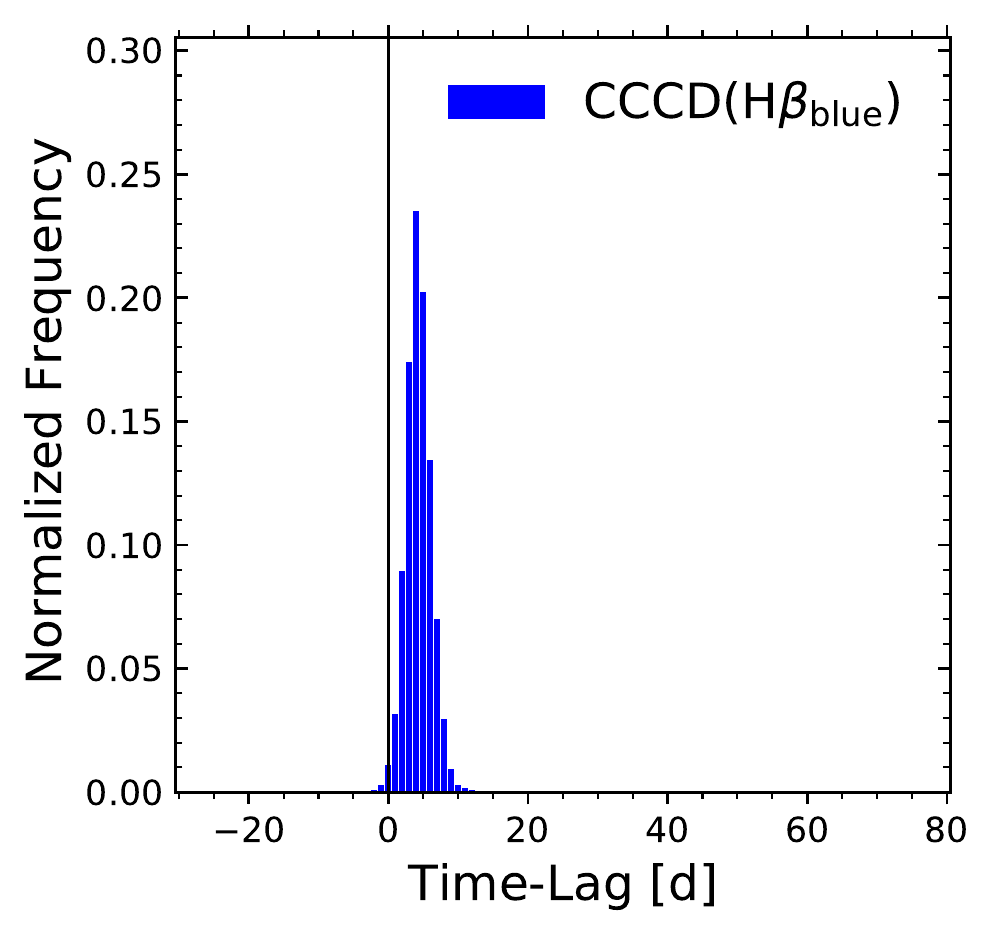}
         \end{minipage}
         \hfill
         \begin{minipage}[t]{0.48\textwidth}
            \includegraphics[width=\textwidth]{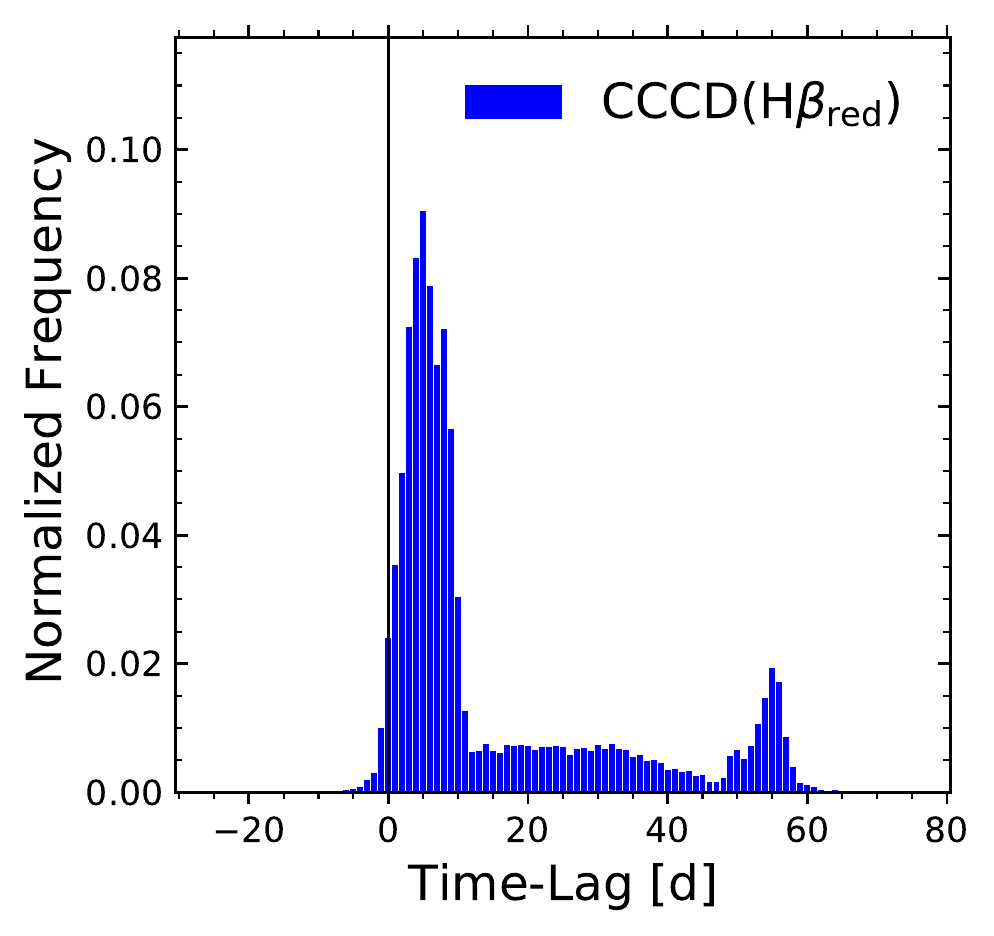}
         \end{minipage}
        \caption{}
    \end{subfigure}
    \vfill
    \begin{subfigure}[t]{0.48\textwidth}
         \centering
         \begin{minipage}[t]{0.48\textwidth}
            \includegraphics[width=\textwidth]{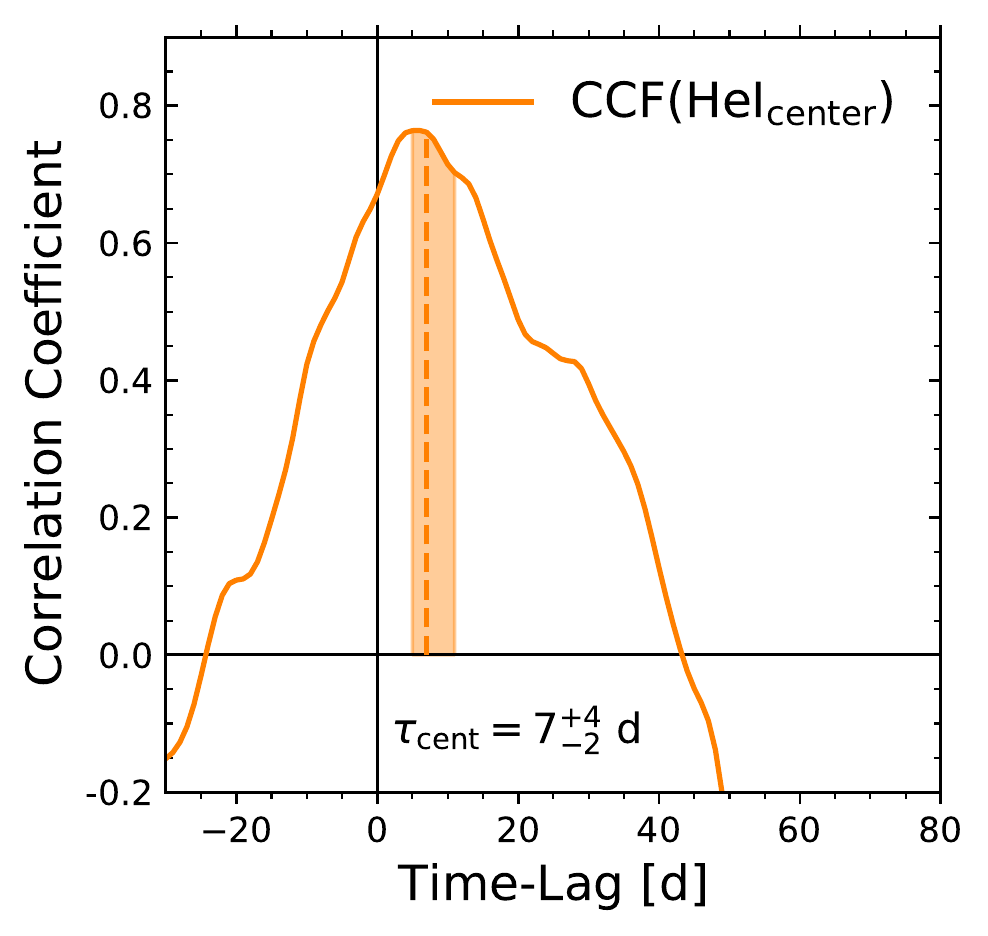}
            \caption{}
         \end{minipage}
         \hfill
         \begin{minipage}[t]{0.48\textwidth}
            \includegraphics[width=\textwidth]{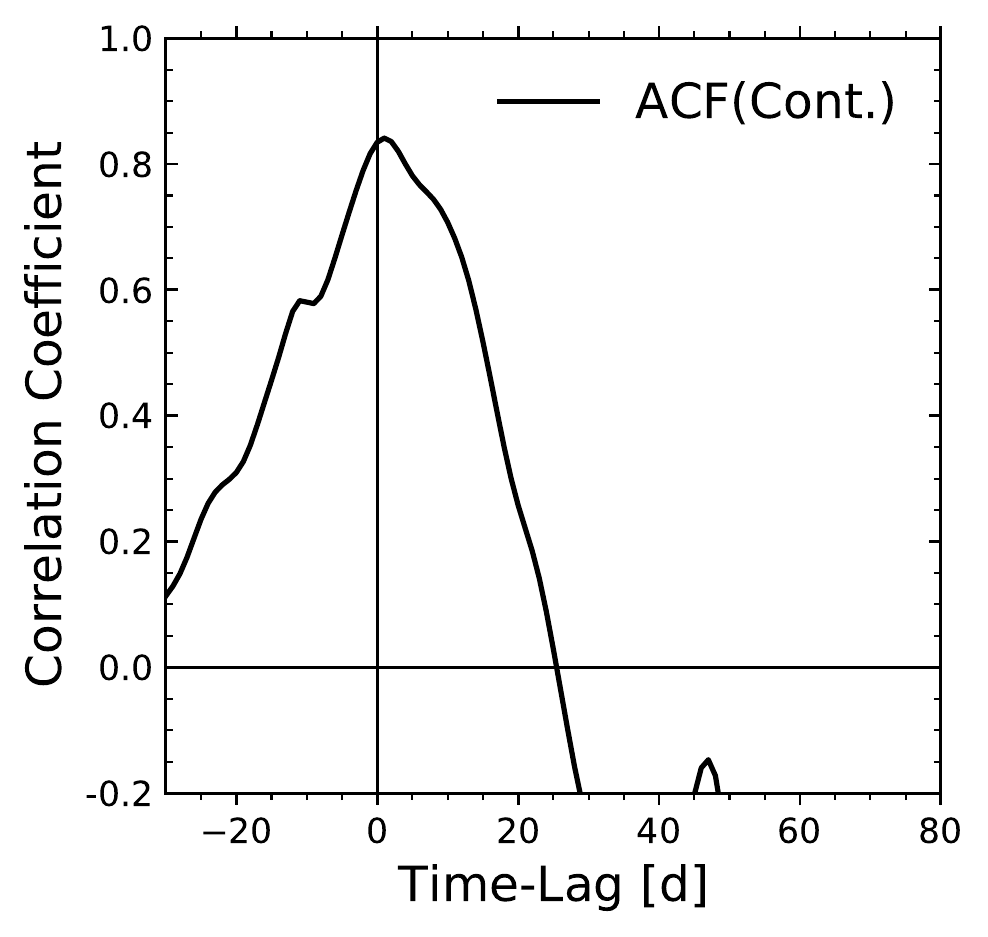}
            \caption{}
         \end{minipage}
    \end{subfigure}
    \hfill
    \begin{subfigure}[b]{0.48\textwidth}
         \centering
         \begin{minipage}[t]{0.48\textwidth}
            \includegraphics[width=\textwidth]{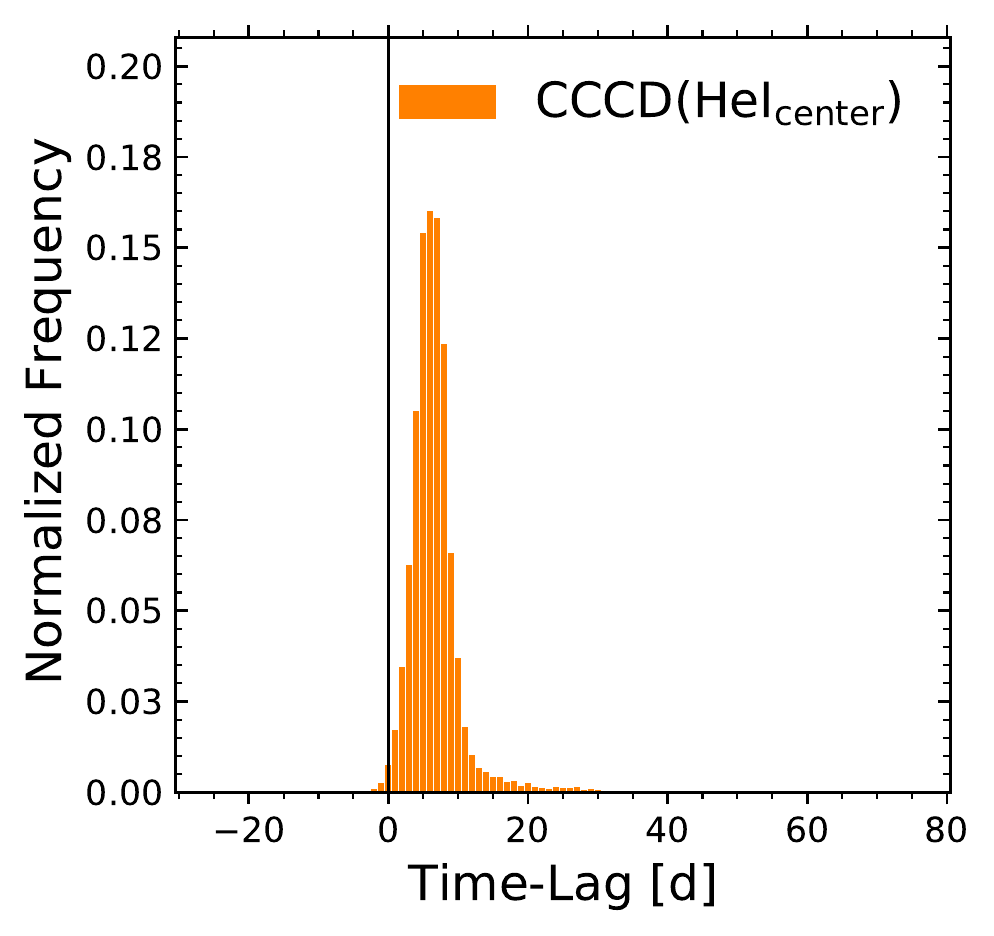}
            \caption{}
         \end{minipage}
         \hfill
         \begin{minipage}[t]{0.48\textwidth}
            \includegraphics[width=\textwidth]{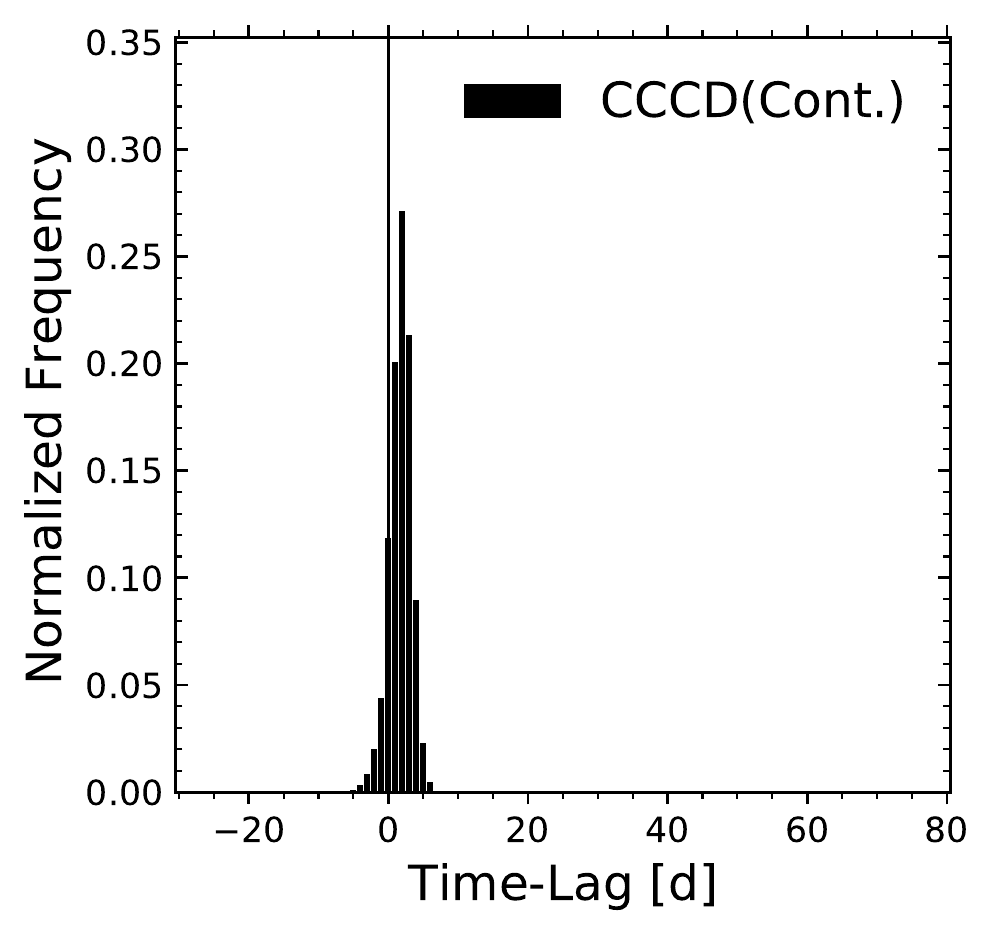}
            \caption{}
         \end{minipage}
    \end{subfigure}
    \caption{Left panel: CCFs of the integrated \Ha{} \textit{(a)} and \Hb{} \textit{(c)} line, of their segments (center, blue, red), and of the  central \ion{He}{I} line \textit{(e)} with respect to the combined continuum at 5180\,\AA{} (rest frame). The time lag, $\tau_{\rm cent}$, is denoted by a dashed line, with the shaded area corresponding to a $\pm1\sigma$ interval.  The CCF of the spectroscopic continuum light curve with respect to the combined (photometric plus spectroscopic) continuum reference light curve is shown in \textit{(f)}.\ Right panel: CCCDs of the integrated \Ha{} \textit{(b)} and \Hb{} \textit{(d)} line, of their segments (center, blue, red), and of the central \ion{He}{I} line \textit{(g)} with respect to the combined continuum at 5180\,\AA{} (rest frame). The boundary between CCCDs of individual peaks is denoted by a dashed gray line.  The CCCD of the spectroscopic continuum light curve with respect to the combined continuum reference light curve is shown in \textit{(h)}.}
    \label{mrk926_CCF_RSS}
\end{figure*}

The mean distances of the broad emission-line regions to the central ionizing source can be determined by correlating the broad emission-line light curves with the light curve of the ionizing continuum flux. Normally, an optical continuum light curve is used as surrogate for the ionizing light curve. For this study, we correlated the integrated light curves as well as the segment light curves (center, blue, red) of the  Balmer lines \Ha{} and \Hb{}, and the \ion{He}{I}\,$\lambda 5876_{\rm center}$ line with the combined 5180\,\AA{} (rest frame) and V-band continuum light curve (see Fig.~\ref{ochmLCcombiWisHet.pdf} in Sect.~\ref{sec:continuum_spectral_variations}). For \Hg{},  the boundary effects mentioned in Sect.~\ref{sec:mean_rms_profiles} prevented a reliable light curve extraction, and therefore no cross-correlation function (CCF) was calculated. The correlation technique we used is a variant of the discrete correlation function \citep[DCF;][]{edelson88}. Instead of classically binning the data points (i.e.,\ using a rectangular weighting function), we performed a weighted averaging of the data points using a Gaussian kernel $b_\tau(t) = (1/\sqrt{2\pi h}) \exp(-(\tau-t)^2/2h^2)$ as a smooth density function \citep[][and references therein]{rehfeld2011}. The width $h$ was chosen such that it was on the order of the mean sampling rate, which in our case translates to $\sim 2$ days. The Pearson correlation coefficient, $\rho_{x,y} = \text{COV}(x,y)/(\sigma_x \sigma_y),$ of two time series, $x$ and $y$, with time steps $t^x$ and $t^y$ then transforms to
\begin{equation}
    \text{CCF}(\tau) = \frac{\sum^{N_x}_{i=1}(x_i-\mu_x) \cdot (z^\tau_i-\mu_{z^\tau})}{N_x\sigma_x\sigma_z},
\end{equation}
with $z^\tau_i = \sum^{N_y}_{j=1}y_j \cdot b_\tau(t^y_j-t^x_i)$. The normalization of the CCF ensures that CCF$(\tau)\leq 1$ for all $\tau$. We note that the light curves were not detrended as the timescale of the campaign does not suffice to remove long-term trends. The derived cross-correlation functions, CCF$\left(\tau\right),$ are shown in Fig.~\ref{mrk926_CCF_RSS}.

We determined the centroids $\tau_\text{cent}$ of the CCFs by using only those parts of the CCFs above 80\% of the peak value. A threshold value of 0.8 $r_\text{max}$ is generally a good choice as has been shown before \citep{peterson04}. We derive the uncertainties of the time lags, $\tau_\text{cent}$, by calculating the cross-correlation lags a large number of times using a model-independent Monte Carlo method known as flux randomization/random subset selection  (FR/RSS). This method has been described by \cite{peterson98}. The resulting cross-correlation centroid distributions (CCCDs) for the integrated H$\alpha$, H$\beta$, and \ion{He}{I} lines as well as for their line segments (center, blue, red) with the combined 5180\,\AA\ (rest frame) and V-band continuum are presented in Fig. \ref{mrk926_CCF_RSS}. Each CCCD was determined after $2\times10^4$ independent runs (i.e., $2\times10^4$ independent subsamples) for each light curve. For the CCFs of the central Balmer line segments showing two distinct peaks, the CCCD was calculated for each peak individually. The final centroid time lags are given in Table~\ref{CCF_1D}. The error intervals correspond to 68\%  ($\pm 1\sigma$) confidence level.

The resulting CCFs and CCCDs are presented in Fig.~\ref{mrk926_CCF_RSS}. The CCFs of the integrated Balmer line light curves and of the central Balmer line component light curves (within $\pm 5000$\, \kms{}) exhibit two discrete peaks. In contrast, the blue and red Balmer segment light curves (i.e.,\ the light curves of the Balmer satellites) are single-peaked. The double-peaked CCFs of the integrated Balmer line light curves show delays of $8^{+2}_{-5}$ and $56^{+3}_{-6}$ days for \Ha{}, and $5^{+2}_{-2}$ and $48^{+9}_{-6}$ days for \Hb{}. The CCFs of the central Balmer line component light curves show delays of $10^{+3}_{-2}$ and $57^{+3}_{-3}$ days for \Ha{}, and $5^{+2}_{-2}$ and $48^{+7}_{-6}$ days for \Hb{}. The interpretation of these double-peaked CCFs is explained in more detail in Sect.~\ref{sec:2D_CCFs}.  The CCFs of the Balmer satellite light curves are single-peaked and show, in comparison to the CCFs of the integrated and central component light curves, much shorter delays of only $3 - 5$ light-days (i.e., the Balmer satellites react promptly to the variations in the continuum). There are indications for a secondary peak in the red Balmer satellite CCF of \Hb{} (see Fig.~\ref{fig:test}); however, the light curve of this wing is relatively noisy due to the strong underlying [\ion{O}{III}] lines (see Sect.~\ref{sec:velo_resolved_discussion} for more information). The CCF of the blue \Hb{} Balmer satellite shows no indication of a secondary peak.

We also calculated the CCFs and CCCDs for the weak HeI\,$\lambda 5876$ line. The outer wings are heavily contaminated by absorption. Therefore, out of precaution we excluded the blue and red wing from the integrated line light curve, leaving only the central profile light curve. In contrast to the Balmer lines, the CCF of the central profile light curve  only shows  a singular peak at a delay of $7^{+4}_{-2}$ days and no evidence for a secondary peak. Instead, the singular peak is very broad and high correlation values are found for delays up to $\sim 40$ days. 

The overall lags determined by the CCF analysis do not depend significantly on the use of either the purely spectroscopic or the combined spectroscopic and photometric continuum light curve. Both driving light curves recover the same responses of the emission lines with only minor variations of $\pm 2$ days at maximum. In comparison to the spectroscopic driving light curve, the use of the combined continuum light curve generates more stable CCFs and narrower, more confined CCCDs.

\begin{table}[h!]
    \centering
    \tabcolsep+3mm
    \caption{Cross-correlation lags of the integrated \Ha{} and \Hb{} line light curves as well as of their line segments (center, blue, red) with respect to the combined 5180\,\AA\ (rest frame) and V-band continuum light curve.}
    
    \begin{tabular}{@{\hspace*{0.15cm}}l@{\hspace*{0.15cm}}|c|c|c|c|c|c}
    \hline \hline 
    \noalign{\smallskip}
    & \multicolumn{6}{c}{Line}\\ 
     & \multicolumn{2}{c|}{\Ha{}} & \multicolumn{2}{c|}{\Hb{}} & \multicolumn{2}{c}{\ion{He}{I}}\\
     \cline{2-7}
    \multirow{2}{*}{Segment}    & \multicolumn{2}{c|}{$\tau$ [days]} & \multicolumn{2}{c|}{$\tau$ [days]} & \multicolumn{2}{c}{$\tau$ [days]}\\
        &{\tiny inner} & {\tiny outer} & {\tiny inner} & {\tiny outer} & {\tiny inner} & {\tiny outer}\\
    \noalign{\smallskip}
    \hline
    \noalign{\smallskip}
    total   & $8^{+2}_{-5}$     & $56^{+3}_{-6}$    & $5^{+2}_{-2}$ & $48^{+9}_{-5}$    & -             & -\\ [0.7ex]
    center  & $10^{+3}_{-2}$    & $57^{+3}_{-3}$    & $5^{+2}_{-2}$ & $48^{+7}_{-6}$    & $7^{+4}_{-2}$ & -\\ [0.7ex]
    blue    & $4^{+1}_{-3}$     & -                 & $5^{+1}_{-3}$ & -                 & -             & -\\ [0.7ex]
    red     & $3^{+2}_{-2}$     & -                 & $6^{+2}_{-5}$ & -                 & -             & -\\
    \noalign{\smallskip}
    \hline 
    \end{tabular}
    \label{CCF_1D}
\end{table}

\subsection{Velocity-resolved CCFs of the \Ha{} and \Hb{} line}\label{sec:2D_CCFs}
In Sect.~\ref{sec:1D_CCFs} we calculated the time lags of the integrated \Ha{}, \Hb{}, and HeI lines as well as of the central and outer wing regions with respect to the combined V-band continuum at 5180\,\AA{} (rest frame). Now,  we investigate the profile variations in the \Ha{} and \Hb{} lines in more detail by calculating the lags of individual line segments. For the weak HeI\,$\lambda 5876$ line, the absorption in the central line segment and in the line wings prohibits a clear determination of a velocity-resolved CCF. The way we proceed has been described before in our studies of line profile variations in Mrk\,110 \citep{kollatschny02, kollatschny03}, Mrk\,926 \citep{kollatschny10}, 3C120 \citep{kollatschny14} and HE\,1136-2304
\citep{kollatschny18}.

\begin{figure*}[!htp]
    \centering
    \begin{subfigure}[t]{0.495\textwidth}
        \centering
        \includegraphics[width=\linewidth]{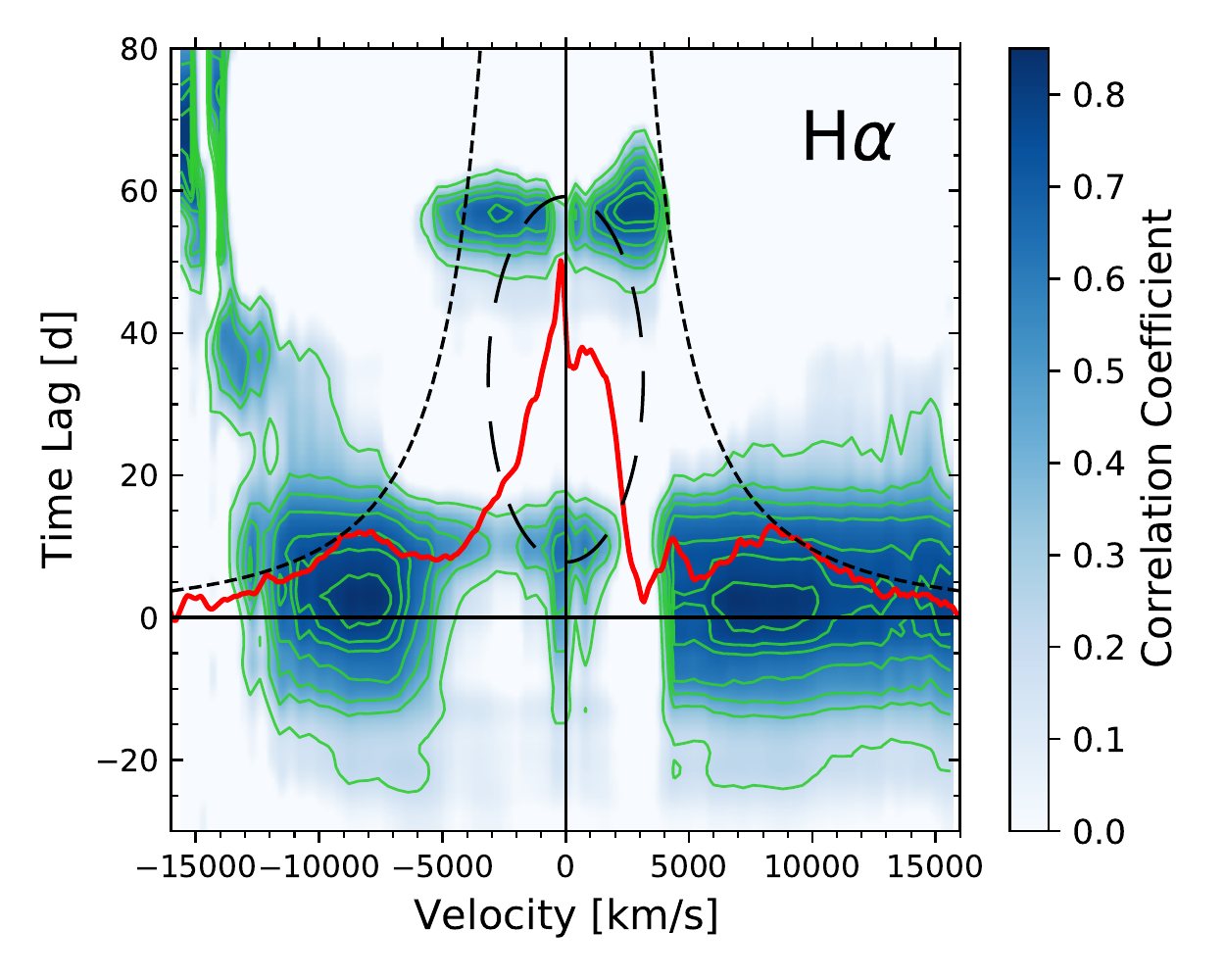}
        \caption{}
        \label{ochm2DCCF_Ha_comb.pdf}
    \end{subfigure}
   \hfill
    \begin{subfigure}[t]{0.495\textwidth}
        \centering
        \includegraphics[width=\linewidth]{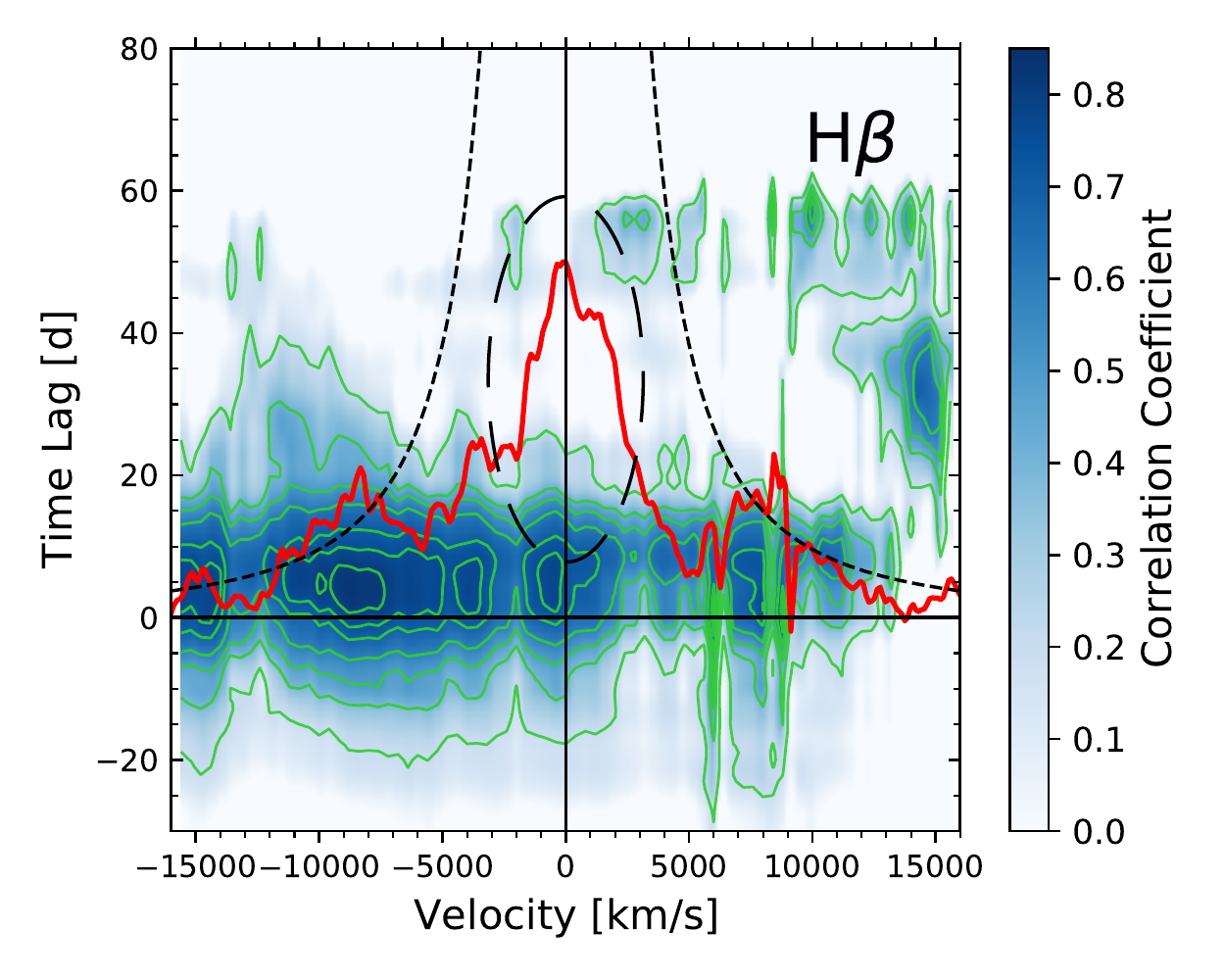}
        \caption{}
        \label{ochm2DCCF_Hb_comb.pdf}
    \end{subfigure}
    \vspace{10pt}
    \caption{Velocity-resolved CCFs($\tau$,$v$) showing the correlation coefficient of the \Ha{} \textit{(a)} and \Hb{} \textit{(b)}  line segment light curves ($\Delta v \sim 400$\,\kms{}) with respect to the combined continuum light curve
    as a function of velocity and time delay (blue scale). Contours of the correlation coefficients are plotted at levels 0.2, 0.4, 0.5, 0.6, 0.7, 0.75, 0.8, and 0.825 (green lines). In each plot, the rms profile of the emission line is shown in red. The dashed curve shows the escape velocities for a Keplerian disk inclined by $\sim 50^{\circ}$ with a central mass of 1.1 $\times 10^{8} M_{\odot}$.  The dashed ellipse corresponds to a circular Keplerian orbit at $R/c = 33.5$\,d inclined by $\sim50^\circ$ to the line of sight mapped onto the velocity-delay plane. The line center ($v =0$\,\kms{}) and the time delay of $\tau = 0$ days are indicated by  a vertical and a horizontal black line, respectively.  The delays computed for the red wing of \Hb{} are disturbed by the [\ion{O}{III}]\,$\lambda$ 4959, 5007 lines, which results in higher-delay residuals not confined within the viral envelope.}
    \label{fig:velocity_resolved_CCFs}
\end{figure*}

We sliced the velocity profiles of the continuum-subtracted Balmer lines into velocity segments with a width of $\Delta~v = 400$ \kms{}. This value of 400 \kms{} corresponds to the spectral resolution of our observations. A central line-segment was integrated from $v = -200$ to $+200$\,\kms{}. Afterward, we measured the intensities of all subsequent line-of-sight velocity segments from $v = -15\,800$ to $+15\,800$\,\kms{} and compiled their light curves. We computed the CCF$\left(\tau\right)$ for each individual line-segment light curve ($\Delta{}v=400$\,\kms{}) of \Ha{} and \Hb{} with respect to the combined 5180\,\AA{} (rest frame) continuum light curve. In this way, we computed velocity-resolved CCFs of \Ha{} and \Hb{} as a function of distance to the line center (blue scale). These velocity-resolved CCFs are shown in
Figs.~\ref{ochm2DCCF_Ha_comb.pdf} and \ref{ochm2DCCF_Hb_comb.pdf}. The green lines in  these figures delineate the contour lines of the correlation coefficient at different levels (0.2, 0.4, 0.5, 0.6, 0.7, 0.75, 0.8, 0.825). The red line shows the rms line profile for comparison.

The \Ha{} line shows two velocity-delay structures in the central segment (within $\pm 5000$\, \kms{}) at  a distance of $57^{+3}_{-3}$ and $10^{+3}_{-2}$ light-days  (see Sect.~\ref{sec:1D_CCFs}). Adopting the interpretation of \citet{horne21} for the velocity-delay maps of the NGC\,5548 STORM campaign,  this structure might be interpreted as the upper and lower half of an ellipse in the velocity-delay plane that might be the signature of a line-emitting ring orbiting the black hole (BH) at a radius $R = 33.5$ light-days. The stronger far side of the annulus extends to $\tau = (R/c)\,(1+\sin i) \approx 57$ days, and the weaker near side of the annulus has a delay  of $\tau = (R/c)\,(1-\sin i)$ $\approx 10$ days. This gives $\sin i \sim 0.71$ and thus $i \sim 45^{\circ}$ as the inclination angle of an assumed thin disk. 

The \Hb{} line also shows two velocity-delay structures in the central segment at $48^{+7}_{-6}$ and $5^{+2}_{-2}$ light-days (see Sect.~\ref{sec:1D_CCFs}). A possible line-emitting ring orbiting the BH therefore has a radius $R = 26.5$ light-days. The weaker far side of the annulus then extends to $\tau = (R/c)\,(1+\sin i) \approx 48$ days, and the stronger near side of the annulus has a delay at $\tau = (R/c)\,(1-\sin i) \approx 5$ days.  This gives $\sin i \sim 0.81$ and thus $i \sim 54^{\circ}$ of the disk. Therefore, we adopt $i \sim 50^{\circ}$ as the mean disk inclination angle.

\subsection{Central black hole mass} \label{sec:bh_mass}

The masses of the central BHs in AGN can be estimated -- based on the assumption that the gas dynamics are dominated by the central massive object -- by evaluating 
\begin{equation}
    M = f\,c\,\tau\,\Delta\,v^{2}\, G^{-1}.
    \label{eq:BH_mass}
\end{equation}
The characteristic velocity $\Delta v$ of the emission-line region can be estimated from the FWHM of the rms profiles or from the line dispersions $\sigma_\text{line}$. Furthermore, the distance $c\,\tau$ of the line-emitting regions from the ionizing source  can be estimated from the delayed response of the line light curves to the continuum variations. Characteristic distances of the individual line-emitting regions are given by the centroid $\tau_\text{cent}$ of the individual CCFs of the emission-line variations relative to the continuum variations \citep[e.g.,][]{koratkar91,kollatschny97}.

The scaling factor $f$ in Eq.~\ref{eq:BH_mass} is on the order of unity and depends on the kinematics, structure, and orientation of the BLR. This scaling factor differs by $\sim0.4$~dex rms from galaxy to galaxy, for example, depending on whether we see the central accretion disk including the BLR from the edge or face-on.  Typically, the scaling factor $f$ is discussed with respect to the line dispersion $\sigma_\text{line}$ (rms). Empirically found $f$ values are $f=5.5$ \citep[e.g.,][]{onken04}, $f=4.31$ \citep[e.g.,][]{grier13b}, or $f=3.6$ \citep{graham11}. The very broad lines in  Mrk\,926 (see Sect.~\ref{sec:mean_rms_profiles}) and the high disk inclination angle of $i \sim 50^{\circ}$ (see Sect.~\ref{sec:2D_CCFs}) indicate that Mrk\,926's orientation toward us is rather edge-on in comparison to other AGN.
Therefore, we adopt an $f$ value of $f=3.6$ in order to calculate the BH mass. We note that the virial factor, $f$, for $\sigma_\text{line}$ and FWHM differs since FWHM/$\sigma_\text{line}$ typically takes values of $\sim 2$ \citep{peterson04}.
We account for that by following the procedure in \citet{kollatschny11, kollatschny13} and, for example, \citet{kollatschny18}: adopting a correction factor in order to obtain the true rotational velocity, $v$, from
the observed FWHM for each line. This reduces the effective $f$ factor of FWHM to $\sim1.2$. We use FWHM (rms), which generally gives more reliable results than FWHM (mean) due to the fact that the rms profiles only shows the varying fraction of the BLR line-emitting gas.
Based on the delays of the central Balmer line regions (see Table~\ref{CCF_1D} and Sect.~\ref{sec:2D_CCFs}) and on the line widths of the rms profiles (FWHM) (see Table~\ref{tab:line_widths}), we derive a weighted mean BH mass (see Table~\ref{bh_masses}) of 
\begin{equation}
    M = (1.1 \pm 0.2) \times 10^{8} M_{\odot}\ \ \ \text{or} \ \ \  \log(M/M_\odot) = 8.04 \pm 0.1. 
\end{equation}
The BH mass based on the HeI\,$\lambda 5876$ line has a larger error due to larger uncertainties in their FWHM.  However, it confirms the BH mass derived from the Balmer lines. The BH mass estimate based on the \Ha{} line dispersion $\sigma_\text{line}$ (mean) is by a factor of $\sim2$ larger than the BH estimate from the FWHM (rms). Considering the larger uncertainties of $\sigma_\text{line}$, it is still in agreement with the BH mass estimates from FWHM (rms).

\begin{table}[h!]
\centering
\tabcolsep+7.0mm
\caption{
 Black hole masses based on the FWHM  
 of the line profiles in the rms spectra and the line dispersion $\sigma_\text{line}$ 
 of the mean \Ha{} profile. The masses were calculated assuming an $f$ value of 3.6.}
\begin{tabular}{lcc}
\hline \hline 
\noalign{\smallskip}
Line                    & $M_{\text{BH, FWHM}}$    &  $M_{\text{BH},\sigma_\text{line}}$ \\
                        &  [$10^8 M_{\odot}$]      &  [$10^8 M_{\odot}$]                 \\
\noalign{\smallskip}
\hline
\noalign{\smallskip}
\Ha{}                   & 1.2 $\pm{}$ 0.4 & 2.8 $\pm{}$ 1.7 \\
\Hb{}                   & 1.1 $\pm{}$ 0.3 & -- \\
\noalign{\smallskip}
\hline
\noalign{\smallskip}
weigh. mean                    & 1.1 $\pm{}$ 0.2 & --\\
\noalign{\smallskip}
\hline\hline 
\noalign{\smallskip}
HeI\,$\lambda 5876$     & 1.0 $\pm{}$ 0.6 & --\\
\noalign{\smallskip}
\hline 
\end{tabular}

\label{bh_masses}
\end{table}

\section{Discussion}\label{sec:discussion}

\subsection {Optical variability}
\subsubsection{Optical variability amplitudes}
It is known that Mrk\,926 varied at least since the beginning of the 1990s \citep{kollatschny06}. A first variability campaign was carried out during the years 2004-2005 \citep{kollatschny10}. Back then, the optical continuum flux density at 5180\,\AA{} varied between 2.4 and 3.6 $\times 10^{-15}$ erg cm$^{-2}$ s$^{-1}$\,\AA$^{-1}$. In July 2010, the continuum flux density was significantly higher and amounted to a value of 6 $\times 10^{-15}$ erg cm$^{-2}$  s$^{-1}$\,\AA$^{-1}$ at 5180\,\AA{} (rest frame) at the beginning of the variability campaign presented here.
Immediately afterward, the optical continuum started to drop, finally reaching less than 50\% of its original intensity (see Fig.~\ref{ochmLCcombiWisHet.pdf} and Table~\ref{HET_cont_intens}) within only 2.5 months. Even more remarkably, the true non-stellar continuum in Mrk\,926 declines to only 30 -- 40\% continuum intensity between the beginning and end of the variability campaign taking into account the constant host galaxy contribution to the flux as determined by means of the FVG method (see Table~\ref{bvrhostflux}).

The variability amplitude in the blue spectral range is higher than in the red (see Table~\ref{variab_statistics} and the rms spectrum in Fig.~\ref{ochmavg_rms_spectra.pdf}). Interestingly, the continuum variation amplitudes in Mrk\,926 are much higher than those observed in other highly variable and even changing-look AGN. For example, during a post-outburst campaign in 2014-2015,  the changing-look AGN HE\,1136-2304 exhibited a fractional variation of $F_\text{V,var}$~=~0.11 and $F_\text{V,var}$~=~0.25 with and without host, respectively \citep{zetzl18}. Mrk\,926 exhibits fractional variation of $F_\text{V,var}$~=~0.22 and $F_\text{V,var}$~=~0.31 with and without host, respectively. The fractional variation depends on the duration of the monitoring campaign, on the examined wavelength, and on the (accurate) subtraction of the host galaxy contribution. A typical value for the fractional variation $F_\text{V,var}$ of the continuum at roughly 5100\,\AA\  is 0.05 to 0.15 for variability periods of 6 to 12 months. For longer campaigns,  typical $F_\text{V,var}$ values are higher and range from 0.1 to 0.25 \citep[][and references therein]{zetzl18}. On timescales of years, a decline of the continuum  to only 20\% of its original value has been observed for some sources \citep[e.g., Fairall 9;][]{kollatschny85}. However, Mrk\,926 was observed for a much shorter period and showed a similarly drastic decline. To summarize, it can be stated that the optical continuum variations in Mrk\,926 are very strong. 

\subsubsection{Signature of an accretion disk in Mrk\,926}\label{sec:signature_AD}
We now intend to test whether the observed continuum spectral index in the optical
is consistent with theoretical predictions from the Shakura \& Sunyaev accretion disk model \citep[SS73;][]{shakura73} .
Specifically, the Shakura \& Sunyaev model predicts $f_\nu \propto \nu^{1/3}$ over the self-similar part of the spectrum, that is, far away from the spectral bands affected by the inner (UV to extreme UV) and outer (far-infrared) disk radii. This holds for Mrk\,926 over the optical range. The spectral index of $\alpha = {1/3}$ for $f_\nu$ then translates to $\beta = {-7/3}$ for $f_\lambda$ as $f_\lambda = f_\nu d\nu/d\lambda \sim  v^{1/3}v^2 \sim\lambda^{-7/3}$. 

\begin{figure}[h!]
\centering
\includegraphics[width=9.8cm,angle=0]{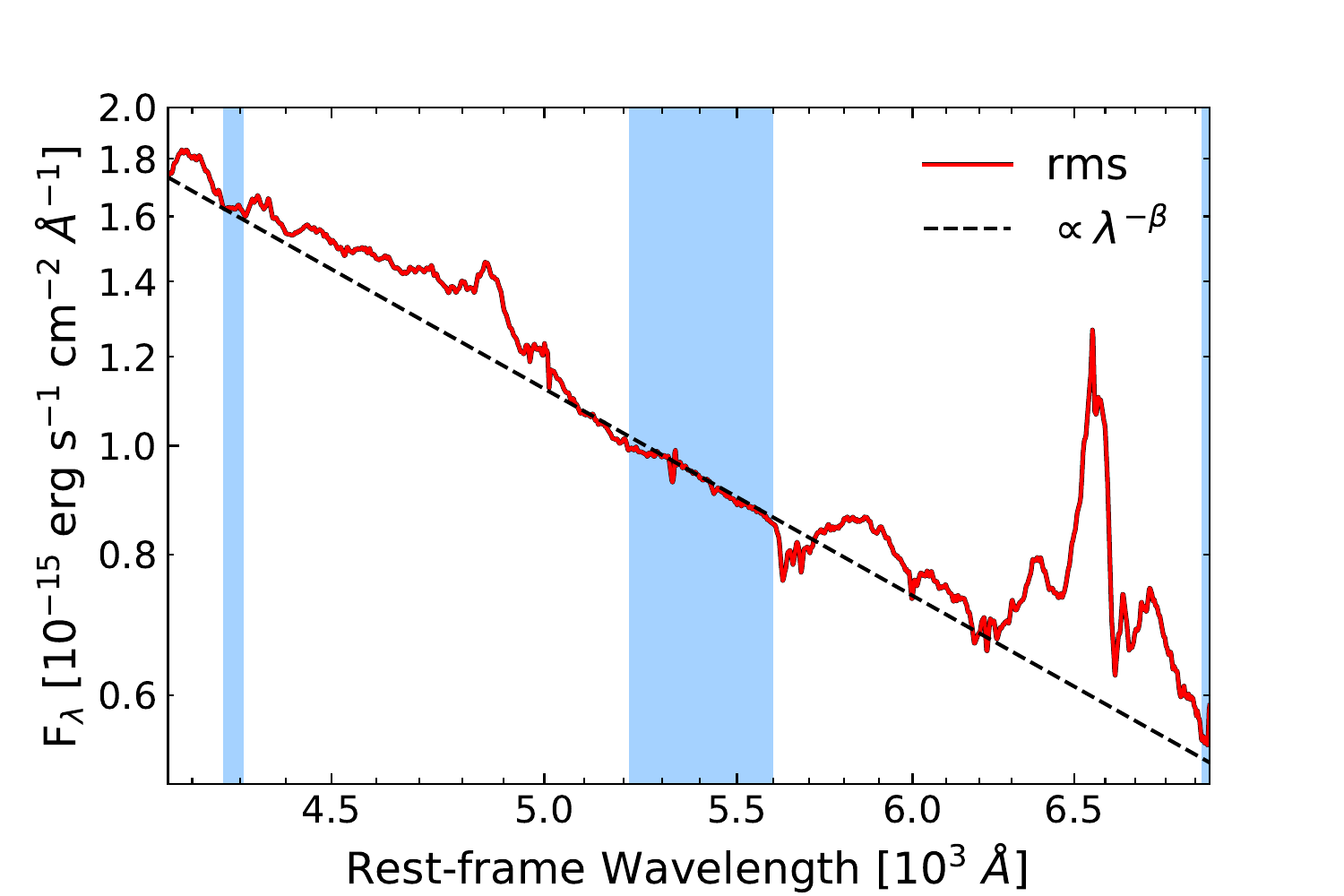}
\caption{RMS spectrum of the HET campaign corrected for Galactic foreground extinction. A power-law model was fitted to the continuum, giving a spectral index $\beta=2.33\pm0.02$.}
\label{fig:power_law}
\end{figure}

Figure\,\ref{fig:power_law} shows the observed rms spectrum corrected for Galactic foreground extinction. A power-law model $F_{\lambda} \propto \lambda^{-\beta}$ was fitted to the continuum (blue shaded areas), giving a spectral index $\beta=2.33\pm0.02$. The rms spectrum of the continuum is thus consistent with a power- law with a spectral index of  $\beta = 7/3$ as predicted by the Shakura \& Sunyaev model.

\subsubsection{Balmer decrement variability}
\begin{figure}[h!]
\centering
\includegraphics[width=0.375\textheight,angle=0]{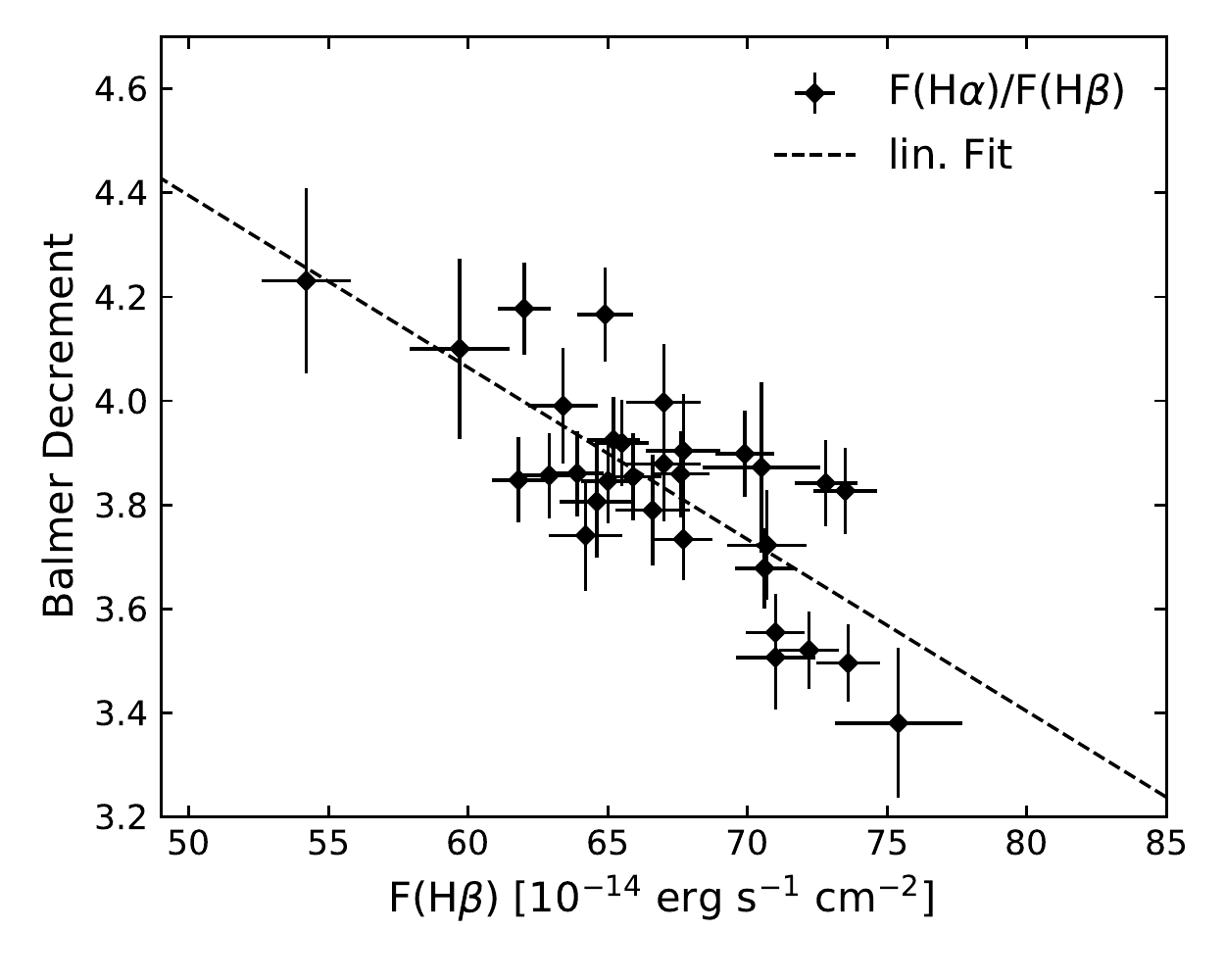}
\caption{Balmer decrement, $F({\rm H}\alpha)/F({\rm H}\beta)$, of the broad-line components versus broad-line \Hb{} intensity. The dashed line on the graph represents the linear regression.}
\label{Balmerdec_vs_Fbeta.pdf}
\end{figure}

The Balmer decrement $F({\rm H}\alpha_{\rm narrow})/F({\rm H}\beta{}_{\rm narrow})$ of the narrow-line components in Mrk\,926 amounts to 2.70 (see Table~\ref{NEL-intensities}). This corresponds to the expected theoretical line ratio (Case B) without any reddening. The lack of reddening toward the narrow-line region (NLR) is consistent with the insignificant reddening implied by the RMS fit to a Shakura \& Sunyaev model (see Sect.~\ref{sec:signature_AD}). In contrast to the constant Balmer decrement of the narrow lines, the Balmer decrement of the broad-line components takes values of 3.4 to 4.2. Figure\,\ref{Balmerdec_vs_Fbeta.pdf} shows the Balmer decrement $F({\rm H}\alpha_{\rm broad})/F({\rm H}\beta_{\rm broad})$  versus broad-line \Hb{} flux.

The Balmer decrement varies as a function of the broad \Hb{} flux. More precisely, the Balmer decrement shows a roughly linear dependence on the \Hb{} flux, and increases with decreasing \Hb{} broad-line intensity. Similar, although even stronger variations in the Balmer decrement have been found before in  highly variable Seyfert galaxies as, for example, NGC\,7603 (\citealt{kollatschny00}) or HE\,1136-2304 (\citealt{kollatschny18}). The broad-line Balmer decrement in these galaxies increased up to values of 7.3 in combination with decreasing H$\beta$ line intensities. These observations might be explained by optical depth effects in the BLR. This is in accordance with the finding that H$\alpha$ often originates at larger distances than H$\beta$. It has been discussed by \citet{korista04} that the radial stratification in the BLR is a result of optical-depth effects of the Balmer lines: the broad-line Balmer decrement decreases in high continuum states and increases in low states.

\subsection {H$\beta$  lag versus optical continuum luminosity}
Now, we test whether Mrk\,926 follows the general trend of the H$\beta$-lag and optical continuum luminosity relationship (R$_\text{BLR}$-L$_\text{AGN}$) for AGN \citep[e.g.,][and references therein]{kaspi00, bentz13, kilerci15, grier17, kollatschny18}. We determined continuum luminosities 
$\log( \lambda\,L_\lambda/{\rm erg~s}^{-1})$ of 43.68 to 
44.13 (1.87 to 5.30 $\times10^{-15}$\,erg\,s$^{-1}$\,cm$^{-2}$\,\AA$^{-1}$) in the optical at 5180\,\AA{} (rest frame) after correction for the contribution of the host galaxy. Furthermore, we derived a mean distance of 26.5 light-days for the H$\beta$ line-emitting region based on the delay of the H$\beta$ line variability with respect to the optical continuum light curve. 
This derived distance of 26.5 light-days for the H$\beta$ line-emitting region and the observed continuum luminosities closely follows the general$R_\text{BLR}$-$L_\text{AGN}$ relation with a slope of $\alpha \sim0.53$ \citep{bentz13}. 

\subsection{Line profiles}\label{sec:mean_rms_profiles_discussion}
\subsubsection{Mean and rms line profiles of the  2010 campaign}
The mean and rms profiles of the Balmer and \ion{He}{i}\,$\lambda 5876$ lines in velocity space are presented in Fig.~\ref{ochmprofmeanrms.pdf}. The mean and rms profiles of \Ha{} and \Hb{} are very similar. They exhibit line widths (FWHM) of $\sim$ 5250 \kms{} (mean profiles) and line widths of $\sim$ 4000 \kms{} (rms profiles) (see Table~\ref{tab:line_widths}). These values are larger than the average. The  \Ha{} FWHM distribution for broad-line quasars -- based on many thousand emission-line galaxies from the Sloan Digital Sky Survey -- shows a clear maximum at about 2000 \kms{} and falls steadily to FWHM values of about 9000 \kms{} \citep{hao05}.
There is an additional inner red emission component to be seen at $\sim$ +1500 \kms{}. Such inner red components have been seen before in other variable AGN, for example Ark\,120 \citep{kollatschny81}, NGC4593 \citep{kollatschny97}, NGC7603 \citep{kollatschny00}, or HE\,1136-2304
\citep{kollatschny18}. Possible explanations for these red components might be disk components, additional radial motions, or opacity components. More comments regarding inner red line components can be found, for example, in \cite{gaskell18} or \cite{du18}. 

The Balmer  lines show symmetric additional outer-line components between $\pm 5000$ and $\pm 13\,000$ \kms{}, which we term Balmer satellites. We note that the red wing of \ion{He}{i}\,$\lambda 5876$ might also indicate the presence of such a line satellite; however, due to the interference by absorption, a clear detection cannot be claimed. In particular, the Balmer satellites components show up in the rms profiles and their relative variability amplitudes are stronger than that of the central components between $-5000$ and $+5000$ \kms{}. The fractional variations in the individual line components are given in Table~\ref{variab_statistics}. Five years before the variability campaign presented here, when Mrk\,926 was by 50\% fainter in the continuum compared to the beginning of our campaign, these outer components in the rms profiles were also present, but are now slightly more pronounced \citep{kollatschny10}. Back in 2005, the central line components were  noticeably broader when the source was in a lower state. The central Balmer line components exhibited FWHM rms of roughly 8500 \kms{} in 2005 in comparison to 4000\,\kms{} in 2010. The uncertainties in the line dispersion $\sigma_\text{line}$ of the broad emission lines determined in Sect.~\ref{sec:mean_rms_profiles} are relatively high. Generally, the line dispersion $\sigma_\text{line}$ is very sensitive to the line widths used for integration. However, in Mrk\,926, the uncertainties in the line widths -- full width at zero intensity (FWZI) -- are high because of the large line widths and the additional outer line-components. Therefore, we did not derive a FWHM/$\sigma$ to determine the height/radius ratio of the BLR as presented in other galaxies \citep[e.g.,][]{kollatschny18}. Typical values for FWHM/$\sigma$ in other galaxies are on the order of two or more. Only in very rare cases the $\sigma_\text{line}$ values are equal or larger than the FWHM values \citep{peterson04, kollatschny11, kollatschny13}. 

\subsubsection{Comparison of the 2010 and 2004-2005 campaigns}
We now compare the Balmer line profiles observed in 2010 with the Balmer line profiles observed during the campaign in 2004-2005. First, the continuum as well as broad-line flux in 2005 was by a factor of about two smaller than in 2010 \citep{kollatschny10}. Second, the comparison of the Balmer line profiles also reveals differences in the mean and rms profiles. A strong blue component was present in the mean profiles in 2005, but nearly disappeared in 2010.  A small residual of this blue component appears to be left in the rms profiles at around $-1500$ \kms{}. The narrow-line subtraction for the campaign of 2010 revealed a red component in the mean profile of \Ha{} at around $\pm 1200$\,\kms{} (see Sect.~\ref{sec:continuum_spectral_variations}) that is also discernible in the \Hb{} profile (see above). Moreover, the rms profiles show a variable red component with the same velocity. This inner red component is in agreement with a red and slightly stronger component already present in the mean and rms profiles of the campaign in 2004/05. In summary, both the mean and rms profiles of the campaigns in 2004/05 and 2010 show blue and red components at mirrored velocities. These components were stronger in 2004/05. Presumably, these symmetric (with respect to line-of-sight velocity) components are the signature of an accretion disk as accretion disks are expected to result in double-peaked profiles \citep[e.g.,][and references therein]{horne86, eracleous03, gezari07, shapovalova13, storchi17}.

The highly variable outer rms   Balmer satellite components were also present in 2005, but are more pronounced in 2010. Such variable outer line-components are very rare and have been seen only in the rms profiles of a few broad-line radio galaxies, for example, 3C390.3, Arp102B, 3C382, or 3C332 \citep{gezari07}. Mrk\,926 shows a radio flux of $18 \pm 5$\,mJy at 5 Ghz \citep{bicay95}. The optical B-band flux is 6.6 mJy \citep{mcalary83}. Therefore, the radio-to-optical luminosity R (F(5 Ghz)/F(B-band)) has a value of 3. This is above the definition for a radio-quiet AGN (R: $0.1 - 1$) and below the value for a radio-loud AGN (R: 10 - 1000) \citep{kellermann89}. However, the highly variable outer components in the rms profiles shown by \cite{gezari07} are almost exclusively connected to the flanks of the double-peaked mean profiles, which are assumed to originate from an accretion disk. This is not the case for Mrk\,926. Instead, the rms components are further out and their position overlaps with the outer line-wings, but not the line-flanks (see Fig.~\ref{ochmprofmeanrms.pdf}).

\subsection{Central black hole mass and Eddington ratio, L/L$_{edd}$}\label{sec:eddington_ratio}

In Sect.~\ref{sec:bh_mass}, we derived a central BH mass of $1.1 \times 10^8 M_{\odot}$. This is a normal BH mass in the typical range of $10^6 - 10^9 M_{\odot}$  \citep[e.g.,][]{woo02, peterson04}. In the following, we derive the Eddington ratios, $L/L_\text{edd}$, for the low and high intensity state of Mrk\,926 during our variability campaign. The optical luminosity of Mrk\,926 during the low state in November 2010 was by a factor of 2.8 lower with respect to the high state in August 2010.
We  derive bolometric luminosities of $L_\text{bol}(\text{max}) = 1.2 \times 10^{45}$erg\,s$^{-1}$ and $L_\text{bol}(\text{min}) = 4.3 \times 10^{44}$erg\,s$^{-1}$ for the high and low state, respectively, based on the monochromatic luminosity L$_{5100}$ and on an average bolometric correction factor of $f_\text{bol}$ = $L_\text{bol}$/$L_{5100}$ = 10 \citep{kaspi00, castello16}.  Although the boundary between Seyfert galaxies and the higher luminosity quasars is not well defined, a bolometric luminosity of $L_\text{bol} = 10^{45 }$erg\,s$^{-1}$ of the central source is often used as a dividing line to distinguish between these two types \citep{netzer13}. This places Mrk\,926 directly at the dividing line, hence being, depending on the time of observation, either a very strong Seyfert 1 galaxy or a quasar. 

The BH mass of $1.1 \times 10^8 M_{\odot}$ corresponds to an Eddington luminosity of $L_\text{edd}  = 1.4 \times 10^{46}$ erg\,s$^{-1}$. Therefore,  the Eddington ratios,  $L/L_\text{edd}$, for the high and low state have comparatively low values, 8 and 3 percent only. Such low Eddington ratios of a few  percent are typical for highly variable objects. For example, the changing-look AGN IRAS\,23226-3843 showed a low Eddington ratio of only one percent \citep{kollatschny20}. This low Eddington ratio is consistent with investigations of \cite{noda18} and \cite{macleod19}. In particular, \cite{noda18} suggested that all changing-look AGN are associated with state transitions at Eddington ratios of a few percent.

\subsection{Structure and kinematics of the BLR}\label{sec:2D_CCF_discussion}

\subsubsection{CCFs of the central, blue, and red line segments} \label{sec:1D_segments_discussion}

The CCFs of the integrated Balmer line light curves and of the central Balmer line component light curves (within $\pm 5000$\, \kms{}) exhibit two discrete peaks, one at a delay of $\sim 10$ days (\Ha{}) and  $\sim 5$ days (\Hb{}), and the outer at a delay of $\sim 56$ days (\Ha{}) and $\sim 48$ days (\Hb{}). The long-delay peaks of \Ha{} and \Hb{} indicate a stratification of the BLR as the CCFs peak at $57^{+3}_{-3}$ and $48^{+7}_{-6}$ days, respectively. The blue and red Balmer segment light curves (i.e.,\ the light curves of the Balmer satellite components) are single-peaked, with delays of $\sim 4$ days for \Ha{} and $\sim 5$ days for \Hb{}. This means that the outer wings respond promptly to the variations in the continuum.

The HeI\,$\lambda 5876$ line shows no double-peaked CCF.  Instead, the CCF of the central line profile light curve shows a very broad peak with high correlations up to $\sim 40$ days.  This broad peak might be caused by two merging peaks (i.e., by two independently responding components exactly as observed for \Ha{} and \Hb{}) but with relatively broad transfer functions that our campaign was not able to resolve.

\subsubsection{Velocity-resolved CCFs of the Balmer lines}\label{sec:velo_resolved_discussion}
The velocity-resolved CCFs of \Ha{} and \Hb{} (Figs.~\ref{ochm2DCCF_Ha_comb.pdf} and \ref{ochm2DCCF_Hb_comb.pdf}) are, in essence, similar to each other. The delays of the central profiles and of the outer wings are all confined within the virial envelope for a Keplerian disk inclined by $\sim 50^{\circ}$ with a central mass of $1.1 \times 10^{8} M_{\odot}$. Signatures outside of the virial envelopes can be attributed to relatively strong narrow-line residuals (e.g., in the red wing of \Hb{}), to overlapping velocity-resolved line signatures (e.g., of \ion{He}{II}\,$\lambda$ 4686 and \Hb{}), or to strong absorption disturbing the segment light curves (e.g., in the outermost blue wing of \Ha{}).  Both  velocity-resolved CCFs show a broad inner delay region with small time lags of $\sim 3 - 10$\, days. The correlation is strongest in the line wings beyond $\pm 5000$\, \kms{}, exactly where we observed the maximum of the additional, varying  Balmer satellites in the rms profiles. These Balmer satellites respond promptly to the continuum variations with delays of only $\sim 3-5$ days.

The \Ha{} line shows two velocity-delay structures in the central segment (within $\pm 5000$\, \kms{}) at $10^{+3}_{-2}$ and $57^{+3}_{-3}$ light-days. This might be interpreted as the upper and lower half of an ellipse in the velocity-delay plane, which might be the signature of a line-emitting ring orbiting the BH at a radius $R = 33.5$  light-days. This interpretation of the two velocity-delay structures being the signature of a line-emitting ring is analogous to that of the velocity-delay maps of NGC\,5548 based on the STORM campaign \citep{horne21}, and is in turn supported by our finding that the line profiles in Mrk\,926 show the signature of an accretion disk (see Sect.~\ref{sec:mean_rms_profiles_discussion}).
Likewise, the \Hb{} line  shows two velocity-delay structures in the central region at $5^{+2}_{-2}$ and $48^{+7}_{-6}$ light-days, although less pronounced. The possible line-emitting ring orbiting the BH has a radius of $R = 26.5$  light-days.  However, the relative strength of the responses of the near and far side of the ring is different in \Ha{} and \Hb{}. This might be caused by optical depths effects. Moreover, \Ha{} shows a strong discontinuity between the signature of the central line profile and the Balmer satellites, which is less pronounced in the velocity-resolved CCF of \Hb{}. This might be due to interference of \Hb{} with the \ion{He}{ii}\,$\lambda 4686$ line. 

Assuming the two Balmer velocity-delay structures to be the signature of an inclined accretion disk, we determined a mean inclination angle of $i \sim 50^{\circ}$. This rather high inclination angle is supported by the remarkably broad emission lines ($\sim 30\,000$\,\kms{} FWZI) in Mrk\,926.

\subsubsection{Origin of the Balmer satellites}
The existence of the additional, fast-response Balmer satellite components in the rms line profiles and in the velocity-resolved CCFs is an indication for an additional line-emitting component in the  nuclear region that  might not be directly connected to the accretion disk. Because of the clear separation between the central and outer components, their differing variability behavior, the clearly distinct time lags, as well as the outer rms component not being connected to the line-flanks, we propose that the Balmer satellite components may originate in a region that is spatially distinct from the rest of the BLR. For example, the outer wing components might be connected with a different spatial region like a hollow outflow cone with a high tilt angle or small-scale central radio jets. Notably, \citet{ulvestad1984} found a slightly resolved radio source for Mrk\,926 with the VLA at 6\,cm. \citet{mundell00} suggested additional extended radio emission on scales between about 2 and 260\,mas based on observations with the VLBA and VLA and the comparison of their radio beams.

\section{Summary}\label{sec:summary}

We present results of a spectroscopic and photometric monitoring campaign of the very broad-line AGN Mrk\,926, which was carried out with the 10m HET telescope and the Wise Observatory in 2010. Our findings can be summarized as follows:

\begin{enumerate}[(1)]

\item Mrk\,926 is a highly variable AGN. The continuum luminosity showed a drastic decrease during our campaign. It dropped to less than 50\% of its original luminosity within only 2.5 months. 

\item Mrk\,926 shows very broad \Ha{} and \Hb{} line profiles, with additional outer  Balmer satellite components ranging from $\pm 5000$ to $\pm 13\,000$ \kms{} that are  clearly discernible in the rms spectra.

\item The Balmer lines show two velocity-delay structures in their central line component (within $\pm 5000$\, \kms{}), at $\sim 10$ and $\sim 57$ light-days (\Ha{}) and at $\sim 5$ and $\sim 48$ light-days (\Hb{}). These structures might be interpreted as the upper and lower halves of an ellipse in the velocity-delay plane, which might be the signature of a line-emitting ring orbiting the BH at radii, $R$, of 33.5 and 26.5 light-days. 

\item The continuum luminosities $\log( \lambda\,L_\lambda/{\rm erg~s}^{-1})$ at 5180\,\AA{} (rest frame) of 43.68 to 44.13 are in good agreement with the established $R_\text{BLR}-L_\text{AGN}$ relation.

\item The derived BH mass of $1.1 \times 10^8 M_{\odot}$ indicates a low Eddington ratio, which decreased from 8 to 3 percent within only 2.5 months.  

\item  Based on the comparison of the variability behavior of the central line component and the outer line wings, we speculate that the outer emission components (the Balmer satellites) originate in a different, spatially distinct region.

\end{enumerate}

Further densely sampled spectroscopic and photometric studies of this highly variable AGN are desirable.

\begin{acknowledgements}
This work has been supported by the DFG grants KO857/35-1 and CH71/34-3. KH acknowledges support from STFC grant ST/M001296/1. DC acknowledges support from ISF grant 2398/19. This paper is based on observations obtained with the Hobby-Eberly Telescope, which is a joint project of the University of Texas at Austin, the Pennsylvania State University,  Ludwig-Maximilians-Universit\"at M\"unchen, and Georg-August-Universit\"at G\"ottingen.
\end{acknowledgements}

\bibliographystyle{aa} 
\bibliography{literature} 

\appendix
\section{Additional figures}
\begin{minipage}{\textwidth}
\centering
    \begin{minipage}{1.0\textwidth}
    \centering
    \includegraphics[width=0.58\textwidth,angle=-90]{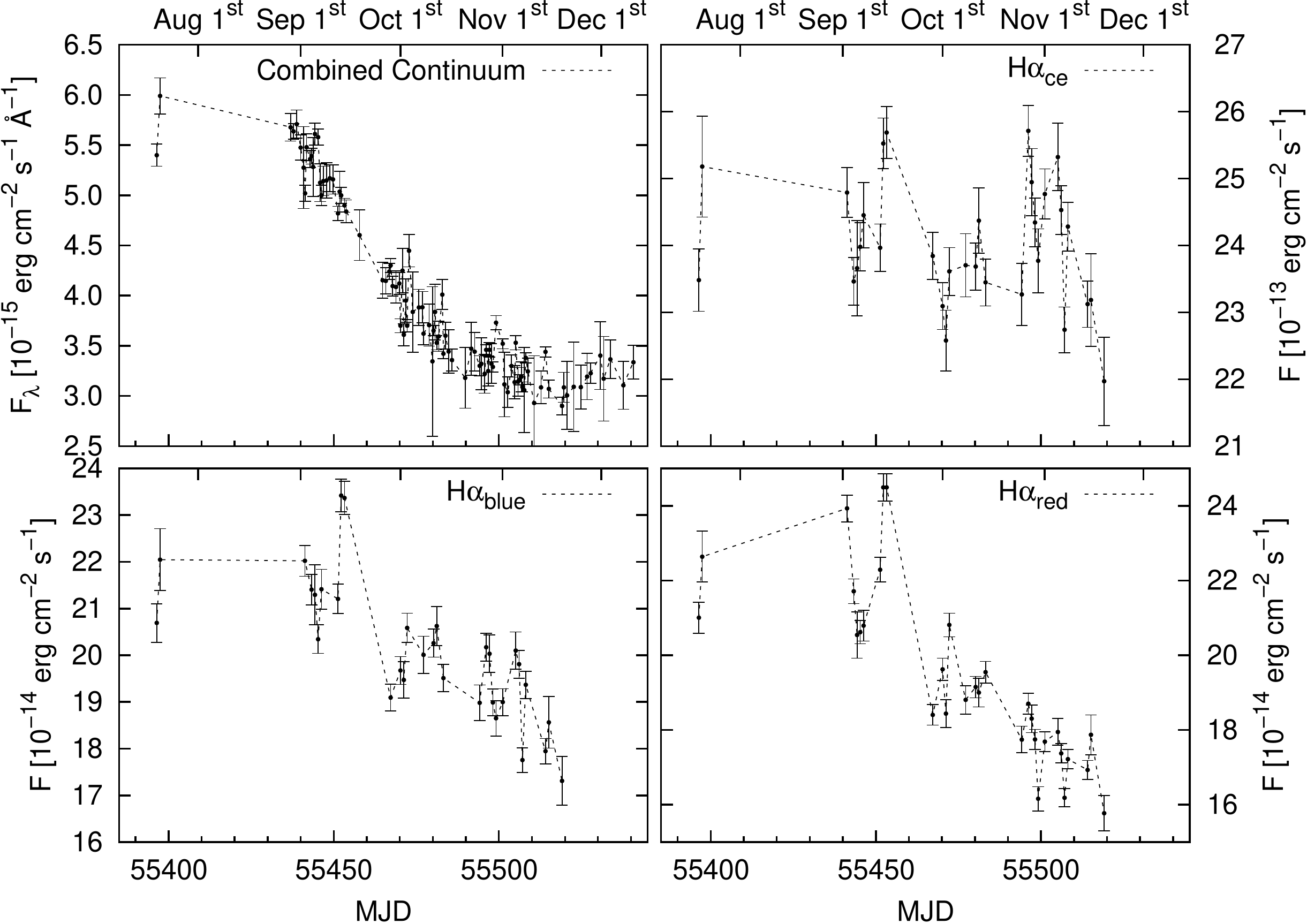}
    \captionof{figure}{Light curves of the combined continuum flux densities at 5420\,\AA\ (observed frame; in units of 10$^{-15}$ erg cm$^{-2}$ s$^{-1}$\,\AA$^{-1}$) as well as of the \Ha{} segment (center, blue, red) light curves.}
   \label{ochmHa_segments_20200929.pdf}
    \end{minipage}  
 \vfill
 \vspace*{0.4cm}
    \begin{minipage}{1.0\textwidth}
    \centering
    \includegraphics[width=0.58\textwidth,angle=-90]{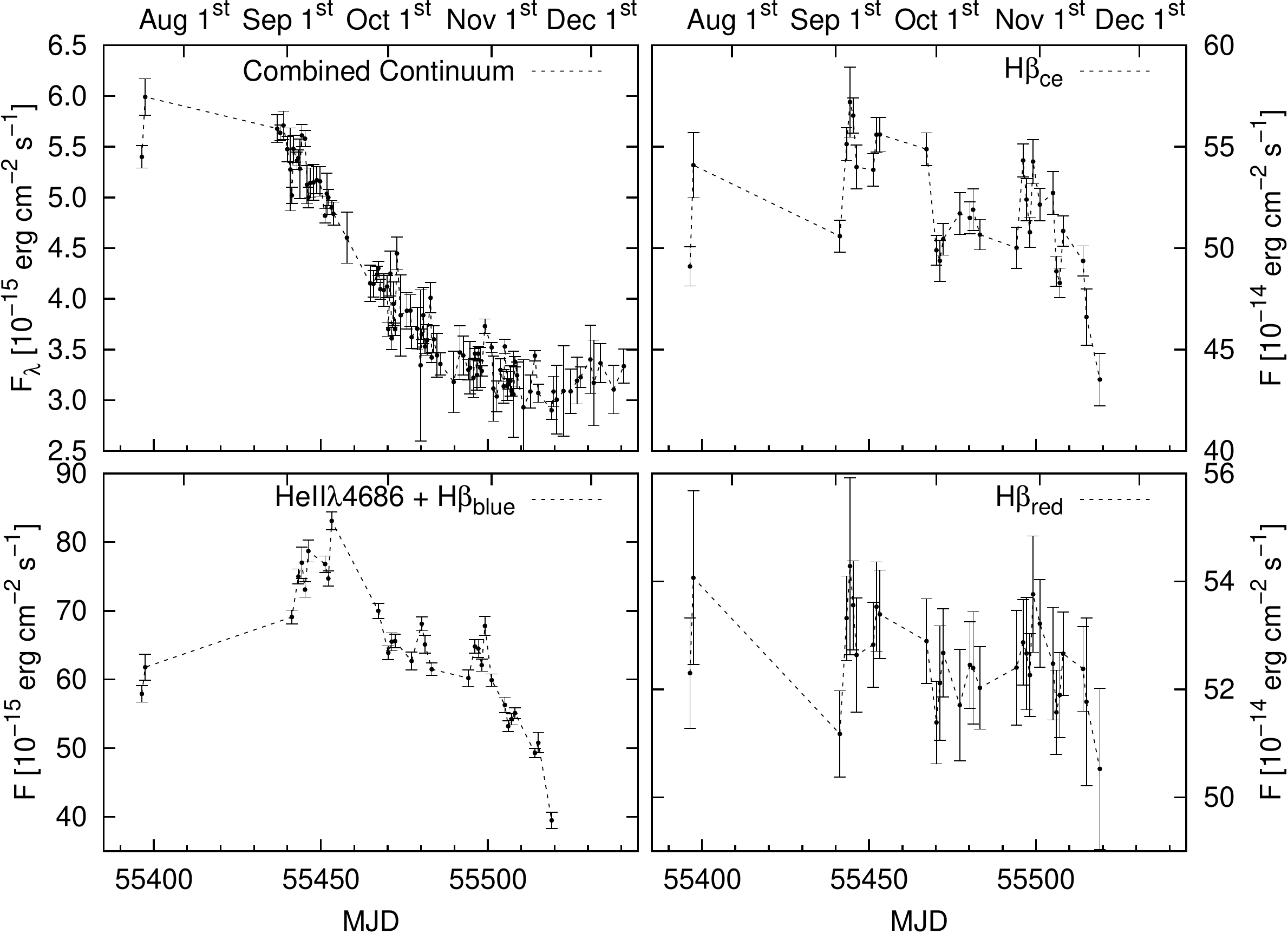}
    \captionof{figure}{Light curves of the combined continuum flux densities at 5420\,\AA\ (observed frame; in units of 10$^{-15}$ erg cm$^{-2}$ s$^{-1}$\,\AA$^{-1}$) as well as of the \Hb{} segment (center, blue, red) light curves.}
   \label{ochmHb_segments_20200929.pdf}
   \end{minipage}
\end{minipage}

\clearpage
\section{Additional tables}
\begin{table}[!htp]
\centering
\tabcolsep1.5mm
\newcolumntype{d}{D{.}{.}{-2}}
\caption{HET continuum flux densities (observed frame).}
\begin{tabular}{cccccccccc}
\noalign{\smallskip}
\hline \hline 
\noalign{\smallskip}
Mod. Julian Date &   \mcr{Cont.~4425\,\AA}    &  \mcr{Cont.~4755\,\AA}    &  \mcr{Cont.~5420\,\AA}   &  \mcr{Cont.~5485\,\AA}   &  \mcr{Cont.~5755\,\AA} &  \mcr{Cont.~6530\,\AA}   &  \mcr{Cont.~7255\,\AA}  \\
(1) & (2) & (3) & (4) & (5) & (6) & (7) & (8)\\
\noalign{\smallskip}
\hline
\noalign{\smallskip}
{ 55396.41} & { 8.12 $\pm$ 0.16} & { 7.07 $\pm$ 0.14} & { 5.40 $\pm$ 0.11} & { 5.39 $\pm$ 0.11}  & { 4.95 $\pm$ 0.10} & { 4.65 $\pm$ 0.09} & { 4.22 $\pm$ 0.08}\\
{ 55397.42} & { 8.73 $\pm$ 0.26} & { 7.73 $\pm$ 0.23} & { 5.99 $\pm$ 0.18} & { 5.94 $\pm$ 0.18}  & { 5.44 $\pm$ 0.16} & { 4.92 $\pm$ 0.15} & { 4.48 $\pm$ 0.13} \\
{ 55441.31} & { 7.49 $\pm$ 0.11} & { 6.61 $\pm$ 0.10} & { 5.02 $\pm$ 0.08} & { 4.99 $\pm$ 0.07}  & { 4.60 $\pm$ 0.07} & { 4.30 $\pm$ 0.06} & { 4.27 $\pm$ 0.06} \\
{ 55443.30} & { 7.44 $\pm$ 0.11} & { 6.85 $\pm$ 0.10} & { 5.39 $\pm$ 0.08} & { 5.41 $\pm$ 0.08}  & { 4.96 $\pm$ 0.07} & { 4.53 $\pm$ 0.07} & { 4.29 $\pm$ 0.06}\\
{ 55444.30} & { 7.73 $\pm$ 0.15} & { 7.04 $\pm$ 0.14} & { 5.61 $\pm$ 0.11} & { 5.66 $\pm$ 0.11}  & { 5.28 $\pm$ 0.11} & { 4.43 $\pm$ 0.09} & { 4.40 $\pm$ 0.09} \\
{ 55445.28} & { 7.76 $\pm$ 0.12} & { 6.99 $\pm$ 0.10} & { 5.58 $\pm$ 0.08} & { 5.62 $\pm$ 0.08}  & { 5.27 $\pm$ 0.08} & { 4.29 $\pm$ 0.06} & { 4.63 $\pm$ 0.07} \\
{ 55446.29} & { 7.43 $\pm$ 0.15} & { 6.59 $\pm$ 0.13} & { 5.00 $\pm$ 0.10} & { 5.00 $\pm$ 0.10}  & { 4.57 $\pm$ 0.09} & { 4.30 $\pm$ 0.09} & { 4.31 $\pm$ 0.09} \\
{ 55451.27} & { 7.09 $\pm$ 0.11} & { 6.26 $\pm$ 0.09} & { 4.82 $\pm$ 0.07} & { 4.84 $\pm$ 0.07}  & { 4.43 $\pm$ 0.07} & { 4.03 $\pm$ 0.06} & { 4.03 $\pm$ 0.06} \\
{ 55452.26} & { 6.25 $\pm$ 0.09} & { 5.95 $\pm$ 0.09} & { 5.00 $\pm$ 0.08} & { 5.03 $\pm$ 0.08}  & { 4.77 $\pm$ 0.07} & { 4.19 $\pm$ 0.06} & { 3.91 $\pm$ 0.06} \\
{ 55453.26} & { 7.04 $\pm$ 0.11} & { 6.19 $\pm$ 0.09} & { 4.90 $\pm$ 0.07} & { 4.90 $\pm$ 0.07}  & { 4.57 $\pm$ 0.07} & { 4.07 $\pm$ 0.06} & { 4.01 $\pm$ 0.06} \\
{ 55467.23} & { 5.41 $\pm$ 0.08} & { 5.03 $\pm$ 0.08} & { 4.30 $\pm$ 0.07} & { 4.37 $\pm$ 0.07}  & { 4.09 $\pm$ 0.06} & { 3.51 $\pm$ 0.05} & { 4.03 $\pm$ 0.06} \\
{ 55470.21} & { 5.32 $\pm$ 0.08} & { 4.69 $\pm$ 0.07} & { 3.70 $\pm$ 0.07} & { 3.73 $\pm$ 0.06}  & { 3.47 $\pm$ 0.05} & { 3.31 $\pm$ 0.05} & { 3.47 $\pm$ 0.05} \\
{ 55471.21} & { 4.98 $\pm$ 0.15} & { 4.48 $\pm$ 0.13} & { 3.61 $\pm$ 0.11} & { 3.65 $\pm$ 0.11}  & { 3.40 $\pm$ 0.10} & { 3.33 $\pm$ 0.10} & { 3.31 $\pm$ 0.10} \\
{ 55472.22} & { 4.57 $\pm$ 0.07} & { 4.34 $\pm$ 0.07} & { 3.70 $\pm$ 0.06} & { 3.77 $\pm$ 0.06}  & { 3.54 $\pm$ 0.05} & { 3.31 $\pm$ 0.05} & { 3.34 $\pm$ 0.05} \\
{ 55477.20} & { 4.83 $\pm$ 0.14} & { 4.46 $\pm$ 0.13} & { 3.62 $\pm$ 0.11} & { 3.67 $\pm$ 0.11}  & { 3.40 $\pm$ 0.10} & { 3.36 $\pm$ 0.10} & { 3.45 $\pm$ 0.10} \\
{ 55480.20} & { 4.40 $\pm$ 0.07} & { 4.23 $\pm$ 0.06} & { 3.65 $\pm$ 0.05} & { 3.74 $\pm$ 0.06}  & { 3.53 $\pm$ 0.05} & { 3.44 $\pm$ 0.05} & { 3.45 $\pm$ 0.05} \\
{ 55481.20} & { 4.59 $\pm$ 0.09} & { 4.21 $\pm$ 0.08} & { 3.53 $\pm$ 0.07} & { 3.56 $\pm$ 0.07}  & { 3.32 $\pm$ 0.07} & { 3.26 $\pm$ 0.07} & { 3.32 $\pm$ 0.07} \\
{ 55483.20} & { 4.76 $\pm$ 0.07} & { 4.23 $\pm$ 0.06} & { 3.42 $\pm$ 0.05} & { 3.45 $\pm$ 0.05}  & { 3.24 $\pm$ 0.05} & { 3.04 $\pm$ 0.05} & { 3.31 $\pm$ 0.05} \\
{ 55494.16} & { 4.12 $\pm$ 0.12} & { 3.78 $\pm$ 0.11} & { 3.30 $\pm$ 0.10} & { 3.35 $\pm$ 0.10}  & { 3.15 $\pm$ 0.09} & { 3.10 $\pm$ 0.09} & { 3.18 $\pm$ 0.10} \\
{ 55496.15} & { 4.23 $\pm$ 0.06} & { 3.98 $\pm$ 0.06} & { 3.46 $\pm$ 0.05} & { 3.51 $\pm$ 0.05}  & { 3.26 $\pm$ 0.05} & { 3.13 $\pm$ 0.05} & { 3.39 $\pm$ 0.05} \\
{ 55497.16} & { 4.25 $\pm$ 0.09} & { 3.95 $\pm$ 0.08} & { 3.46 $\pm$ 0.07} & { 3.50 $\pm$ 0.07}  & { 3.29 $\pm$ 0.07} & { 3.26 $\pm$ 0.06} & { 3.31 $\pm$ 0.07} \\
{ 55498.17} & { 4.28 $\pm$ 0.06} & { 3.88 $\pm$ 0.06} & { 3.29 $\pm$ 0.05} & { 3.35 $\pm$ 0.05}  & { 3.16 $\pm$ 0.05} & { 3.03 $\pm$ 0.05} & { 3.31 $\pm$ 0.05} \\
{ 55499.16} & { 4.36 $\pm$ 0.09} & { 4.18 $\pm$ 0.08} & { 3.73 $\pm$ 0.07} & { 3.84 $\pm$ 0.08}  & { 3.62 $\pm$ 0.07} & { 3.42 $\pm$ 0.07} & { 3.48 $\pm$ 0.07} \\
{ 55501.14} & { 4.07 $\pm$ 0.06} & { 3.89 $\pm$ 0.06} & { 3.52 $\pm$ 0.05} & { 3.60 $\pm$ 0.05}  & { 3.45 $\pm$ 0.05} & { 3.02 $\pm$ 0.05} & { 3.46 $\pm$ 0.05} \\
{ 55505.12} & { 4.26 $\pm$ 0.09} & { 4.03 $\pm$ 0.08} & { 3.53 $\pm$ 0.07} & { 3.57 $\pm$ 0.07}  & { 3.41 $\pm$ 0.07} & { 3.12 $\pm$ 0.06} & { 3.34 $\pm$ 0.07} \\
{ 55506.12} & { 4.05 $\pm$ 0.06} & { 3.72 $\pm$ 0.06} & { 3.16 $\pm$ 0.05} & { 3.18 $\pm$ 0.05}  & { 2.96 $\pm$ 0.04} & { 3.07 $\pm$ 0.05} & { 3.10 $\pm$ 0.05} \\
{ 55507.11} & { 4.27 $\pm$ 0.06} & { 3.81 $\pm$ 0.06} & { 3.09 $\pm$ 0.05} & { 3.14 $\pm$ 0.05}  & { 2.92 $\pm$ 0.04} & { 2.86 $\pm$ 0.04} & { 3.04 $\pm$ 0.05} \\
{ 55508.11} & { 4.13 $\pm$ 0.06} & { 3.92 $\pm$ 0.06} & { 3.38 $\pm$ 0.05} & { 3.46 $\pm$ 0.05}  & { 3.25 $\pm$ 0.05} & { 3.13 $\pm$ 0.05} & { 3.31 $\pm$ 0.05} \\
{ 55514.12} & { 4.26 $\pm$ 0.06} & { 3.93 $\pm$ 0.06} & { 3.44 $\pm$ 0.05} & { 3.52 $\pm$ 0.05}  & { 3.34 $\pm$ 0.05} & { 2.86 $\pm$ 0.04} & { 3.32 $\pm$ 0.05} \\
{ 55515.12} & { 4.01 $\pm$ 0.12} & { 3.57 $\pm$ 0.11} & { 3.07 $\pm$ 0.09} & { 3.12 $\pm$ 0.09}  & { 2.91 $\pm$ 0.09} & { 2.89 $\pm$ 0.09} & { 2.95 $\pm$ 0.09} \\
{ 55519.10} & { 3.93 $\pm$ 0.12} & { 3.52 $\pm$ 0.11} & { 2.90 $\pm$ 0.09} & { 2.96 $\pm$ 0.09}  & { 2.72 $\pm$ 0.08} & { 2.76 $\pm$ 0.08} & { 2.96 $\pm$ 0.09} \\
\noalign{\smallskip}
\hline 
\noalign{\smallskip}
\end{tabular}
\tablefoot{Continuum flux densities in units of 10$^{-15}$\,erg\,s$^{-1}$\,cm$^{-2}$\,\AA$^{-1}$.}
\label{HET_cont_intens}
\end{table}

\begin{table*}
\centering
\tabcolsep4.3mm
\newcolumntype{d}{D{.}{.}{-2}}
\caption{
Integrated broad-line fluxes -- including the narrow components -- of the Balmer and helium lines
for different epochs.}
\begin{tabular}{ccrcrc}
\noalign{\smallskip}
\hline \hline 
\noalign{\smallskip}
Mod. Julian Date &  \mcc{\Ha{}}    &   \mcc{\Hb{}}    &   \mcc{\Hg{}}   &  \mcc{\ion{He}{I}$_\textrm{center}$} & \mcc{\ion{He}{II} + \Hb{}$_{\rm blue}$}   \\
(1) & (2) & \mcc{(3)} & (4) & \mcc{(5)} & (6) \\
\noalign{\smallskip}
\hline
\noalign{\smallskip}
{ 55396.41} & { 2765. $\pm$ 55.} & { 1072. $\pm$ 21.} & { 152.7 $\pm$ 3.1} & 85.4   $\pm$  2.2 &  { 57.9 $\pm$ 1.2} \\
{ 55397.42} & { 2965. $\pm$ 89.} & { 1143. $\pm$ 34.} & { 163.6 $\pm$ 4.9} & 93.0   $\pm$  1.4 &  { 61.8 $\pm$ 1.9} \\
{ 55441.31} & { 2939. $\pm$ 44.} & { 1087. $\pm$ 17.} & { 136.6 $\pm$ 2.1} & 96.6   $\pm$  1.5 &  { 69.1 $\pm$ 1.0} \\ 
{ 55443.30} & { 2777. $\pm$ 42.} & { 1160. $\pm$ 17.} & { 129.6 $\pm$ 1.9} & 105.0  $\pm$  2.7 &  { 75.0 $\pm$ 1.1}\\
{ 55444.30} & { 2784. $\pm$ 84.} & { 1192. $\pm$ 36.} & { 129.7 $\pm$ 3.9} & 115.8  $\pm$  1.8 &  { 77.0 $\pm$ 2.3}\\
{ 55445.28} & { 2808. $\pm$ 42.} & { 1174. $\pm$ 18.} & { 147.3 $\pm$ 2.2} & 109.1  $\pm$  2.8 &  { 73.1 $\pm$ 1.1}\\
{ 55446.29} & { 2867. $\pm$ 57.} & { 1145. $\pm$ 23.} & { 146.6 $\pm$ 2.9} & 93.7   $\pm$  1.5 &  { 78.7 $\pm$ 1.6}\\
{ 55451.27} & { 2832. $\pm$ 42.} & { 1144. $\pm$ 17.} & { 157.6 $\pm$ 2.4} & 98.8   $\pm$  2.0 &  { 76.8 $\pm$ 1.2}\\
{ 55452.26} & { 3032. $\pm$ 45.} & { 1166. $\pm$ 18.} & { 138.2 $\pm$ 2.1} & 132.2  $\pm$  2.0 &  { 74.7 $\pm$ 1.1}\\
{ 55453.26} & { 3048. $\pm$ 46.} & { 1173. $\pm$ 18.} & { 160.6 $\pm$ 2.4} & 121.5  $\pm$  1.9 &  { 83.1 $\pm$ 1.3}\\
{ 55467.23} & { 2759. $\pm$ 41.} & { 1148. $\pm$ 17.} & { 155.5 $\pm$ 2.3} & 104.6  $\pm$  1.6 &  { 70.0 $\pm$ 1.1}\\
{ 55470.21} & { 2702. $\pm$ 41.} & { 1077. $\pm$ 16.} & { 152.4 $\pm$ 2.3} & 92.4   $\pm$  1.4 &  { 63.9 $\pm$ 1.0}\\
{ 55471.21} & { 2637. $\pm$ 53.} & { 1080. $\pm$ 22.} & { 156.6 $\pm$ 3.1} & 99.4   $\pm$  1.5 &  { 65.5 $\pm$ 1.3}\\
{ 55472.22} & { 2775. $\pm$ 42.} & { 1097. $\pm$ 17.} & { 155.0 $\pm$ 2.3} & 106.2  $\pm$  1.6 &  { 65.6 $\pm$ 1.0}\\
{ 55477.20} & { 2759. $\pm$ 55.} & { 1104. $\pm$ 22.} & { 163.5 $\pm$ 3.3} & 98.8   $\pm$  2.0 &  { 62.7 $\pm$ 1.3}\\
{ 55480.20} & { 2763. $\pm$ 41.} & { 1115. $\pm$ 17.} & { 164.4 $\pm$ 2.5} & 105.5  $\pm$  1.6 &  { 68.1 $\pm$ 1.0}\\
{ 55481.20} & { 2834. $\pm$ 57.} & { 1108. $\pm$ 22.} & { 173.4 $\pm$ 3.5} & 101.3  $\pm$  1.6 &  { 65.1 $\pm$ 1.3}\\
{ 55483.20} & { 2735. $\pm$ 41.} & { 1088. $\pm$ 16.} & { 178.4 $\pm$ 2.7} & 93.7   $\pm$  1.5 &  { 61.5 $\pm$ 0.9}\\
{ 55494.16} & { 2694. $\pm$ 54.} & { 1084. $\pm$ 22.} & { 164.6 $\pm$ 3.3} & 89.8   $\pm$  1.8 &  { 60.2 $\pm$ 1.2}\\
{ 55496.15} & { 2960. $\pm$ 44.} & { 1137. $\pm$ 17.} & { 183.5 $\pm$ 2.8} & 92.2   $\pm$  1.4 &  { 64.8 $\pm$ 1.0}\\
{ 55497.16} & { 2878. $\pm$ 58.} & { 1115. $\pm$ 22.} & { 184.7 $\pm$ 3.7} & 92.7   $\pm$  1.9 &  { 64.5 $\pm$ 1.3} \\
{ 55498.17} & { 2802. $\pm$ 42.} & { 1093. $\pm$ 16.} & { 187.8 $\pm$ 2.8} & 85.0   $\pm$  1.7 &  { 62.1 $\pm$ 0.9}\\
{ 55499.16} & { 2725. $\pm$ 55.} & { 1148. $\pm$ 23.} & { 192.1 $\pm$ 3.8} & 93.1   $\pm$  2.4 &  { 67.8 $\pm$ 1.4}\\
{ 55501.14} & { 2844. $\pm$ 43.} & { 1114. $\pm$ 17.} & { 181.2 $\pm$ 2.7} & 92.3   $\pm$  1.4 &  { 59.9 $\pm$ 0.9}\\
{ 55505.12} & { 2913. $\pm$ 58.} & { 1108. $\pm$ 22.} & { 190.3 $\pm$ 3.8} & 94.7   $\pm$  2.4 &  { 56.3 $\pm$ 1.1}\\
{ 55506.12} & { 2825. $\pm$ 42.} & { 1058. $\pm$ 16.} & { 166.1 $\pm$ 2.5} & 85.1   $\pm$  1.3 &  { 53.2 $\pm$ 0.8}\\
{ 55507.11} & { 2613. $\pm$ 39.} & { 1056. $\pm$ 16.} & { 180.3 $\pm$ 2.7} & 73.1   $\pm$  1.1 &  { 54.2 $\pm$ 0.8}\\
{ 55508.11} & { 2794. $\pm$ 42.} & { 1090. $\pm$ 16.} & { 177.6 $\pm$ 2.7} & 86.3   $\pm$  1.3 &   { 55.1 $\pm$ 0.8}\\
{ 55514.12} & { 2661. $\pm$ 40.} & { 1067. $\pm$ 16.} & { 159.9 $\pm$ 2.4} & 85.8   $\pm$  1.3 &   { 49.3 $\pm$ 0.7}\\
{ 55515.12} & { 2683. $\pm$ 80.} & { 1035. $\pm$ 31.} & { 148.3 $\pm$ 4.4} & 79.7   $\pm$  1.2 &   { 50.8 $\pm$ 1.5}\\
{ 55519.10} & { 2528. $\pm$ 76.} & { 980. $\pm$ 29.}  & { 143.8 $\pm$ 4.3} & 66.8   $\pm$  1.0 &   { 39.5 $\pm$ 1.2}\\
\noalign{\smallskip}
\hline 
\noalign{\smallskip}
\end{tabular}
\tablefoot{
Line fluxes in units of 10$^{-15}$\,erg\,s$^{-1}$\,cm$^{-2}$.
}
\label{em_integline_intens}
\end{table*}

\begin{sidewaystable*}
\newcolumntype{d}{D{.}{.}{-2}}
\caption{HET continuum light curves used for the FVG method (observed frame).}
\resizebox{\textwidth}{!}{
\begin{tabular}{ccccccccccccc}
\noalign{\smallskip}
\hline \hline 
\noalign{\smallskip}
Mod. &   \mcr{Cont.~4425\,\AA}   &  \mcr{Cont.~4425\,\AA}     &  \mcr{Cont.~4425\,\AA} &  \mcr{Cont.~4425\,\AA}   &  \mcr{Cont.~5420\,\AA}    &  \mcr{Cont.~5420\,\AA}  &  \mcr{Cont.~5420\,\AA} &  \mcr{Cont.~5420\,\AA}  &  \mcr{Cont.~6530\,\AA}    &  \mcr{Cont.~6530\,\AA}  &  \mcr{Cont.~6530\,\AA} &  \mcr{Cont.~6530\,\AA}  \\
Julian Date&    &       &       [mJy]   & [mJy] & &     &       [mJy]   & [mJy] & &       &       [mJy]   & [mJy] \\&     &       ext. corr.      &       &       ext. corr.   &       &       ext. corr.      &       &       ext. corr.      &       &       ext. corr.   &       &       ext. corr.\\
(1) & (2) & (3) & (4) & (5) & (6) & (7) & (8) & (9) & (10) & (11) & (12) & (13)\\
\noalign{\smallskip}
\hline
\noalign{\smallskip}
55396.41        &       8.12    $\pm$   0.16    &       9.31    $\pm$   0.19    &       5.30    $\pm$   0.11    &       6.08    $\pm$   0.12    &       5.40    $\pm$   0.11    &       5.99    $\pm$   0.13    &       5.29    $\pm$   0.11    &       5.87    $\pm$   0.12    &       4.65    $\pm$   0.09    &       5.05    $\pm$   0.10    &       6.61    $\pm$   0.13    &       7.18    $\pm$   0.14    \\
55397.42        &       8.73    $\pm$   0.26    &       10.01   $\pm$   0.30    &       5.70    $\pm$   0.17    &       6.54    $\pm$   0.20    &       5.99    $\pm$   0.18    &       6.65    $\pm$   0.20    &       5.87    $\pm$   0.18    &       6.51    $\pm$   0.20    &       4.92    $\pm$   0.15    &       5.34    $\pm$   0.17    &       7.00    $\pm$   0.22    &       7.60    $\pm$   0.24    \\
55441.31        &       7.49    $\pm$   0.11    &       8.59    $\pm$   0.13    &       4.89    $\pm$   0.08    &       5.61    $\pm$   0.09    &       5.02    $\pm$   0.08    &       5.57    $\pm$   0.09    &       4.92    $\pm$   0.08    &       5.46    $\pm$   0.09    &       4.30    $\pm$   0.06    &       4.67    $\pm$   0.07    &       6.12    $\pm$   0.09    &       6.64    $\pm$   0.10    \\
55443.30        &       7.44    $\pm$   0.11    &       8.53    $\pm$   0.13    &       4.86    $\pm$   0.08    &       5.57    $\pm$   0.09    &       5.39    $\pm$   0.08    &       5.98    $\pm$   0.09    &       5.28    $\pm$   0.08    &       5.86    $\pm$   0.09    &       4.53    $\pm$   0.07    &       4.92    $\pm$   0.08    &       6.44    $\pm$   0.10    &       6.99    $\pm$   0.11    \\
55444.30        &       7.73    $\pm$   0.15    &       8.87    $\pm$   0.18    &       5.05    $\pm$   0.10    &       5.79    $\pm$   0.12    &       5.61    $\pm$   0.11    &       6.23    $\pm$   0.13    &       5.50    $\pm$   0.11    &       6.10    $\pm$   0.12    &       4.43    $\pm$   0.09    &       4.81    $\pm$   0.10    &       6.30    $\pm$   0.13    &       6.84    $\pm$   0.14    \\
55445.28        &       7.76    $\pm$   0.12    &       8.90    $\pm$   0.14    &       5.07    $\pm$   0.08    &       5.81    $\pm$   0.09    &       5.58    $\pm$   0.08    &       6.19    $\pm$   0.09    &       5.47    $\pm$   0.08    &       6.07    $\pm$   0.09    &       4.29    $\pm$   0.06    &       4.66    $\pm$   0.07    &       6.10    $\pm$   0.09    &       6.62    $\pm$   0.10    \\
55446.29        &       7.43    $\pm$   0.15    &       8.52    $\pm$   0.18    &       4.85    $\pm$   0.10    &       5.57    $\pm$   0.12    &       5.00    $\pm$   0.10    &       5.55    $\pm$   0.12    &       4.90    $\pm$   0.10    &       5.44    $\pm$   0.11    &       4.30    $\pm$   0.09    &       4.67    $\pm$   0.10    &       6.12    $\pm$   0.13    &       6.64    $\pm$   0.14    \\
55451.27        &       7.09    $\pm$   0.11    &       8.13    $\pm$   0.13    &       4.63    $\pm$   0.08    &       5.31    $\pm$   0.09    &       4.82    $\pm$   0.07    &       5.35    $\pm$   0.08    &       4.72    $\pm$   0.07    &       5.24    $\pm$   0.08    &       4.03    $\pm$   0.06    &       4.37    $\pm$   0.07    &       5.73    $\pm$   0.09    &       6.22    $\pm$   0.10    \\
55452.26        &       6.25    $\pm$   0.09    &       7.17    $\pm$   0.11    &       4.08    $\pm$   0.06    &       4.68    $\pm$   0.07    &       5.00    $\pm$   0.08    &       5.55    $\pm$   0.09    &       4.90    $\pm$   0.08    &       5.44    $\pm$   0.09    &       4.19    $\pm$   0.06    &       4.55    $\pm$   0.07    &       5.96    $\pm$   0.09    &       6.47    $\pm$   0.10    \\
55453.26        &       7.04    $\pm$   0.11    &       8.08    $\pm$   0.13    &       4.60    $\pm$   0.08    &       5.27    $\pm$   0.09    &       4.90    $\pm$   0.07    &       5.44    $\pm$   0.08    &       4.80    $\pm$   0.07    &       5.33    $\pm$   0.08    &       4.07    $\pm$   0.06    &       4.42    $\pm$   0.07    &       5.79    $\pm$   0.09    &       6.28    $\pm$   0.10    \\
55467.23        &       5.41    $\pm$   0.08    &       6.21    $\pm$   0.10    &       3.53    $\pm$   0.06    &       4.05    $\pm$   0.06    &       4.30    $\pm$   0.07    &       4.77    $\pm$   0.08    &       4.21    $\pm$   0.07    &       4.68    $\pm$   0.08    &       3.51    $\pm$   0.05    &       3.81    $\pm$   0.06    &       4.99    $\pm$   0.08    &       5.42    $\pm$   0.08    \\
55470.21        &       5.32    $\pm$   0.08    &       6.10    $\pm$   0.10    &       3.47    $\pm$   0.06    &       3.99    $\pm$   0.06    &       3.70    $\pm$   0.07    &       4.11    $\pm$   0.08    &       3.63    $\pm$   0.07    &       4.02    $\pm$   0.08    &       3.31    $\pm$   0.05    &       3.59    $\pm$   0.06    &       4.71    $\pm$   0.08    &       5.11    $\pm$   0.08    \\
55471.21        &       4.98    $\pm$   0.15    &       5.71    $\pm$   0.18    &       3.25    $\pm$   0.10    &       3.73    $\pm$   0.12    &       3.61    $\pm$   0.11    &       4.01    $\pm$   0.13    &       3.54    $\pm$   0.11    &       3.93    $\pm$   0.12    &       3.33    $\pm$   0.10    &       3.61    $\pm$   0.11    &       4.74    $\pm$   0.15    &       5.14    $\pm$   0.16    \\
55472.22        &       4.57    $\pm$   0.07    &       5.24    $\pm$   0.09    &       2.98    $\pm$   0.05    &       3.42    $\pm$   0.06    &       3.70    $\pm$   0.06    &       4.11    $\pm$   0.07    &       3.63    $\pm$   0.06    &       4.02    $\pm$   0.07    &       3.31    $\pm$   0.05    &       3.59    $\pm$   0.06    &       4.71    $\pm$   0.08    &       5.11    $\pm$   0.08    \\
55477.20        &       4.83    $\pm$   0.14    &       5.54    $\pm$   0.17    &       3.15    $\pm$   0.10    &       3.62    $\pm$   0.11    &       3.62    $\pm$   0.11    &       4.02    $\pm$   0.13    &       3.55    $\pm$   0.11    &       3.94    $\pm$   0.12    &       3.36    $\pm$   0.10    &       3.65    $\pm$   0.11    &       4.78    $\pm$   0.15    &       5.19    $\pm$   0.16    \\
55480.20        &       4.40    $\pm$   0.07    &       5.05    $\pm$   0.09    &       2.87    $\pm$   0.05    &       3.30    $\pm$   0.06    &       3.65    $\pm$   0.05    &       4.05    $\pm$   0.06    &       3.58    $\pm$   0.05    &       3.97    $\pm$   0.06    &       3.44    $\pm$   0.05    &       3.73    $\pm$   0.06    &       4.89    $\pm$   0.08    &       5.31    $\pm$   0.08    \\
55481.20        &       4.59    $\pm$   0.09    &       5.27    $\pm$   0.11    &       3.00    $\pm$   0.06    &       3.44    $\pm$   0.07    &       3.53    $\pm$   0.07    &       3.92    $\pm$   0.08    &       3.46    $\pm$   0.07    &       3.84    $\pm$   0.08    &       3.26    $\pm$   0.07    &       3.54    $\pm$   0.08    &       4.64    $\pm$   0.10    &       5.03    $\pm$   0.11    \\
55483.20        &       4.76    $\pm$   0.07    &       5.46    $\pm$   0.09    &       3.11    $\pm$   0.05    &       3.57    $\pm$   0.06    &       3.42    $\pm$   0.05    &       3.80    $\pm$   0.06    &       3.35    $\pm$   0.05    &       3.72    $\pm$   0.06    &       3.04    $\pm$   0.05    &       3.30    $\pm$   0.06    &       4.32    $\pm$   0.08    &       4.69    $\pm$   0.08    \\
55494.16        &       4.12    $\pm$   0.12    &       4.73    $\pm$   0.14    &       2.69    $\pm$   0.08    &       3.09    $\pm$   0.09    &       3.30    $\pm$   0.10    &       3.66    $\pm$   0.12    &       3.23    $\pm$   0.10    &       3.59    $\pm$   0.11    &       3.10    $\pm$   0.09    &       3.36    $\pm$   0.10    &       4.41    $\pm$   0.13    &       4.79    $\pm$   0.14    \\
55496.15        &       4.23    $\pm$   0.06    &       4.85    $\pm$   0.07    &       2.76    $\pm$   0.04    &       3.17    $\pm$   0.05    &       3.46    $\pm$   0.05    &       3.84    $\pm$   0.06    &       3.39    $\pm$   0.05    &       3.76    $\pm$   0.06    &       3.13    $\pm$   0.05    &       3.40    $\pm$   0.06    &       4.45    $\pm$   0.08    &       4.83    $\pm$   0.08    \\
55497.16        &       4.25    $\pm$   0.09    &       4.88    $\pm$   0.11    &       2.78    $\pm$   0.06    &       3.18    $\pm$   0.07    &       3.46    $\pm$   0.07    &       3.84    $\pm$   0.08    &       3.39    $\pm$   0.07    &       3.76    $\pm$   0.08    &       3.26    $\pm$   0.06    &       3.54    $\pm$   0.07    &       4.64    $\pm$   0.09    &       5.03    $\pm$   0.10    \\
55498.17        &       4.28    $\pm$   0.06    &       4.91    $\pm$   0.07    &       2.80    $\pm$   0.04    &       3.21    $\pm$   0.05    &       3.29    $\pm$   0.05    &       3.65    $\pm$   0.06    &       3.22    $\pm$   0.05    &       3.58    $\pm$   0.06    &       3.03    $\pm$   0.05    &       3.29    $\pm$   0.06    &       4.31    $\pm$   0.08    &       4.68    $\pm$   0.08    \\
55499.16        &       4.36    $\pm$   0.09    &       5.00    $\pm$   0.11    &       2.85    $\pm$   0.06    &       3.27    $\pm$   0.07    &       3.73    $\pm$   0.07    &       4.14    $\pm$   0.08    &       3.65    $\pm$   0.07    &       4.06    $\pm$   0.08    &       3.42    $\pm$   0.07    &       3.71    $\pm$   0.08    &       4.86    $\pm$   0.10    &       5.28    $\pm$   0.11    \\
55501.14        &       4.07    $\pm$   0.06    &       4.67    $\pm$   0.07    &       2.66    $\pm$   0.04    &       3.05    $\pm$   0.05    &       3.52    $\pm$   0.05    &       3.91    $\pm$   0.06    &       3.45    $\pm$   0.05    &       3.83    $\pm$   0.06    &       3.02    $\pm$   0.05    &       3.28    $\pm$   0.06    &       4.30    $\pm$   0.08    &       4.66    $\pm$   0.08    \\
55505.12        &       4.26    $\pm$   0.09    &       4.89    $\pm$   0.11    &       2.78    $\pm$   0.06    &       3.19    $\pm$   0.07    &       3.53    $\pm$   0.07    &       3.92    $\pm$   0.08    &       3.46    $\pm$   0.07    &       3.84    $\pm$   0.08    &       3.12    $\pm$   0.06    &       3.39    $\pm$   0.07    &       4.44    $\pm$   0.09    &       4.82    $\pm$   0.10    \\
55506.12        &       4.05    $\pm$   0.06    &       4.65    $\pm$   0.07    &       2.65    $\pm$   0.04    &       3.03    $\pm$   0.05    &       3.16    $\pm$   0.05    &       3.51    $\pm$   0.06    &       3.10    $\pm$   0.05    &       3.44    $\pm$   0.06    &       3.07    $\pm$   0.05    &       3.33    $\pm$   0.06    &       4.37    $\pm$   0.08    &       4.74    $\pm$   0.08    \\
55507.11        &       4.27    $\pm$   0.06    &       4.90    $\pm$   0.07    &       2.79    $\pm$   0.04    &       3.20    $\pm$   0.05    &       3.09    $\pm$   0.05    &       3.43    $\pm$   0.06    &       3.03    $\pm$   0.05    &       3.36    $\pm$   0.06    &       2.86    $\pm$   0.04    &       3.10    $\pm$   0.05    &       4.07    $\pm$   0.06    &       4.42    $\pm$   0.07    \\
55508.11        &       4.13    $\pm$   0.06    &       4.74    $\pm$   0.07    &       2.70    $\pm$   0.04    &       3.09    $\pm$   0.05    &       3.38    $\pm$   0.05    &       3.75    $\pm$   0.06    &       3.31    $\pm$   0.05    &       3.68    $\pm$   0.06    &       3.13    $\pm$   0.05    &       3.40    $\pm$   0.06    &       4.45    $\pm$   0.08    &       4.83    $\pm$   0.08    \\
55514.12        &       4.26    $\pm$   0.06    &       4.89    $\pm$   0.07    &       2.78    $\pm$   0.04    &       3.19    $\pm$   0.05    &       3.44    $\pm$   0.05    &       3.82    $\pm$   0.06    &       3.37    $\pm$   0.05    &       3.74    $\pm$   0.06    &       2.86    $\pm$   0.04    &       3.10    $\pm$   0.05    &       4.07    $\pm$   0.06    &       4.42    $\pm$   0.07    \\
55515.12        &       4.01    $\pm$   0.12    &       4.60    $\pm$   0.14    &       2.62    $\pm$   0.08    &       3.00    $\pm$   0.09    &       3.07    $\pm$   0.09    &       3.41    $\pm$   0.10    &       3.01    $\pm$   0.09    &       3.34    $\pm$   0.10    &       2.89    $\pm$   0.09    &       3.14    $\pm$   0.10    &       4.11    $\pm$   0.13    &       4.46    $\pm$   0.14    \\
55519.10        &       3.93    $\pm$   0.12    &       4.51    $\pm$   0.14    &       2.57    $\pm$   0.08    &       2.94    $\pm$   0.09    &       2.90    $\pm$   0.09    &       3.22    $\pm$   0.10    &       2.84    $\pm$   0.09    &       3.15    $\pm$   0.10    &       2.76    $\pm$   0.08    &       3.00    $\pm$   0.09    &       3.93    $\pm$   0.12    &       4.26    $\pm$   0.13    \\

\noalign{\smallskip}
\hline 
\noalign{\smallskip}
\end{tabular}}
\tablefoot{
Unless stated otherwise, the flux is given in 10$^{-15}$\,erg\,s$^{-1}$\,cm$^{-2}$\,\AA$^{-1}$. In order to calculate the light curves corrected for Galactic extinction, we adopted Galactic extinction values of $A_{\rm B}=0.149$, $A_{\rm V}=0.113$, and $A_{\rm R}=0.089$ from \citet{schlafly11} (NED).\\
}
\label{tab:extinct_corr_continua}
\end{sidewaystable*}

\begin{table*}
\newcolumntype{d}{D{.}{.}{-2}}
\caption{Combined HET 5420\,\AA\ (5180\,\AA{} in rest frame) and Wise Observatory V-band light curve.}

\resizebox{\textwidth}{!}{
\begin{tabular}{ccc|ccc}
\noalign{\smallskip}
\hline \hline 
\noalign{\smallskip}

Mod. &   Comb. V band   &  \mcr{Telescope} & Mod. &   Comb. V band   &  \mcr{Telescope}    \\
Julian Date&    [10$^{-15}$\,erg\,s$^{-1}$\,cm$^{-2}$\,\AA$^{-1}$]      &       & Julian Date&    [10$^{-15}$\,erg\,s$^{-1}$\,cm$^{-2}$\,\AA$^{-1}$]      &       \\
(1) & (2) & (3) & (1) & (2) & (3)\\
\noalign{\smallskip}
\hline
\noalign{\smallskip}
55396.41        &       5.40    $\pm$   0.11    &       HET     &       55481.80        &       3.59    $\pm$   0.15    &       Wise    \\
55397.42        &       5.99    $\pm$   0.18    &       HET     &       55482.85        &       4.01    $\pm$   0.15    &       Wise    \\
55436.92        &       5.68    $\pm$   0.14    &       Wise    &       55483.20        &       3.42    $\pm$   0.05    &       HET     \\
55437.86        &       5.64    $\pm$   0.08    &       Wise    &       55483.82        &       3.60    $\pm$   0.13    &       Wise    \\
55438.80        &       5.71    $\pm$   0.14    &       Wise    &       55484.80        &       3.44    $\pm$   0.21    &       Wise    \\
55439.96        &       5.48    $\pm$   0.13    &       Wise    &       55485.81        &       3.36    $\pm$   0.11    &       Wise    \\
55440.90        &       5.28    $\pm$   0.41    &       Wise    &       55489.79        &       3.18    $\pm$   0.30    &       Wise    \\
55441.31        &       5.02    $\pm$   0.08    &       HET     &       55491.66        &       3.47    $\pm$   0.26    &       Wise    \\
55441.80        &       5.48    $\pm$   0.14    &       Wise    &       55492.67        &       3.44    $\pm$   0.19    &       Wise    \\
55442.88        &       5.36    $\pm$   0.09    &       Wise    &       55494.16        &       3.30    $\pm$   0.10    &       HET     \\
55443.30        &       5.39    $\pm$   0.08    &       HET     &       55494.67        &       3.32    $\pm$   0.26    &       Wise    \\
55443.78        &       5.28    $\pm$   0.29    &       Wise    &       55495.68        &       3.22    $\pm$   0.19    &       Wise    \\
55444.30        &       5.61    $\pm$   0.11    &       HET     &       55496.15        &       3.46    $\pm$   0.05    &       HET     \\
55445.28        &       5.58    $\pm$   0.08    &       HET     &       55496.75        &       3.25    $\pm$   0.15    &       Wise    \\
55445.92        &       5.13    $\pm$   0.19    &       Wise    &       55497.16        &       3.46    $\pm$   0.07    &       HET     \\
55446.29        &       5.00    $\pm$   0.10    &       HET     &       55497.73        &       3.32    $\pm$   0.20    &       Wise    \\
55446.84        &       5.14    $\pm$   0.15    &       Wise    &       55498.17        &       3.29    $\pm$   0.05    &       HET     \\
55447.81        &       5.15    $\pm$   0.18    &       Wise    &       55499.16        &       3.73    $\pm$   0.07    &       HET     \\
55448.81        &       5.17    $\pm$   0.13    &       Wise    &       55501.14        &       3.52    $\pm$   0.05    &       HET     \\
55449.78        &       5.16    $\pm$   0.15    &       Wise    &       55501.69        &       3.11    $\pm$   0.32    &       Wise    \\
55451.27        &       4.82    $\pm$   0.07    &       HET     &       55502.71        &       3.04    $\pm$   0.15    &       Wise    \\
55451.79        &       5.04    $\pm$   0.20    &       Wise    &       55503.79        &       3.30    $\pm$   0.11    &       Wise    \\
55452.26        &       5.00    $\pm$   0.08    &       HET     &       55504.79        &       3.14    $\pm$   0.17    &       Wise    \\
55453.26        &       4.90    $\pm$   0.07    &       HET     &       55505.12        &       3.53    $\pm$   0.07    &       HET     \\
55453.78        &       4.84    $\pm$   0.11    &       Wise    &       55505.76        &       3.14    $\pm$   0.14    &       Wise    \\
55457.80        &       4.60    $\pm$   0.25    &       Wise    &       55506.12        &       3.16    $\pm$   0.05    &       HET     \\
55464.79        &       4.15    $\pm$   0.18    &       Wise    &       55506.80        &       3.19    $\pm$   0.15    &       Wise    \\
55465.76        &       4.15    $\pm$   0.13    &       Wise    &       55507.11        &       3.09    $\pm$   0.05    &       HET     \\
55466.85        &       4.24    $\pm$   0.08    &       Wise    &       55507.72        &       3.06    $\pm$   0.43    &       Wise    \\
55467.23        &       4.30    $\pm$   0.07    &       HET     &       55508.11        &       3.38    $\pm$   0.05    &       HET     \\
55467.84        &       4.09    $\pm$   0.10    &       Wise    &       55508.77        &       3.25    $\pm$   0.13    &       Wise    \\
55468.83        &       4.09    $\pm$   0.16    &       Wise    &       55510.66        &       2.93    $\pm$   0.47    &       Wise    \\
55469.84        &       4.12    $\pm$   0.11    &       Wise    &       55512.75        &       3.09    $\pm$   0.16    &       Wise    \\
55470.21        &       3.70    $\pm$   0.07    &       HET     &       55514.12        &       3.44    $\pm$   0.05    &       HET     \\
55470.81        &       4.25    $\pm$   0.22    &       Wise    &       55515.12        &       3.07    $\pm$   0.09    &       HET     \\
55471.21        &       3.61    $\pm$   0.11    &       HET     &       55519.10        &       2.90    $\pm$   0.09    &       HET     \\
55471.84        &       3.95    $\pm$   0.22    &       Wise    &       55519.69        &       3.09    $\pm$   0.15    &       Wise    \\
55472.22        &       3.70    $\pm$   0.06    &       HET     &       55520.65        &       3.01    $\pm$   0.34    &       Wise    \\
55472.80        &       4.45    $\pm$   0.16    &       Wise    &       55522.67        &       3.09    $\pm$   0.45    &       Wise    \\
55473.92        &       3.84    $\pm$   0.40    &       Wise    &       55524.79        &       3.09    $\pm$   0.22    &       Wise    \\
55475.76        &       3.88    $\pm$   0.18    &       Wise    &       55526.71        &       3.19    $\pm$   0.23    &       Wise    \\
55476.85        &       3.88    $\pm$   0.16    &       Wise    &       55527.76        &       3.23    $\pm$   0.10    &       Wise    \\
55477.20        &       3.62    $\pm$   0.11    &       HET     &       55530.71        &       3.40    $\pm$   0.34    &       Wise    \\
55478.89        &       3.71    $\pm$   0.21    &       Wise    &       55531.70        &       3.17    $\pm$   0.42    &       Wise    \\
55479.88        &       3.34    $\pm$   0.75    &       Wise    &       55533.75        &       3.37    $\pm$   0.19    &       Wise    \\
55480.20        &       3.65    $\pm$   0.05    &       HET     &       55537.67        &       3.11    $\pm$   0.24    &       Wise    \\
55480.69        &       3.84    $\pm$   0.28    &       Wise    &       55540.72        &       3.34    $\pm$   0.17    &       Wise    \\
55481.20        &       3.53    $\pm$   0.07    &       HET     &               &                               &               \\
\noalign{\smallskip}
\hline 
\noalign{\smallskip}
\end{tabular}}
\label{tab:combined_continuum}
\end{table*}

\end{document}